\documentclass[fleqn,usenatbib,a4paper]{mnras}

\usepackage{amsmath}
\usepackage{amssymb}

\usepackage{mathptmx}
\usepackage{txfonts}

\usepackage[T1]{fontenc}
\usepackage{ae,aecompl}

\usepackage{graphicx}	
\usepackage{pdflscape}
\usepackage{breakurl}

\newcommand{\kms}{$\text{km}\:\text{s}^{-1}\,$}
\newcommand{\lam}{$\lambda$}

\newcommand{\caii}{[Ca\,{\sc ii}]\ }
\newcommand{\oi}{[O\,{\sc i}]\ }

\newcommand{\caiinos}{[Ca\,{\sc ii}]}  
\newcommand{\oinos}{[O\,{\sc i}]}      

\title[Resolving the kinematics of the disks around Galactic {B[e]} supergiants]
{Resolving the kinematics of the disks around Galactic B[e] supergiants}

\author[G. Maravelias et al.]{
G.~Maravelias$^{1,2}$\thanks{E-mail: 
grigorios.maravelias@uv.cl}, 
M.~Kraus$^{2,3}$, 
L.~S.~Cidale$^{4,5,1}$, 
M.~Borges Fernandes$^{6}$, \newauthor 
M.~L.~Arias$^{4,5}$, 
M.~Cur\'e$^1$,
G.~Vasilopoulos$^7$
\\ 
$^1$ Instituto de F\'isica y Astronom\'ia, Facultad de Ciencias, Universidad de Valpara\'iso, Av. Gran Breta\~na 1111, \\ Casilla 5030, Valpara\'iso, Chile \\
$^2$ Astronomick\'y \'ustav, Akademie v\v{e}d \v{C}esk\'e republiky, Fri\v{c}ova 298, 251\,65 Ond\v{r}ejov, Czech Republic\\
$^3$ Tartu Observatory, University of Tartu, 61602, T\~oravere, Tartumaa, Estonia \\
$^4$ Instituto de Astrof\'sica de La Plata, CCT La Plata, CONICET-UNLP, Paseo del Bosque s/n, \\ La Plata, B1900FWA, Buenos Aires, Argentina \\
$^5$ Departamento de Espectroscop\'ia, Facultad de Ciencias Astron\'omicas y Geof\'isicas, Universidad Nacional de La Plata, \\ Paseo del Bosque s/n, La Plata, B1900FWA, Buenos Aires, Argentina \\
$^6$ Observat\'orio Nacional, Rua General Jos\'e Cristino 77, 20921-400 S\~ao Cristov\~ao, Rio de Janeiro, Brazil \\
$^7$ Max-Planck-Institut f\"ur extraterrestrische Physik,Giessenbachstra{\ss}e, 85748 Garching, Germany
}

\date{Accepted XXX. Received YYY; in original form ZZZ}

\pubyear{2015}

\begin{document}
\label{firstpage}
\pagerange{\pageref{firstpage}--\pageref{lastpage}}
\maketitle

\begin{abstract}
B[e] Supergiants are luminous evolved massive stars. The mass-loss during this phase creates a complex circumstellar environment with atomic, molecular, and dusty regions usually found in rings or disk-like structures. For a better comprehension of the mechanisms behind the formation of these rings, detailed knowledge about their structure and dynamics is essential. To address that, we obtained high-resolution optical and near-infrared spectra for 8 selected Galactic B[e] Supergiants, for which CO emission has been detected. Assuming Keplerian rotation for the disk, we combine the kinematics obtained from the CO bands in the near-IR with those obtained by fitting the forbidden emission \oi \lam5577, \oi \lam\lam6300,6363, and \caii \lam\lam7291,7323 lines in the optical to probe the disk structure. We find that the emission originates from multiple ring structures around all B[e] Supergiants, with each one of them displaying a unique combination of rings regardless of whether the object is part of a binary system. The confirmed binaries display spectroscopic variations of their line intensities and profiles as well as photometric variability, whereas the ring structures around the single stars are stable. 

\end{abstract}

\begin{keywords}
stars: circumstellar matter -- stars: early-type -- stars: massive -- stars: supergiants -- stars: winds, outflows 

\end{keywords}



\section{Introduction}
Massive stars have an exceptionally important impact on their stellar environment and their host galaxies. They lose mass from the start of their lives via strong stellar winds. As stars evolve off the main sequence they pass through several phases of intense or even episodic mass loss before they explode as supernovae. Particularly, one such phase is composed by the B[e] Supergiants (B[e]SGs, for a review see \citealt{Kraus2017}).

B[e]SGs are luminous ($\textrm{log}\,L_*/L_{\sun}\gtrsim4$) B-type stars that do not exhibit significant variability \citep{Lamers1998}. They display complex circumstellar environments (CSEs). The presence of strong stellar winds is indicated by the P Cygni profiles in their Balmer lines. Moreover, their optical spectra are composed by narrow, low-excitation permitted and forbidden emission lines from singly ionized and neutral metals, while their UV spectra exhibit broad absorption features of higher-excitation levels of highly ionized metals. Additionally, they show a strong infrared excess due to hot circumstellar dust.

The presence of such a complex CSE has been puzzling. \cite{Zickgraf1985} proposed a model consisting of a two-component wind: (a) a low-density, fast, line-driven polar wind, where the emission of high-excitation lines originates, (b) a high-density, slow, equatorial outflowing disk, in which the line emission of permitted and forbidden low-excitation lines originate, and in which dust is formed further away. Indications for the presence of such dusty disks come from a variety of observational properties, such as the intrinsic polarization and the CO emission in IR spectra (see \cite{Kraus2017} for more details). The best proof has been provided by interferometry, which actually resolved these disks \citep[][for a review]{deWit2014}. Contemporaneously, it confirmed the earlier findings by \cite{Liermann2010} that the disks are actually detached from the central star \citep{Wheelwright2012, DomicianodeSouza2011, Millour2011}, and that the disks move in Keplerian rotation rather than following the outflowing scenario \citep[e.g.][]{Marchiano2012, Wheelwright2012b}.  Further support comes from high-resolution IR spectroscopy where CO \citep[e.g.][]{Cidale2012, Wheelwright2012b} and SiO \citep{Kraus2015} emission features have been detected. As CO and SiO form under different physical conditions (temperature and density) the information derived by modeling their features can help us probe the structure of the molecular disk. This method has been extended to optical spectra by modeling forbidden emission lines \citep[e.g.][]{Aret2012}, which has unveiled multiple rings around B[e]SGs \citep{Maravelias2017}, which are not necessary homogeneous \citep{Kraus2016, Torres2018}.

Even though observations have helped us acquire a more detailed understanding of the B[e]SGs, we still lack a proper description of how their CSE (and consequently the B[e]SGs) form. During the years, a number of models has been proposed that each can explain some of their observed properties. An approach is to create an asymmetric stellar wind in single stars using fast rotation \citep[e.g.][]{Maeder2001, Kraus2006}, the bi-stability mechanism \citep[e.g.][]{Pelupessy2000, Petrov2016}, slow-wind solutions \citep[e.g.][]{Cure2004, Cure2005} or the magneto-rotational instability mechanism \citep[e.g.][]{Krticka2015, Kurfurst2014}. Alternative scenarios with single stars include mass-loss events triggered by non-radial pulsations and/or other instabilities, or the presence of objects (e.g. minor bodies or planets) that clear their paths creating single/multiple stable ring structures \citep{Kraus2016}. A different forming channel is through binaries, either from mass transfer \citep[e.g.][]{Miroshnichenko2007,Millour2011,Wheelwright2012b} or even from mergers \citep[e.g.][]{Podsiadlowski2006}. Support for the importance of a companion comes from the high fraction of binaries among the OB populations ($\geqslant50-70\%$; \citealt{Sana2012, Sana2013, Dunstall2015}), as well as a significant estimate for the fraction of mergers ($\sim8\%$; \citealt{deMink2014}). Despite that, the number of confirmed binaries among the B[e]SG population is fairly low (6 out of 31 objects; \citealt{Kraus2017}).

To better understand the B[e]SGs we need to study in detail their CSEs and how these are formed. To address that we have initiated a campaign to study the CSEs for a large number of Galactic and Magellanic Cloud B[e]SGs, using high-resolution optical and infrared spectroscopy. The current work is focusing on a sub-sample of 8 selected Galactic B[e]SGs that display CO emission features. For all of these we have obtained/collected optical spectra and derived their kinematical information from forbidden line modeling. 

The paper is structured as follows: in Section \ref{sec:observations+data} we present our sample, the observed and archival spectra obtained, and the data processing we followed. In Section \ref{sec:kinematicalmodel} we describe the model we used, and in Section \ref{sec:results} we present our results. We discuss our results in Section \ref{sec:discussion}, and our conclusions are given in Section \ref{sec:conclusion}.

\section{Observations and data processing}
\label{sec:observations+data}

\subsection{Sample}
Our sample consists of 8 Galactic B[e]SG sources (out of a total population of 16 Galactic examples; \citealt{Kraus2017}) that show CO emission features. For these we have compiled a large collection of high-resolution optical and near-IR spectra, acquired with the same instrument (per band) at different epochs. This enables us to build the most homogeneous dataset for these sources at this resolution.  

In Table \ref{tab:sample} we give some basic properties for our sample: columns 1 and 2 refer to the most usable identifiers, columns 3 and 4 correspond to their coordinates, columns 5 to 8 provide indicative photometric magnitudes in the optical (\textit{V}) and near-IR (\textit{J, H, K}), respectively. Column 9 refers to the binarity status of each source, while further details (e.g. eccentricity, orbital period) for each binary are provided in the sections that follow.

\begin{table*}
\centering
\caption{Sample of stars.}
\label{tab:sample}
\begin{tabular}{lcccrrrrc} 
\hline \hline
\multicolumn{1}{c}{Star} & other common    & RA (J2000)   & Dec (J2000)  & \multicolumn{1}{c}{\textit{V}} & \multicolumn{1}{c}{\textit{J}} & \multicolumn{1}{c}{\textit{H}} & \multicolumn{1}{c}{\textit{K}} & Binarity status\\
     & identifiers     & (hh mm ss.s) & (dd mm ss.s) & (mag)      & (mag)      & (mag)      & (mag)      & \\
 \multicolumn{1}{c}{[1]} & [2]  & [3] &  [4] &  \multicolumn{1}{c}{[5]} &  \multicolumn{1}{c}{[6]} &  \multicolumn{1}{c}{[7]} &  \multicolumn{1}{c}{[8]} & \multicolumn{1}{c}{[9]} \\
\hline
CPD-52 9243 & Hen 3-1138 & 16 07 02.0 & -53 03 45.8  & 10.25     & 6.31       & 5.39       & 4.44        &  -- \\
CPD-57 2874 & Hen 3-394, WRAY 15-535  & 10 15 22.0 & -57 51 42.7  & 10.20     & 5.76       & 4.96       & 4.28        & -- \\
HD 327083   & Hen 3-1359 & 17 15 15.4 & -40 20 06.7  & 9.67      & 5.64       & 4.57       & 3.69        & yes$^1$ \\
HD 62623    & 3 Puppis, l Puppis  & 07 43 48.5  & -28 57 17.4  & 3.93      & 3.40       & 3.07       & 2.34        & yes$^2$ \\ 
GG Car      & HD 94878, CPD-59 2855  & 10 55 58.9  & -60 23 33.4  & 8.70      & 6.76      & 5.94      & 4.97       & yes$^3$ \\
MWC 137     & V1380 Ori   & 06 18 45.5  & +15 16 52.3 & 11.95      & 8.76       & 7.84       & 6.62         & -- \\
HD 87643    & Hen 3-365, MWC 198  & 10 04 30.3  & -58 39 52.1 & 9.50       & 6.22       & 4.76       & 3.65         & yes$^4$\\
Hen 3-298   & WRAY 15-406  & 09 36 44.4  & -53 28 00.0  & 11.4      & 7.85    & 6.84    & 5.70  & -- \\
\hline & 
\end{tabular}

\flushleft{All data retrieved from Simbad (accessed on March 13, 2017), except for HD 327083 obtained from  \cite{Miroshnichenko2003} and the IR data for Hen 3-298 obtained from \cite{Cutri2003}. \textit{References:} $^1$\cite{Wheelwright2012}, $^2$\cite{Millour2011}, $^3$\cite{Marchiano2012}, $^4$\cite{Millour2009}}

\end{table*}

\begin{table*}
\centering
\caption{Observing log} 
\label{tab:obslog}
 \begin{tabular}{ccccccccc}
 \hline \hline
 Object & Date UT  & $\textrm{N} \times t_{\textrm{exp}}$ & SNR & Instrument/Observatory & Resolution & Wavelength range \\
        & (yyyy-mm-dd) & (s)  & &   &   &  \\
\multicolumn{1}{c}{[1]} & [2] & [3] & [4] & [5] & [6] & [7]  \\
 \hline
 CPD-52 9243  & 1999-04-19 & 1x3600 & 100 & FEROS/ESO-La Silla & 48000 & 3600-9200 \AA \\
              & 1999-06-25 & 1x2400 & 80 & FEROS/ESO-La Silla & 48000 & 3600-9200 \AA \\
              & 2000-03-28 & 1x1800 & 50 & FEROS/ESO-La Silla & 48000 & 3600-9200 \AA \\
              & 2000-06-11 & 1x3600 & 75 & FEROS/ESO-La Silla & 48000 & 3600-9200 \AA \\
              & 2005-04-21 & 1x1800, 1x300 & 125 & FEROS/MPG-La Silla & 48000 & 3600-9200 \AA \\
              & 2010-04-06 & 2x20  & 140 & CRIRES/ESO-Paranal & 50000 & 2.276-2.326 $\mu\textrm{m}$ \\
              & 2015-05-13 & 2x600 & 120 & FEROS/MPG-La Silla & 48000 & 3600-9200 \AA \\
              & 2015-10-11 & 2x700 & 85 & FEROS/MPG-La Silla & 48000 & 3600-9200 \AA \\
              & 2016-04-13 & 2x700 & 120 & FEROS/MPG-La Silla & 48000 & 3600-9200 \AA \\
              & 2016-08-02 & 2x1000 & 100 & FEROS/MPG-La Silla & 48000 & 3600-9200 \AA \\

 CPD-57 2874  & 2008-12-22 & 2x1000 & 100 & FEROS/MPG-La Silla & 48000 & 3600-9200 \AA \\
              & 2009-12-02 & 2x32  & 140 & CRIRES/ESO-Paranal & 50000 & 2.276-2.326 $\mu\textrm{m}$ \\
              & 2015-05-13 & 2x600 & 100 & FEROS/MPG-La Silla & 48000 & 3600-9200 \AA \\
              & 2016-01-13 & 2x700 & 90 & FEROS/MPG-La Silla & 48000 & 3600-9200 \AA \\
              & 2016-03-13 & 2x900 & 60 & FEROS/MPG-La Silla & 48000 & 3600-9200 \AA \\

 HD 327083    & 1999-06-25 & 1x1800 & 90 & FEROS/ESO-La Silla & 48000 & 3600-9200 \AA \\
              & 2010-06-28 & 20x20 & 280 & CRIRES/ESO-Paranal & 50000 & 2.276-2.326 $\mu\textrm{m}$ \\
              & 2015-05-11 & 2x600 & 70 & FEROS/MPG-La Silla & 48000 & 3600-9200 \AA \\
              & 2015-10-12 & 2x500 & 80 & FEROS/MPG-La Silla & 48000 & 3600-9200 \AA \\
              & 2015-10-15 & 2x500 & 80 & FEROS/MPG-La Silla & 48000 & 3600-9200 \AA \\
              & 2016-04-13 & 2x500 & 75 & FEROS/MPG-La Silla & 48000 & 3600-9200 \AA \\
              & 2016-07-28 & 2x600 & 70 & FEROS/MPG-La Silla & 48000 & 3600-9200 \AA \\

 HD 62623     & 2008-12-21 & 2x120 & 230 & FEROS/MPG-La Silla & 48000 & 3600-9200 \AA \\
              & 2009-11-29 & 2x32 & 100 & CRIRES/ESO-Paranal & 50000 & 2.276-2.326 $\mu\textrm{m}$ \\
              & 2010-05-03 & 3x300 & 100 & FEROS/MPG-La Silla & 48000 & 3600-9200 \AA \\
              & 2012-04-06 & 8x10 & 500 & GNIRS/GEMINI-Mauna Kea & 18000 & 2.28-2.35 $\mu\textrm{m}$ \\
              & 2013-05-09 & 1x120 & 200 & FEROS/MPG-La Silla & 48000 & 3600-9200 \AA \\
              & 2014-11-29 & 2x10 & 90 & FEROS/MPG-La Silla & 48000 & 3600-9200 \AA \\
              & 2015-05-10 & 2x10 & 120 & FEROS/MPG-La Silla & 48000 & 3600-9200 \AA \\
              & 2015-10-12 & 2x20 & 40 & FEROS/MPG-La Silla & 48000 & 3600-9200 \AA \\

 GG Car       & 1999-04-18 & 1x900 & 130 & FEROS/ESO-La Silla & 48000 & 3600-9200 \AA \\
              & 2000-02-23 & 1x1200 & 100 & FEROS/ESO-La Silla & 48000 & 3600-9200 \AA \\
              & 2008-12-22 & 2x900 & 100 & FEROS/MPG-La Silla & 48000 & 3600-9200 \AA \\
              & 2009-06-09 & 1x810 & 110 & FEROS/MPG-La Silla & 48000 & 3600-9200 \AA \\
              & 2009-12-02 & 2x32  & 100 & CRIRES/ESO-Paranal & 50000 & 2.276-2.326 $\mu\textrm{m}$ \\
              & 2011-03-23 & 1x700 & 110 & FEROS/MPG-La Silla & 48000 & 3600-9200 \AA \\
              & 2015-05-13 & 4x120 & 70 & FEROS/MPG-La Silla & 48000 & 3600-9200 \AA \\
              & 2015-11-23 & 4x120 & 75 & FEROS/MPG-La Silla & 48000 & 3600-9200 \AA \\
              & 2015-11-26 & 4x120 & 70 & FEROS/MPG-La Silla & 48000 & 3600-9200 \AA \\

 MWC 137      & 2009-11-03 & 5x32   & 30 & CRIRES/ESO-Paranal & 50000 & 2.276-2.326 $\mu\textrm{m}$ \\
              & 2015-12-05 & 3x1200 & 70 & FEROS/MPG-La Silla & 48000 & 3600-9200 \AA \\
              & 2016-02-28 & 3x1000 & 90 & FEROS/MPG-La Silla & 48000 & 3600-9200 \AA \\

 HD 87643     & 1999-04-18 & 1x1800, 1x900, 1x360 & 60 & FEROS/ESO-La Silla & 48000 & 3600-9200 \AA \\
              & 2000-02-23 & 1x1200 & 55 & FEROS/ESO-La Silla & 48000 & 3600-9200 \AA \\
              & 2009-12-02 & 2x32  & 120 & CRIRES/ESO-Paranal & 50000 & 2.276-2.326 $\mu\textrm{m}$ \\
              & 2015-05-12 & 7x200 & 10 & FEROS/MPG-La Silla & 48000 & 3600-9200 \AA \\
              & 2015-10-13 & 2x400 & 75 & FEROS/MPG-La Silla & 48000 & 3600-9200 \AA \\
              & 2016-04-13 & 2x400 & 70 & FEROS/MPG-La Silla & 48000 & 3600-9200 \AA \\
              
 Hen 3-298    & 2005-04-19 & 1x3600, 1x600 & 60 & FEROS/MPG-La Silla & 48000 & 3600-9200 \AA \\
              & 2009-12-02 & 4x16   & 100 & CRIRES/ESO-Paranal & 50000 & 2.276-2.326 $\mu\textrm{m}$ \\
              & 2015-05-11 & 2x1100 & 20 & FEROS/MPG-La Silla & 48000 & 3600-9200 \AA \\
              & 2015-11-26 & 3x900 & 40 & FEROS/MPG-La Silla & 48000 & 3600-9200 \AA \\
              & 2015-12-06 & 2x1100 & 60 & FEROS/MPG-La Silla & 48000 & 3600-9200 \AA \\
              & 2016-01-12 & 2x1300 & 60 & FEROS/MPG-La Silla & 48000 & 3600-9200 \AA \\

 \hline& & 
 \end{tabular}
\end{table*}

\subsection{Optical data}

We used the Fiber-fed Extended Range Optical Spectrograph (FEROS, \citealt{Kaufer1999}) a bench-mounted echelle spectrograph. FEROS provides high-resolution spectra ($R\sim48000$) with a wide spectral coverage ($\sim$3600-9200 \AA). The spectrograph was attached to the 1.52 m ESO telescope for observations performed in 1999-2002, and to the 2.2 m MPG telescope later on (both telescopes located at the European Southern Observatory in La Silla, Chile). The targets have been observed systematically during the 2014-2016 period and they are supplemented with data taken scarcely since 1999, through various programs of our team and the archive\footnote{The ESO Science Archive Facility, which includes Phase 3 (fully reduced) data for FEROS, found at \burl{http://archive.eso.org/wdb/wdb/adp/phase3_spectral/form?collection_name=FEROS}}. The spectrograph is fed with light from two individual fibers with a 2\arcsec \, field-of-view each. We used the Object-Sky (OBJSKY) mode which permits simultaneous acquisition of object and sky spectra. An observing log of the optical observations can be found in Table \ref{tab:obslog}. For each source (column 1) we provide the date of the observation (column 2), the number of exposures and the exposure time (column 3), and the SNR (column 4) as derived from a region around $7100\,\AA$ in the optical and around $2.293\, \mu\textrm{m}$  in the near-IR spectra. We also give the instrument used (column 5) along with the resolution (column 6) and the wavelength coverage (column 7). 

For our work we used the FEROS pipeline products which we further processed (using the PyRAF command language). We first remove the barycentric correction applied to the data by the FEROS pipeline, and we combine spectra to obtain higher SNR. To remove the telluric lines we employ the standard IRAF \texttt{telluric} task using standard star data, either from observations taken at the same night or using templates whenever standard stars observations were not available (mainly from 1999 and 2000). We subtract the sky from the object spectrum whenever the [OI] sky lines contaminate the emission features of interest - not possible in all cases due to the lack of sky spectra (in those the lines have been manually removed). After the telluric and sky removal we correct back all final spectra for the barycentric velocity and their corresponding systematic radial velocities. Since these are unknown we opted for correcting with the values that center the [OI] \lam6300 line. As a last step, we select the spectral lines we are interested and we normalize their intensity with local continuum (which is calculated at their central wavelength through a linear fit of continuum regions located at the red and blue parts with respect to the lines; \citealt{Maravelias2014}).

\subsection{Infrared data}

For the IR observations we have used the CRyogenic high-resolution InfraRed Echelle Spectrograph
(CRIRES; \citealt{Kaeufl2004}), equipped on a 8.2 m telescope of ESO-VLT (Paranal, Chile). This instrument can obtain high-resolution ($R\sim50000$) spectra with a range of $2.277-2.325\,\mu\textrm{m}$ in the $K-$band. In order to remove the sky and detector glow we performed observations with a standard nodding on-slit strategy. Each science target was followed by the observation of a standard star to correct for telluric lines. The ESO/CRIRES pipeline (v2.3.4) was used to reduce the data, and the final spectra were corrected for heliocentric velocity. Due to quality issues we use only the part of the spectrum around the first CO bandhead. This allows us to derive concrete results regarding the kinematics for the CO region, but not about the temperature and the column density (at least two bandheads are needed; \citealt{Kraus2009}). 

For HD 62623 we present an additional spectrum using the Gemini Near-Infrared Spectrograph (GNIRS) mounted on the 8.1 m Gemini North telescope (Mauna Kea, Hawaii-US). The observations were performed in long-slit (single order) mode ($R\sim18000$), using the 110 l mm$^{-1}$ grating and the 0.10 arcsec slit, centered at $2.318\,\mu\textrm{m}$. Observations were taken with an ABBA nod pattern along the slit in order to remove sky emission. All the steps of the reduction process were made using IRAF software package tasks. Reduction steps include AB pairs subtraction, flat field correction, telluric correction, and wavelength calibration. Each science target was followed by the observation of a standard (B-type) star to correct for telluric lines. After applying the corrections for heliocentric and systemic velocities, the continuum was used to normalize the data, and finally, it was subtracted to obtain a pure emission spectrum. The observing log for the IR spectra is included in Table \ref{tab:obslog}.

\section{Modeling}
\label{sec:kinematicalmodel}

The various emission features (in the optical and the near-IR) form in regions of different physical conditions (e.g. temperature and density). In the Keplerian rotation scenario each region will display a different rotational velocity according to its position within the disk. The optical \oi \lam5577, \oi \lam\lam6300,6363, and \caii \lam\lam7291,7323 emission lines would form closer to the star than the molecular emission. Moreover, in a standard disk scenario in which density and temperature decrease with distance from the central star, the \caii and the \oi \lam5577 lines occupy similar disk regions close to the star whereas the \oi doublet lines form further out \citep{Kraus2007, Kraus2010,Aret2012}. Hot molecular emission originates from disk regions between the atomic gas and the dust, where the gas temperatures range from 5000 down to 2000 K. In this region, first CO forms, with its first-overtone band emission $(\sim2.29\,\mu\textrm{m})$ tracing the inner edge of the molecular disk \citep{Liermann2010, Oksala2013}, while SiO forms at slightly lower temperatures hence farther out (first-overtone band emission around $4\,\mu\textrm{m}$; \citealt{Kraus2015}). Thus, the kinematical information derived from these emission features can probe the disk's structure. 

For the model computations of the first CO bandhead observed in our IR data, we utilize the previously developed code by \cite{Kraus2000}. It computes a CO emission spectrum assuming that the material is in Local Thermodynamic Equilibrium and located within a narrow rotating ring $(v_{\rm rot})$. Depending on our knowledge of the inclination angle of the system, the $v_{\rm rot}$ corresponds to the de-projected velocity, else we refer to the line-of-sight velocities that are lower limits of the real rotational velocities. This velocity is convolved with a Gaussian component ($v_{\rm g}$) which is a combination of a typical thermal velocity ($v_{\rm thermal}\sim1-2$ \kms) and some random internal motion of the gas ($v_{\rm turbulence}$ typically $\sim1-2$~\kms). The last step of the calculations is another convolution of the model with a single Gaussian component that represents the instrument's spectral resolution $(R)$.

To model the optical emission lines we assume that the atomic gas follows the distribution of the molecular gas, i.e. we consider emission from narrow rotating rings for the \oi \lam5577 and \lam\lam6300,6363, and the \caii \lam\lam7291,7323 lines. We are using the same thermal velocity, as our approach is based purely on the kinematics, but we adjust the instrument's spectral resolution to the optical one ($R\sim5.5-6.5$ \kms, for FEROS). For some cases in our narrow ring approach the model required a value for $v_{\rm turbulence}$ that was (much) larger than the typical value of 1-2 \kms that can be associated with turbulence. In these cases, we interpret the extra Gaussian velocity component needed for the fit as indication for rings with finite (extended) width, and we determine the typical ring-width as: $v_{\rm turbulence}=\sqrt{<v_{\rm g}>^2-R^2-v_{\rm thermal}^2}$, where $<v_{\rm g}>$ is the mean value from all measurements corresponding to the same ring from all epochs available. In this case we convolve the model of the emission line directly with a Gaussian profile that combines the thermal velocity, the turbulence, and the instrument's spectral resolution. This is allowed because the optical forbidden lines are optically thin, in contrast to the (partially) optical thick lines and the line blending in the near-IR region for the CO emission spectrum.

In order to determine the rotational velocities we follow this strategy:

\begin{enumerate}

\item We first determine the rotational velocity of the CO ring, since the CO emission features are a direct evidence of gas presence. 

\item The rotational velocity (and the Gaussian component) of CO is our first guess for the corresponding model of each optical line. If these initial guesses are not sufficient to reproduce the observed profile then we adjust their values accordingly. 

\item A symmetrical broadening of the spectral line is the result of integrating its flux over a complete ring. To account for asymmetries we calculate the line flux over partial rings (see also \citealt{Kraus2016} for the difference between full and partial rings). This indicates the presence of inhomogeneities within these rings.

\item It is possible though that a single rotational velocity cannot fit the line. In that case we add more rings up to the successful fit of the observed profile. 

\item In some cases we need to combine multiple and partial rings. We do this in order to remain consistent with our approach, however, we point that this could not be the only scenario (see Section \ref{sec:discussion}). 

\end{enumerate}

To select the optimum fitting parameters for each line we create a number of models changing one parameter ($v_{\rm rot}$, $v_{\rm g}$, and fluxes in multiple ring-models) at a time, and we visually compare them with the observed profiles. During this process we have implemented a $\chi^2$ measurement to direct us towards the best solution and help us address degeneracy problems, e.g. selecting among multiple ring options the one with the smallest reduced $\chi^2$ value. Quoting those values could be rather misleading as due to the combination of high-quality data (very small errorbars), the telluric/sky residuals (although our best effort to correct for these), and the continuum determination (including the presence of some features at the wings of some lines) the reduced $\chi^2$ values become quite large. The corresponding errors are derived by the maximum and minimum values that provide an acceptable fit. 

We point here that the modeling process refers to the line profiles originating from the emitting regions of the CSE, which means that the gaps between the individual rings are due to gas-free or low-density regions. Alternative models have been tested in the past. An outflow scenario can fit the profiles, but it fails to reproduce the spectra of these sources, e.g. the lack of the {[O\,{\sc ii}]} \lam7319 line \citep{Kraus2010}. A viscous disk scenario underestimates significantly the \oi line intensities \citep[see][]{Porter2003,Kraus2007}.

\section{Results}
\label{sec:results}

For each source in our sample we have obtained spectra from various epochs. This allows us to examine the physical properties of their disks and their variations if present. In the following paragraphs we describe our results per object, summarized in Table \ref{tab:results}. For each source (column 1) we provide the inclination angle (column 2), whenever known to obtain the de-projected velocities, the stellar mass estimates (column 3), the rotational velocities and ring radii (according to mass) of the forming regions for the \oi \lam5577 line (columns 4 and 5), the \caii \lam7291 line (columns 6 and 7), the \oi \lam6300 line (columns 8 and 9), the first CO bandhead (columns 10 and 11), and the first SiO bandhead (columns 12 and 13). In the last column (14) of Table \ref{tab:results} we present the systemic radial velocity derived for each source by  identifying the necessary shift to center the  observed profiles of the \oi \lam6300 line compared to a symmetric model, which is added to the barycentric correction performed by FEROS pipeline. For each star the rotational velocities are shown in a decreasing order, i.e. as we progressively move away from the star. The rotational velocities of the optical lines are the averaged values over all epochs, and their corresponding
errors are the result of error propagation from the individual measurements. We present more detailed information for the fits we have performed for each epoch and star in Appendix \ref{app:individualkinematics}. Considering the near-IR spectra we have re-processed all original CRIRES data (for which \citealt{Muratore2012} presented preliminary fits for some objects) and we provide rotational velocities corrected for the line-of-sight whenever the inclination angle is known. We present all final fits for the first CO overtone bandhead emission (at $\sim2.3\,\mu\textrm{m}$) in Fig. \ref{fig:ir_spectra}, and we discuss these CO features (henceforth) in relation with the results obtained from the optical lines.

\subsection{CPD-52 9243}

The CO emission features have been detected in the near-IR spectra already by the works of \cite{Whitelock1983} and \cite{McGregor1988}. \cite{Cidale2012} using high-resolution sprectroscopy and interferometry have found that \linebreak

\begin{figure*}
	\includegraphics[scale=0.9]{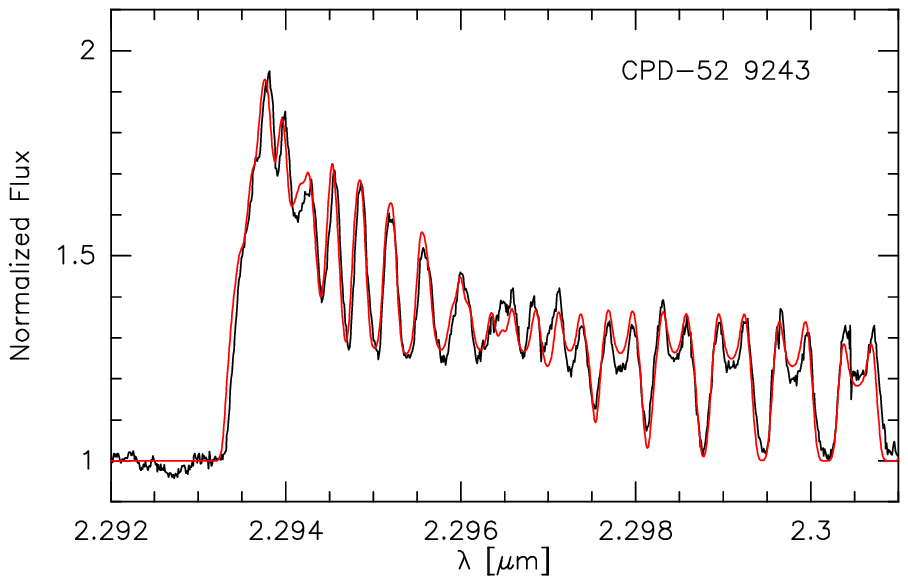} \includegraphics[scale=0.9]{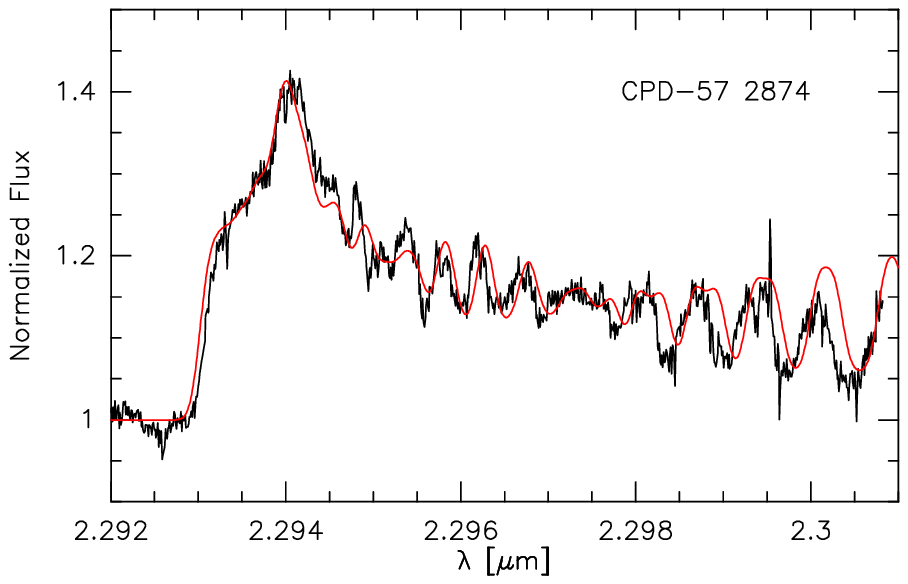} \\
	\includegraphics[scale=0.9]{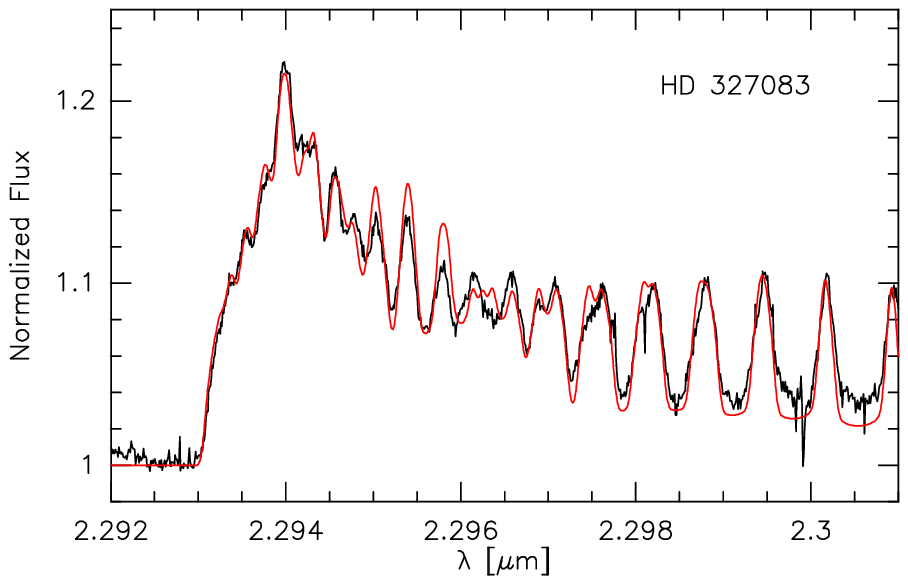} \includegraphics[scale=0.9]{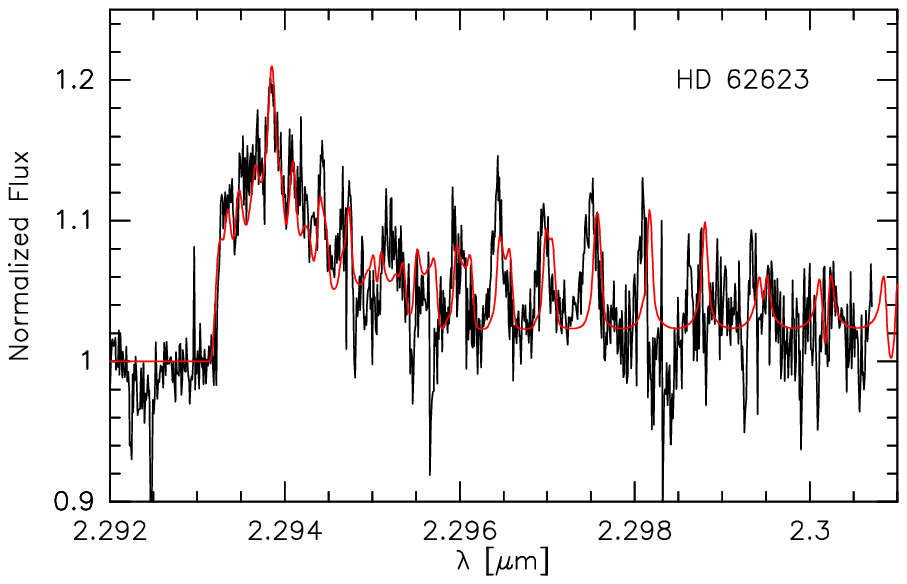} \\ 
	\includegraphics[scale=0.9]{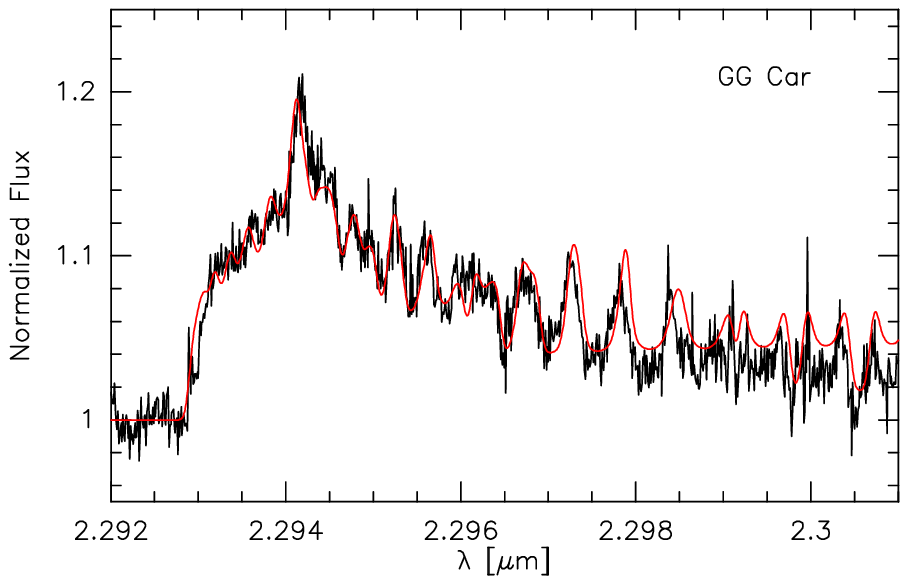} \includegraphics[scale=0.9]{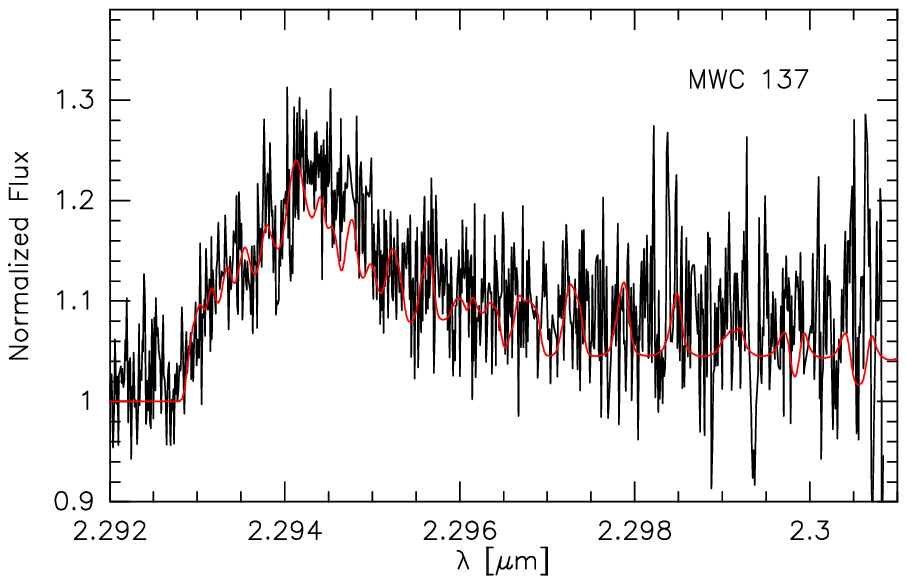} \\ \includegraphics[scale=0.9]{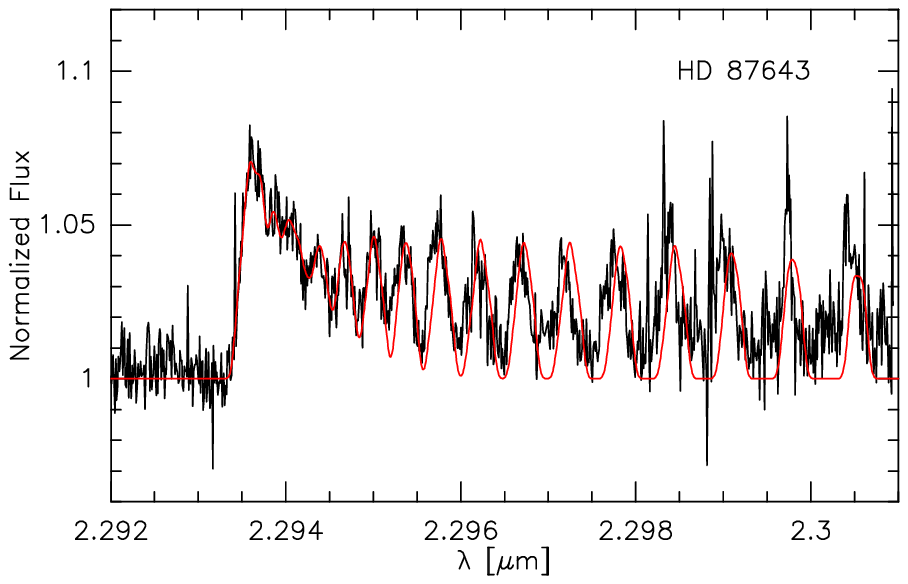} \includegraphics[scale=0.9]{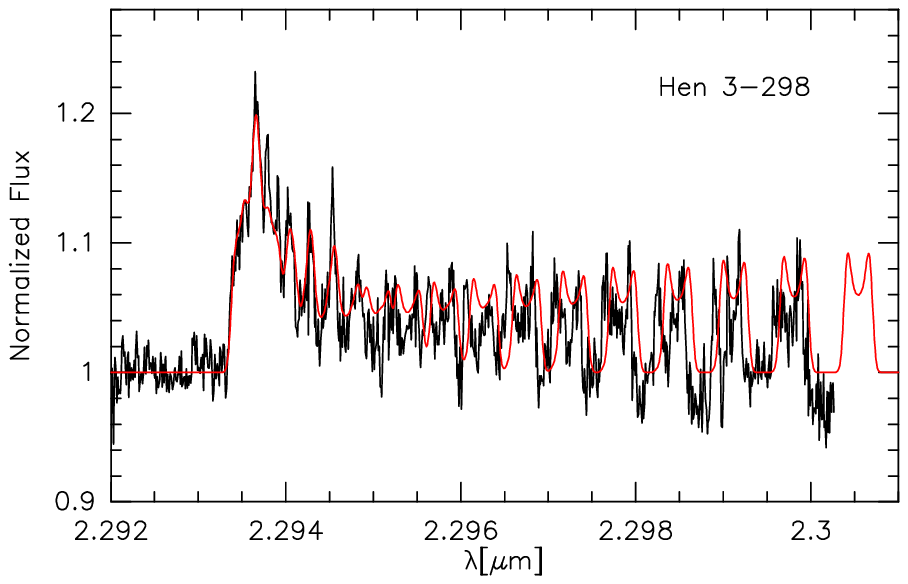} \\
    \caption{The near-IR CRIRES spectra of our sample (black lines), showing the first CO bandhead and their corresponding models (red lines). [See text for more details regarding each source, and Table \ref{tab:obslog} for observing dates.] }
    \label{fig:ir_spectra}
\end{figure*}

\begin{landscape}
	\begin{table}

	\caption{The summary of the kinematics identified in our sample. For each source we present the number of identified rings per line as we move further away from the star (i.e. in a decreasing order of rotational velocities). }
	\label{tab:results}
	\begin{tabular}{lccccccccccccc}
\hline \hline
Star & $i^a$ & Mass & \multicolumn{2}{c}{\oi5577} & \multicolumn{2}{c}{\caii7291} & \multicolumn{2}{c}{\oi6300} & \multicolumn{2}{c}{CO} & \multicolumn{2}{c}{SiO} & systemic \\
 &  &  & $v_{\rm rot}$ & R  & $v_{\rm rot}$ & R & $v_{\rm rot}$ & R & $v_{\rm rot}$ & R & $v_{\rm rot}$ & R & RV$^b$ \\
     & (\degr) & ($M_{\sun}$) & (\kms) & (AU) & (\kms) & (AU) & (\kms) & (AU) & (\kms) & (AU) & (\kms) & (AU) & (\kms) \\
\multicolumn{1}{c}{[1]} & [2] & [3] & [4] & [5] & [6] & [7] & [8] & [9] & [10] & [11] & [12] & [13] & [14]  \\
\hline 
CPD-52 9243 & 46\degr$^1$ & 17.4-18.6$^{1,c}$ & -- & -- & 48.9$\pm$3.0 & 6.7$\pm$0.8 & 51.4$\pm$4.8 & 6.0$\pm$1.1 &        & &      & & -49 \\
            &           & &    & & 30.4$\pm$2.9 & 17.3$\pm$3.3 & 32.1$\pm$4.6  & 15.5$\pm$4.4 &  36$\pm$1$^1$ & 12.3$\pm$0.7 & 35.5$\pm1^2$ & 12.7$\pm$0.7 &  \\
			\\

CPD-57 2874 & 30\degr$^3$ & 15-20$^{3.c}$ & -- & -- & 210.5$\pm$15.0 & 0.35$\pm$0.05 &  & &    & &  & &  3 \\
            &           & &    & & 158.3$\pm$6.2 & 0.62$\pm$0.05 & 166.3$\pm$11.4 & 0.56$\pm$0.08 &              & &      & &   \\ 
            &           & &    & & 108.1$\pm$8.1 & 1.3$\pm$0.2 & 120.3$\pm$6.2 & 1.1$\pm$0.1 & 130$\pm$1$^4$ & 0.92$\pm$0.01 & 110$\pm$1$^2$ & 1.28$\pm$0.02 &  \\
            &           & &    & &               & & 88.3$\pm$7.1 & 2.0$\pm$0.3 &              & &      & &  \\
            &           & &    & &               & & 44.5$\pm$4.5 & 7.8$\pm$1.6 &              & &      & &  \\
			\\

HD 327083   & 50\degr$^5$ & 25$^6$ & -- & -- & 75.4$\pm$3.5 & 3.9$\pm$0.4 & 74.7$\pm$4.3 & 4.0$\pm$0.5 & 86$\pm$1$^4$ & 3.00$\pm$0.07 & 78$\pm$1$^2$ & 3.65$\pm$0.09 & -26 \\
			&             &        &    & &  & & 61.7$\pm$4.1 & 5.8$\pm$0.8 & & & & & \\
			\\              

HD 62623    & 52\degr$^7$ & 9-10.5$^{8,c}$ & 57.1$\pm$7.0 & 2.7$\pm$0.7 & 60.8$\pm$2.4 & 2.3$\pm$0.2 & 61.7$\pm$3.2 & 2.3$\pm$0.2 & 53$\pm$1$^4$ & 3.1$\pm$0.1 & 48$\pm$1$^2$ & 3.8$\pm$0.2 & 27 \\
            &           & &       &                & 41.8$\pm$2.6 & 5.0$\pm$0.6 & 39.5$\pm$4.3 & 4.7$\pm$1.8 &      & &    & &   \\
			\\

GG Car      & 63\degr$\pm$9\degr$^9$ & 38$^{10}$ & -- & -- & 95.8$\pm$5.4 & 3.7$\pm$0.4 & 84.3$\pm$16.3 & 4.7$\pm$1.8 & 91.5$\pm$7.5$^{10}$ & 4.0$\pm$0.7 & -- & -- & -22 \\
            &           & &    & &              & & 29.9$\pm$3.4 & 37.7$\pm$8.5 &             & &    & &   \\
			\\

MWC 137     & unknown   & 10-15$^{11,c}$ & -- & -- & -- & -- &   &   & 84$\pm2^{12,d}$ & 1.57$\pm$0.07 & -- & -- & 42 \\
            &           & &    & &              & & 68.0$\pm2.8^d$ & 2.4$\pm$0.2 &             & &    & &  \\
            &           & &    & &              & & 46.8$\pm1.4^d$ & 5.1$\pm$0.3 &             & &    & &  \\
            &           & &    & &              & & 31.0$\pm1.4^d$ & 11.5$\pm$1.0 &             & &    & & \\
            &           & &    & &              & & 20.3$\pm0.7^d$ & 26.9$\pm$1.9 &             & &    & & \\
			\\

HD 87643    & unknown   & 25$^{13}$ & -- & -- & & & 89.1$\pm9.0^d$ & 2.5$\pm$0.5 &             & &    & &  -3 \\
            &           & &    & &              & & 61.0$\pm5.3^d$ & 5.2$\pm$0.9 &             & &    & &   \\
            &           & &    & &              & & 28.1$\pm2.9^d$ & 24.7$\pm$5.1 &             & &    & &  \\
            &           & &    & &  10.6$\pm1.3^d$ & 173$\pm$43 & 9.7$\pm1.7^d$ & 207$\pm$72 & 11$\pm1^{4,d}$ & 161$\pm$29 & -- & -- & \\

HD 87643$^e$    & 7.4\degr$^4$   & 25$^{13}$ & -- & -- & & & 691$\pm70$ & 0.046$\pm$0.009 &             & &    & &  -3 \\
            &           & &    & &              & & 474$\pm41$ & 0.10$\pm$0.02 &             & &    & &   \\
            &           & &    & &              & & 218$\pm23$ & 0.47$\pm$0.1 &             & &    & &  \\
            &           & &    & &  82.3$\pm10$ & 3.3$\pm$0.8 & 75$\pm13$ & 3.9$\pm$1.4 & 85$\pm8^{4}$ & 3.1$\pm$0.6 & -- & -- & \\
            \\

Hen 3-298   & unknown   & 20$^{14}$ & 23.2$\pm2.7^d$ & 33.0$\pm$7.7 & 21.5$\pm1.1^d$ & 38.4$\pm$3.9 & 18.7$\pm1.1^{4,d}$ & 50.8$\pm$6.0 & 19$\pm1^{4,d}$ & 49.1$\pm$5.2 & -- & --  &  81 \\
\hline \hline
\end{tabular}

\flushleft{
\textit{Notes:}
$^a$ Inclination angle of the system rotation axis with respect to the line of sight ($i = 90\degr$ means edge-on view of the disk); $^b$ The systemic radial velocity for each star, as derived by centering the \oi \lam6300 line; $^c$ For the sources with a range of stellar mass estimates we used an average value for the determination of the corresponding ring radii; $^d$ Rotational velocities as projected to the line-of-sight; $^e$ For HD 87643 we present our results for both the line-of-sight velocities and the velocities derived after constraining the inclination angle (see text for more details).    

\textit{References:}
$^1$ \cite{Cidale2012};
$^2$ \cite{Kraus2015};
$^3$ \cite{DomicianodeSouza2011};
$^4$ This work;
$^5$ Marchiano et al. (in preparation);
$^6$ \cite{Wheelwright2012};
$^7$ \cite{Millour2011}; 
$^8$ \cite{Aret2016};
$^9$ \cite{Marchiano2012};
$^{10}$ \cite{Kraus2013};
$^{11}$ \cite{Mehner2016};
$^{12}$ \cite{Muratore2015};
$^{13}$ \cite{Oudmaijer1998};
$^{14}$ \cite{Oksala2013}.
}

	\end{table}
\end{landscape}

\noindent the CO emission originates from a detached rotating ring at 36 \kms (Fig. \ref{fig:ir_spectra}). Additionally, \cite{Kraus2015} have discovered SiO emission from a rotating ring at 35.5 \kms. Since their velocities are the same, it becomes apparent that the total molecular emission originates from a common region. 

\begin{figure*}
	\includegraphics[scale=0.3]{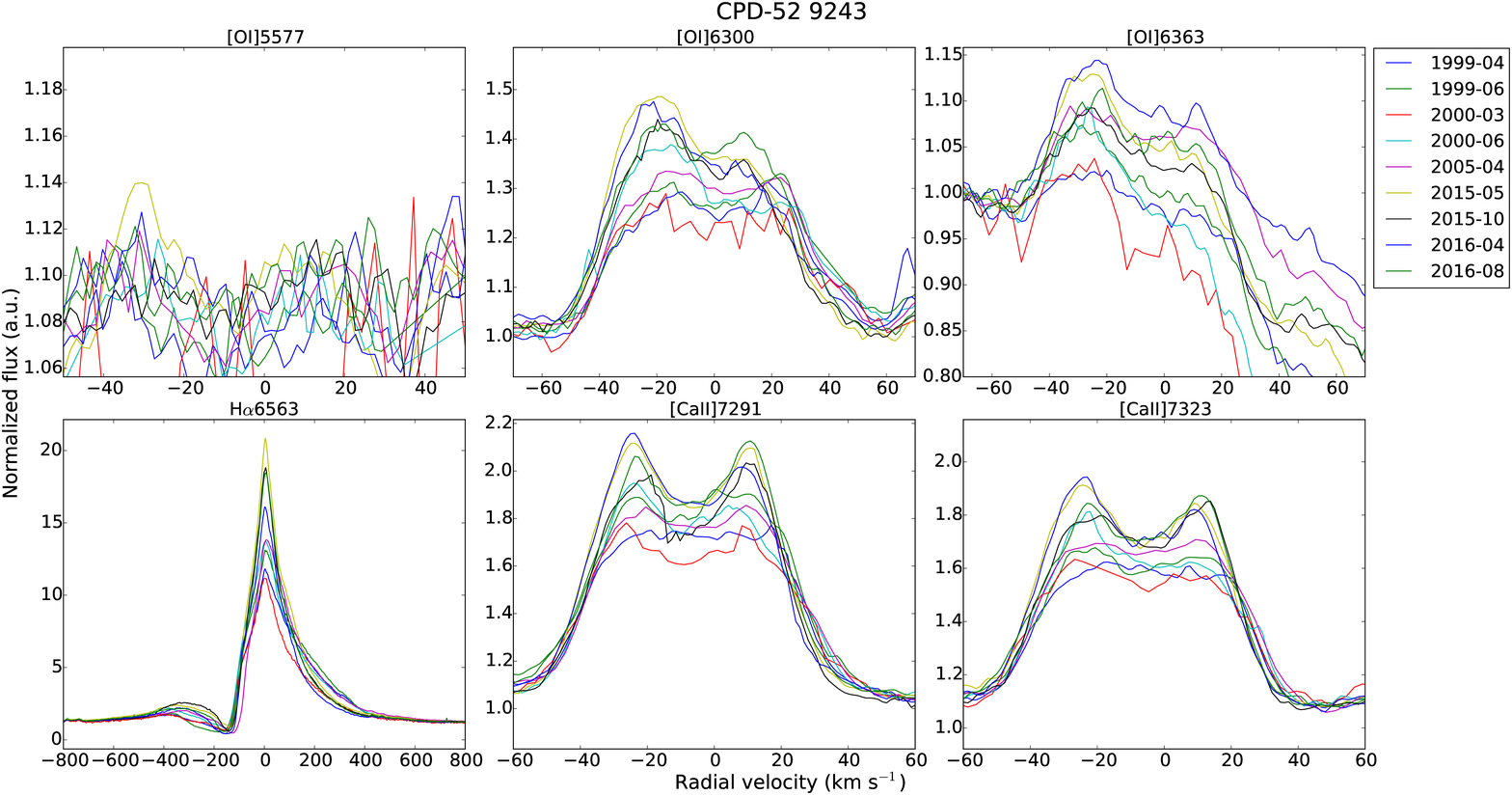}
    \includegraphics[scale=0.3]{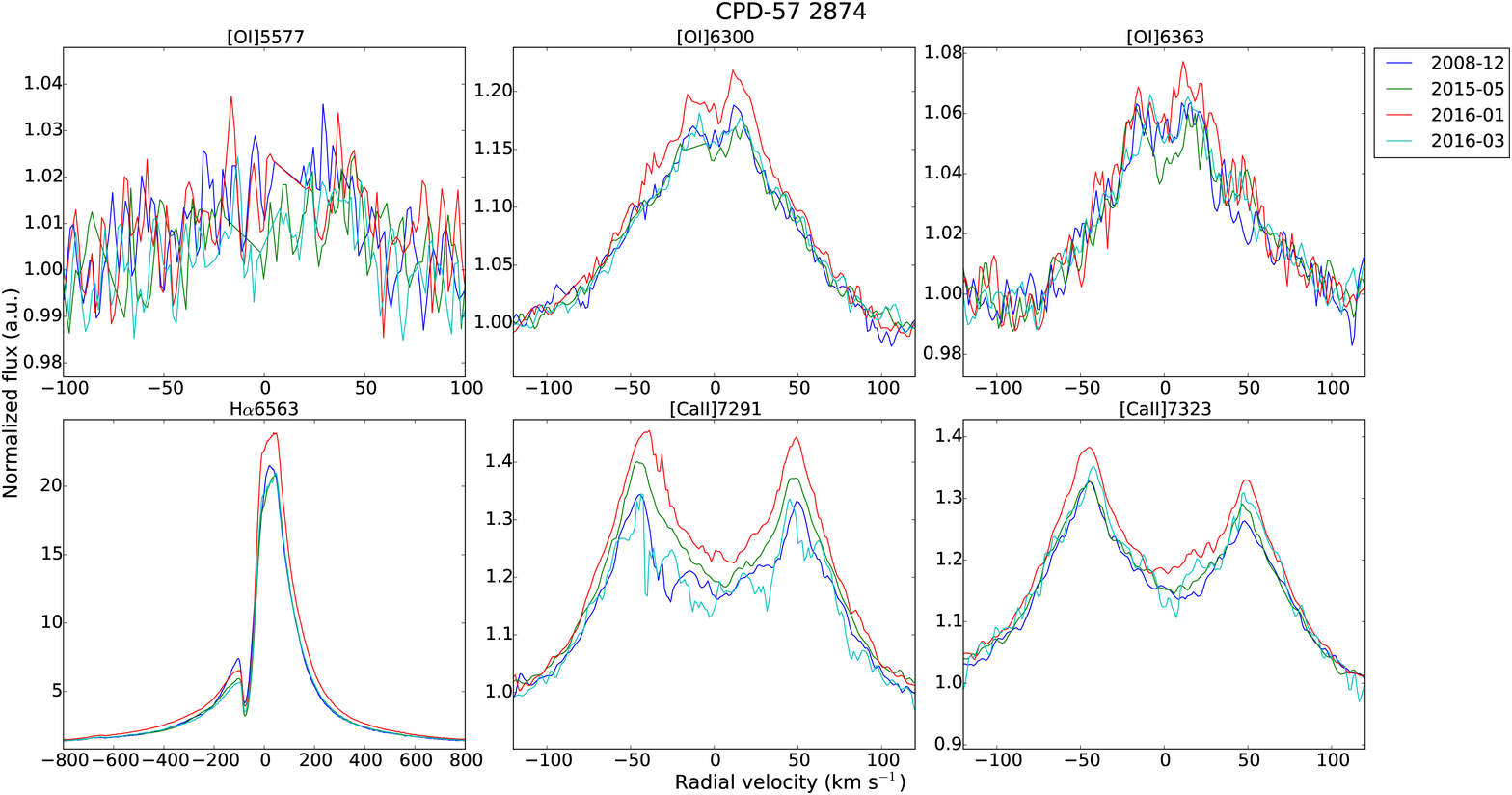}
    \caption{Profiles of the \oi \lam\lam5577,6300,6363, the H$\alpha$ \lam6563, and the \caii \lam\lam7291,7323 lines as derived from all available FEROS observations for CPD-52 9243 (1999-2016) and CPD-57 2874 (2008-2016). [See text for more details.] }
    \label{fig:variability-1}
\end{figure*}

\begin{figure*}
	\textbf{CPD-52 9243}\\
	\includegraphics[scale=0.3, trim=15 0 30 0,clip]{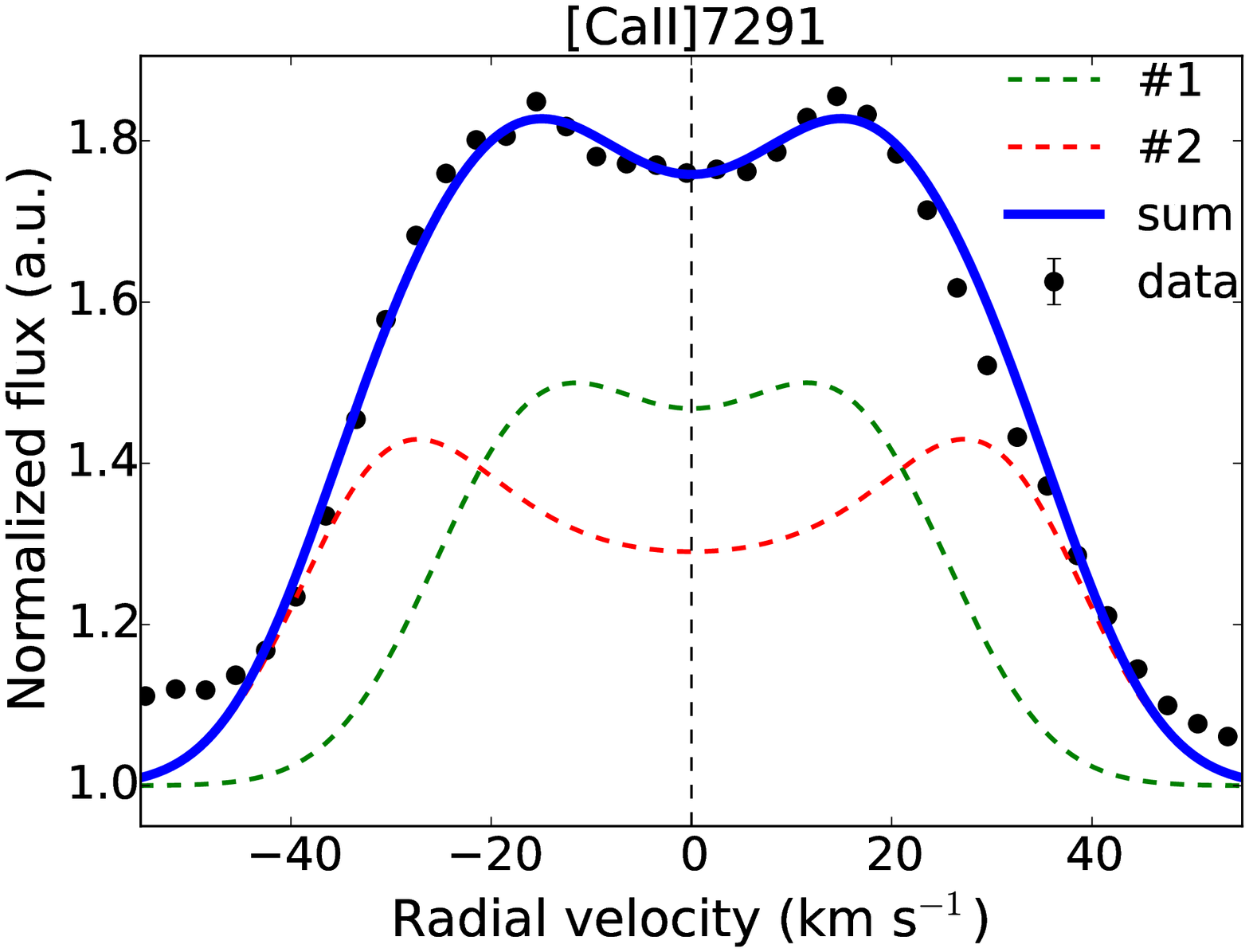} 
	\includegraphics[scale=0.3, trim=40 0 25 0,clip]{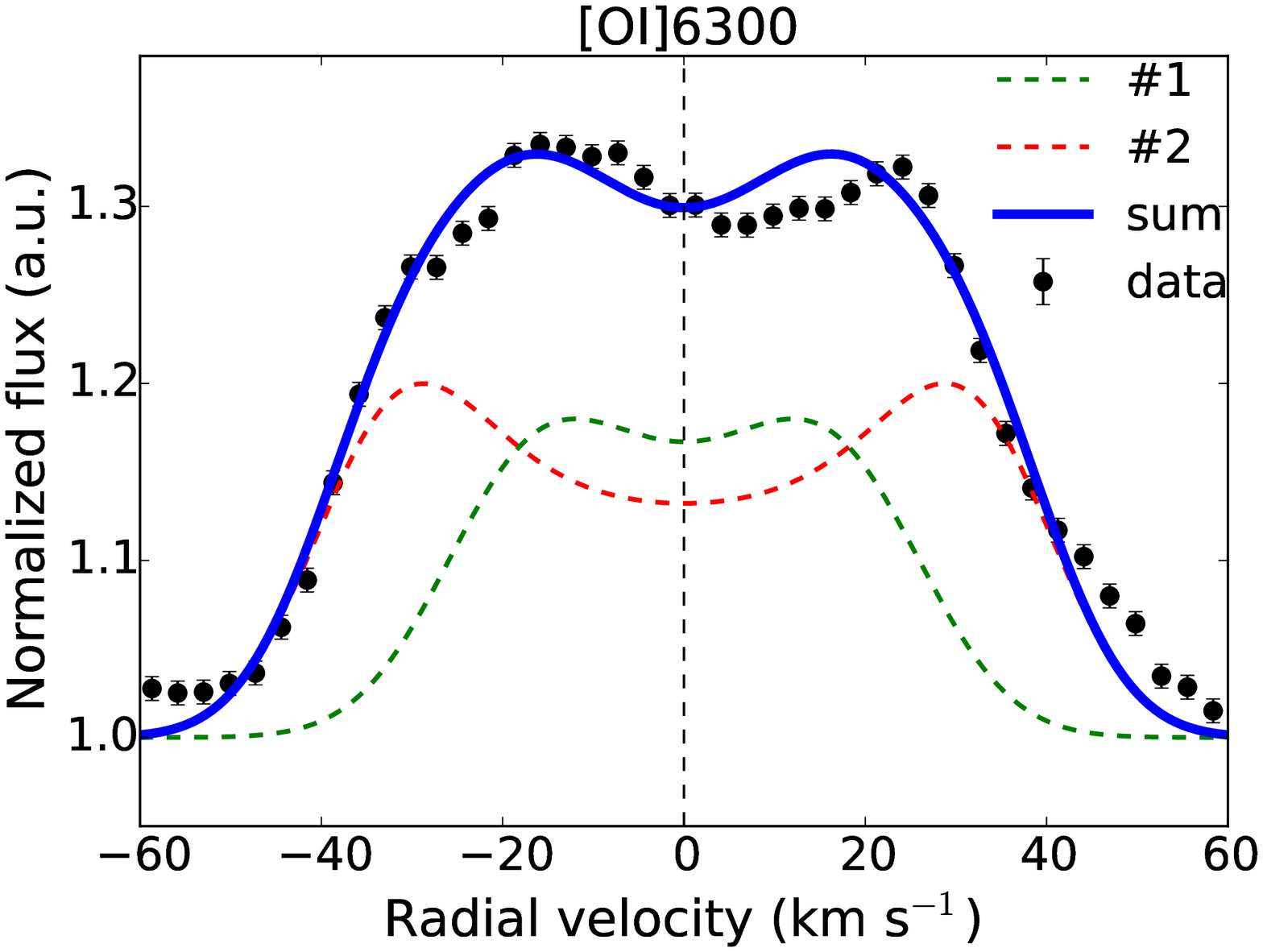}
	\includegraphics[scale=0.3, trim=40 0 25 0,clip]{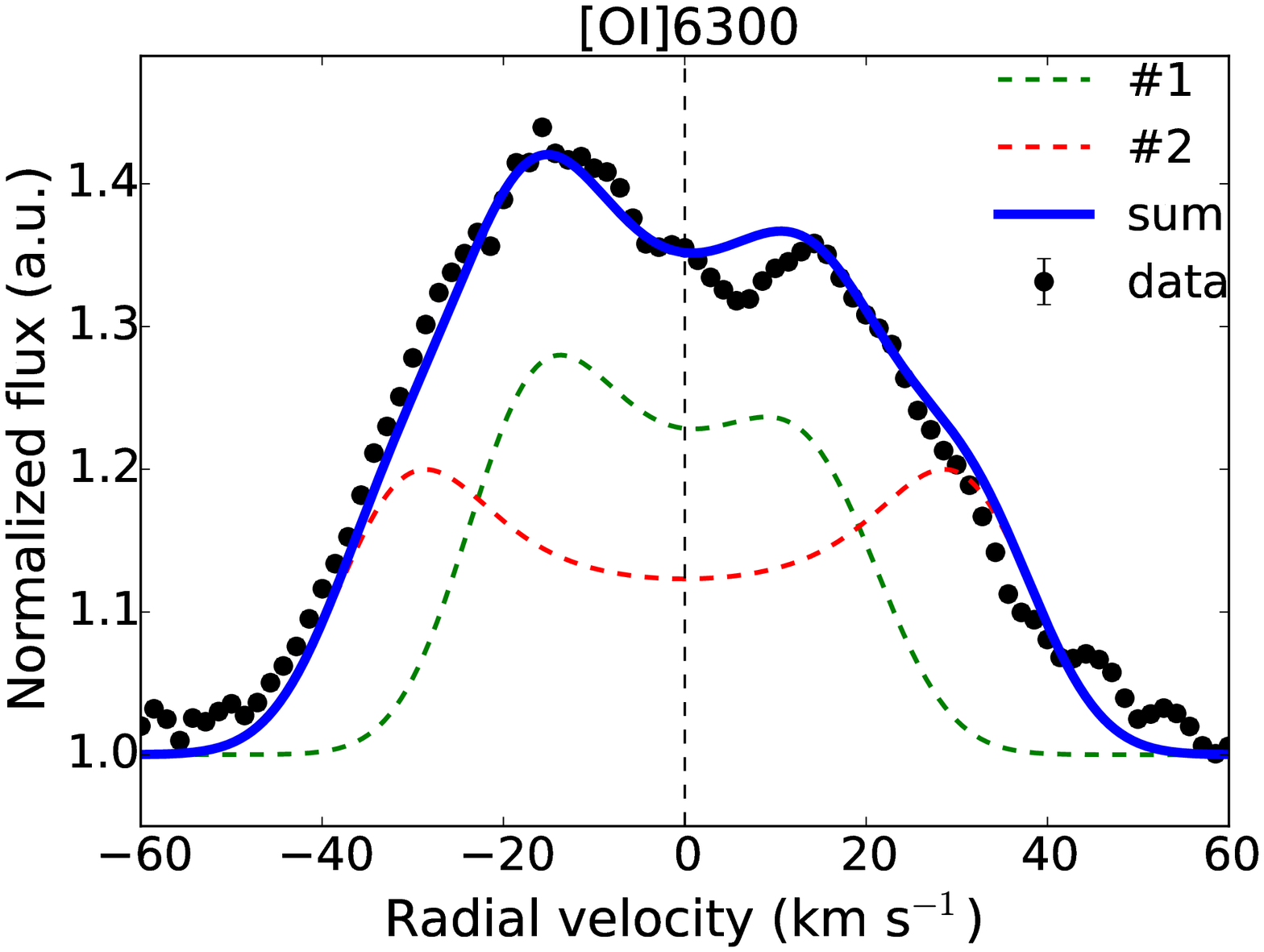} \\
	\textbf{CPD-57 2874}\\
	\includegraphics[scale=0.3, trim=15 0 20 0,clip]{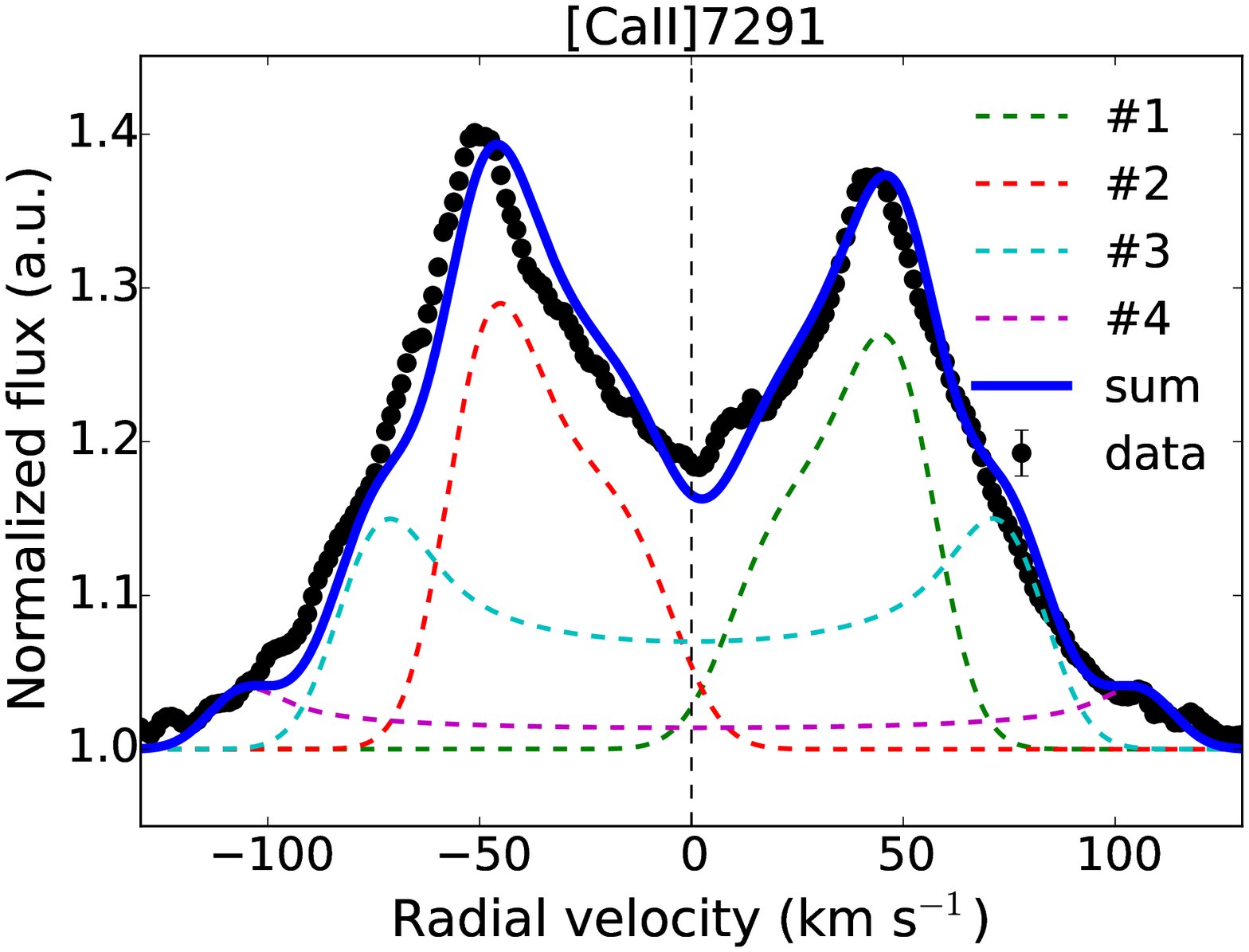} 
	\includegraphics[scale=0.3, trim=40 0 25 0,clip]{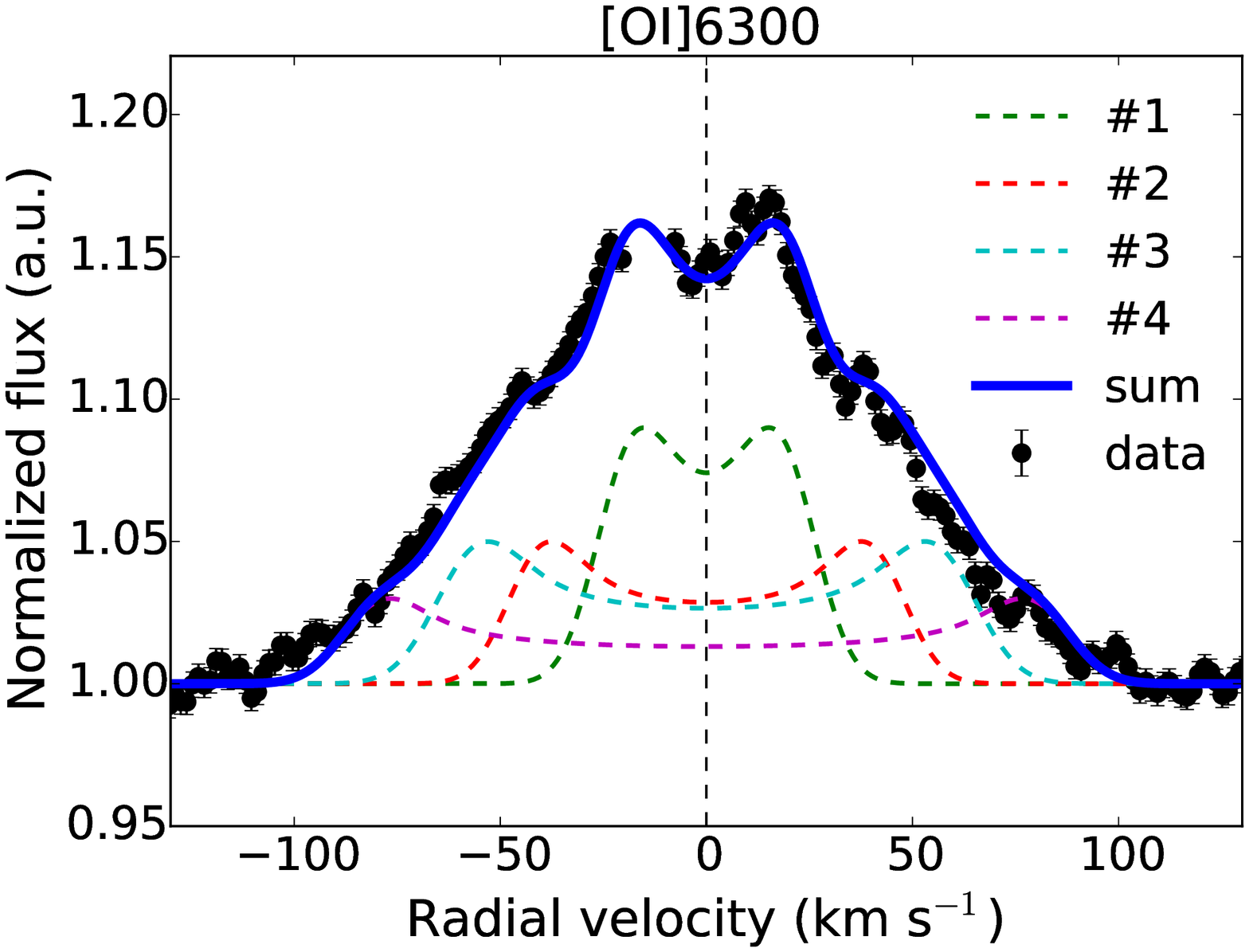} \\
	\textbf{HD 327083}\\	
	\includegraphics[scale=0.3, trim=15 0 10 0,clip]{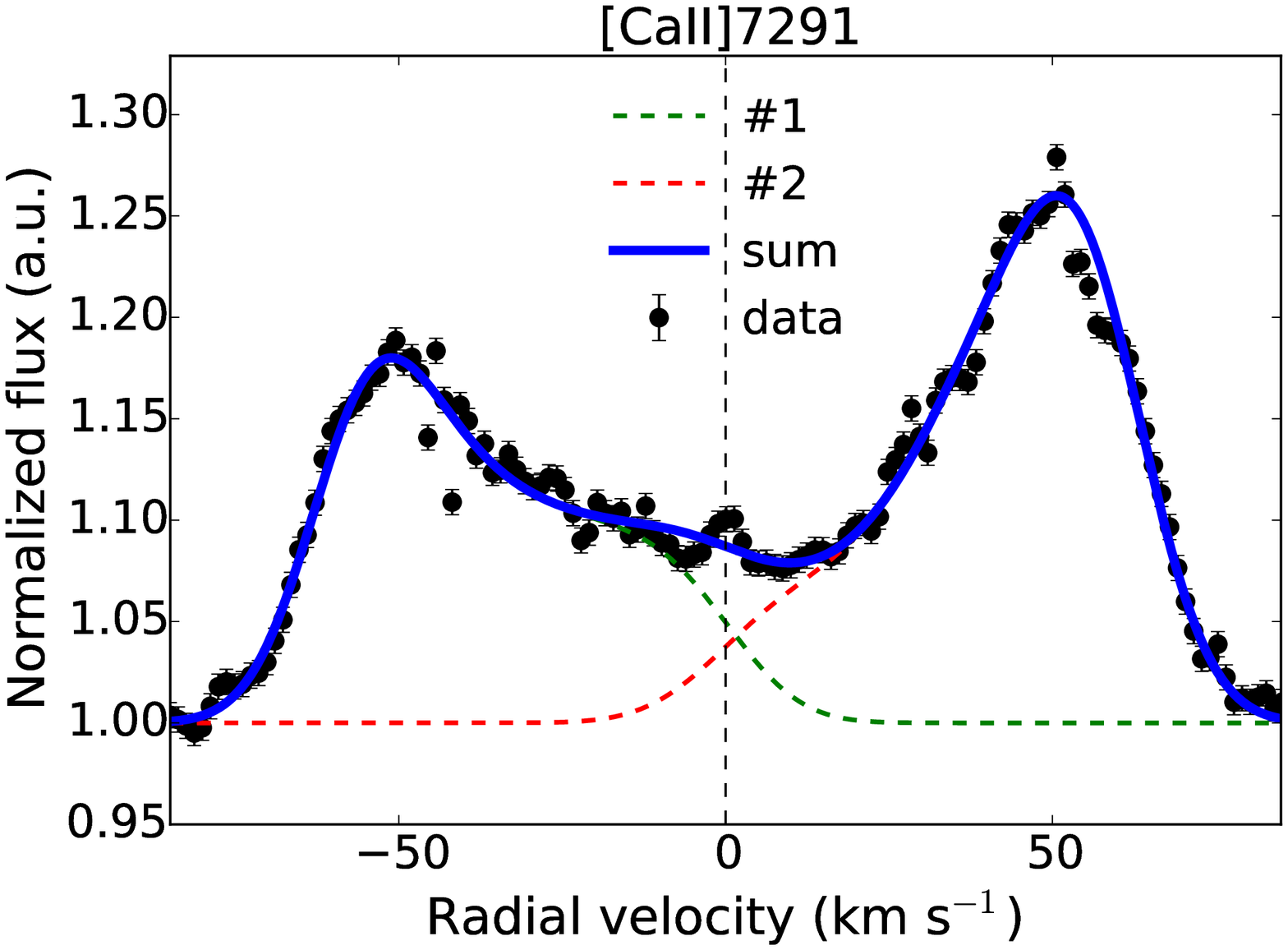} 
	\includegraphics[scale=0.3, trim=40 0 10 0,clip]{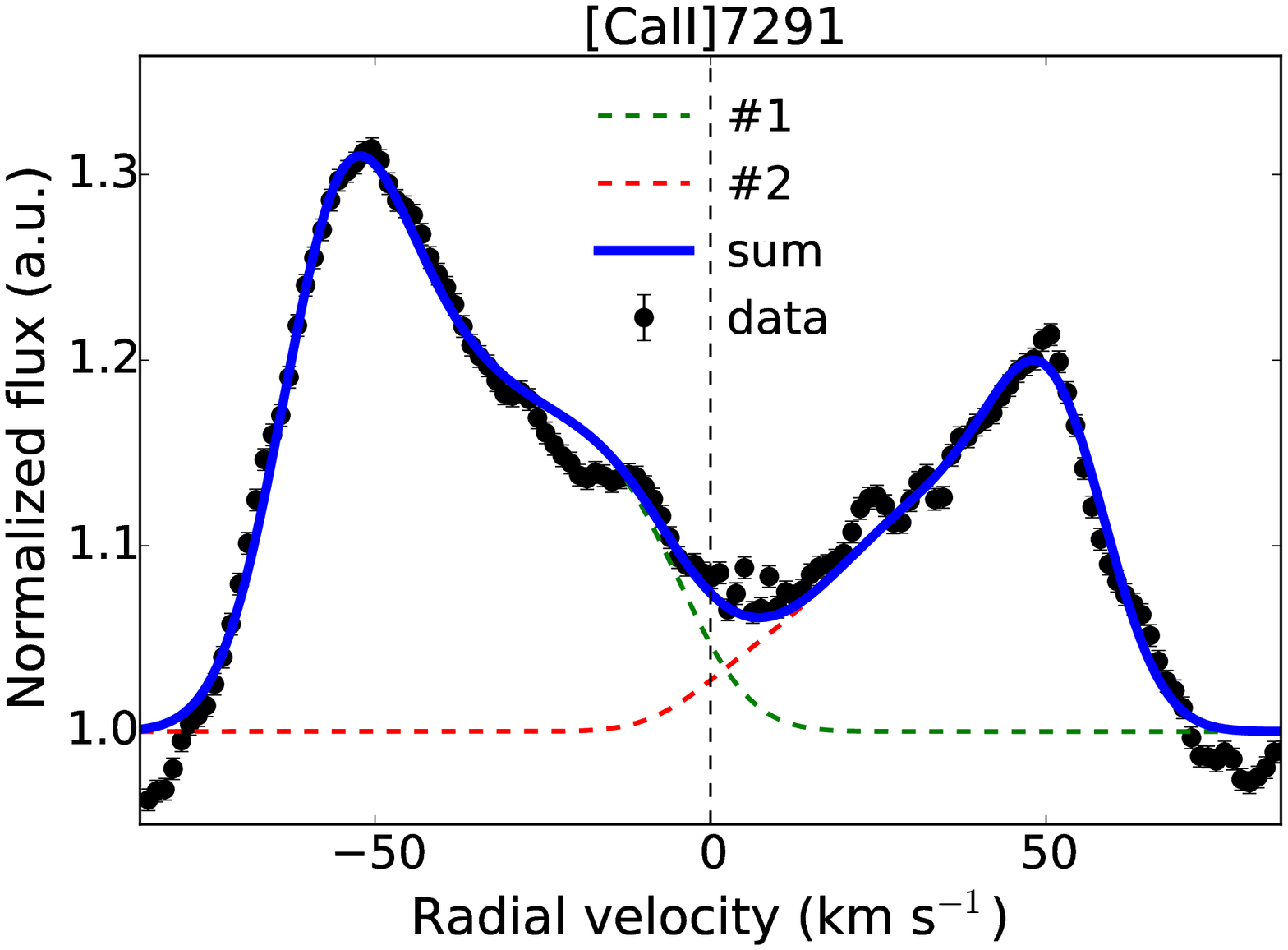}
	\includegraphics[scale=0.3, trim=40 0 25 0,clip]{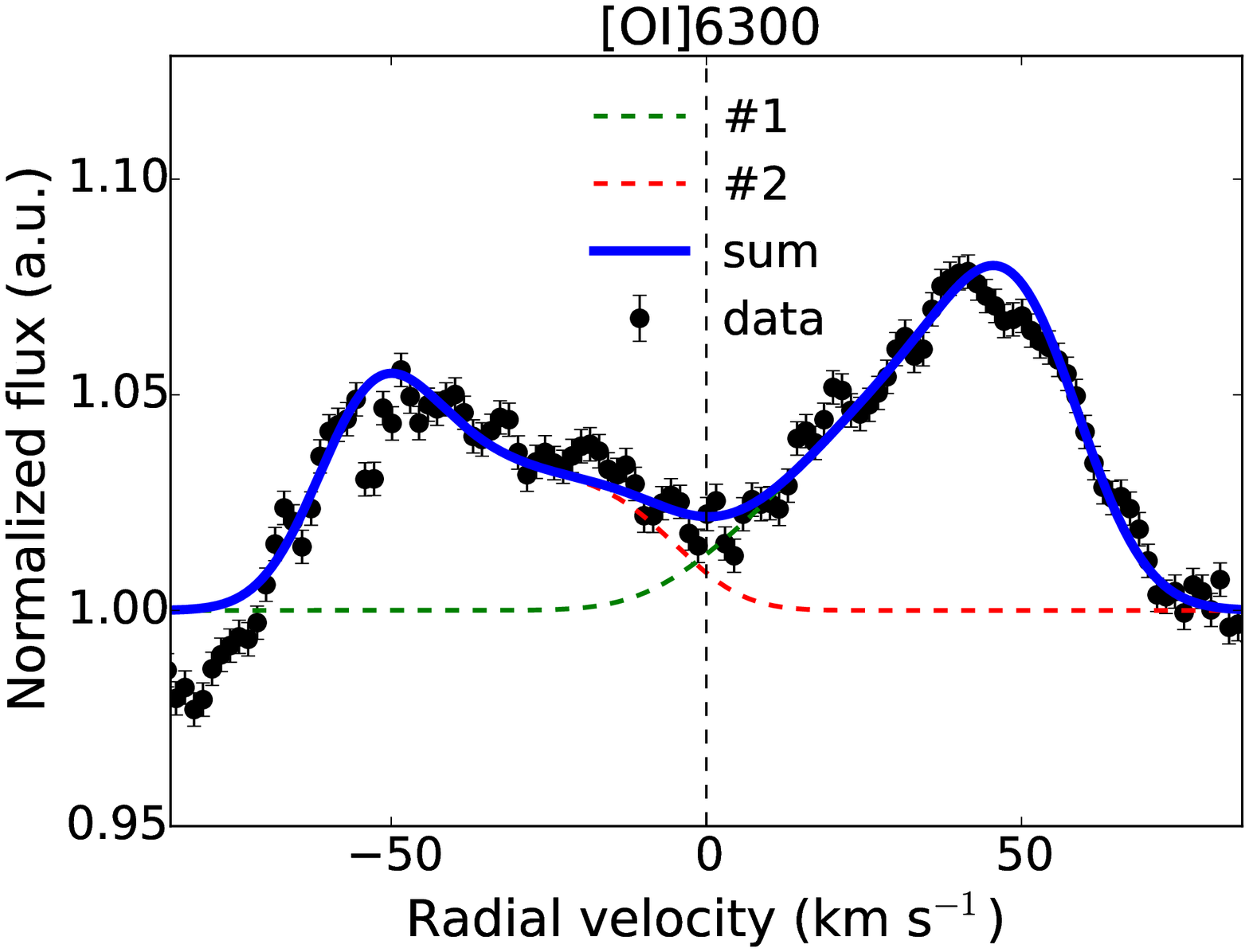}\\
	\textbf{HD 62623}\\
	\includegraphics[scale=0.3, trim=0 0 30 0,clip]{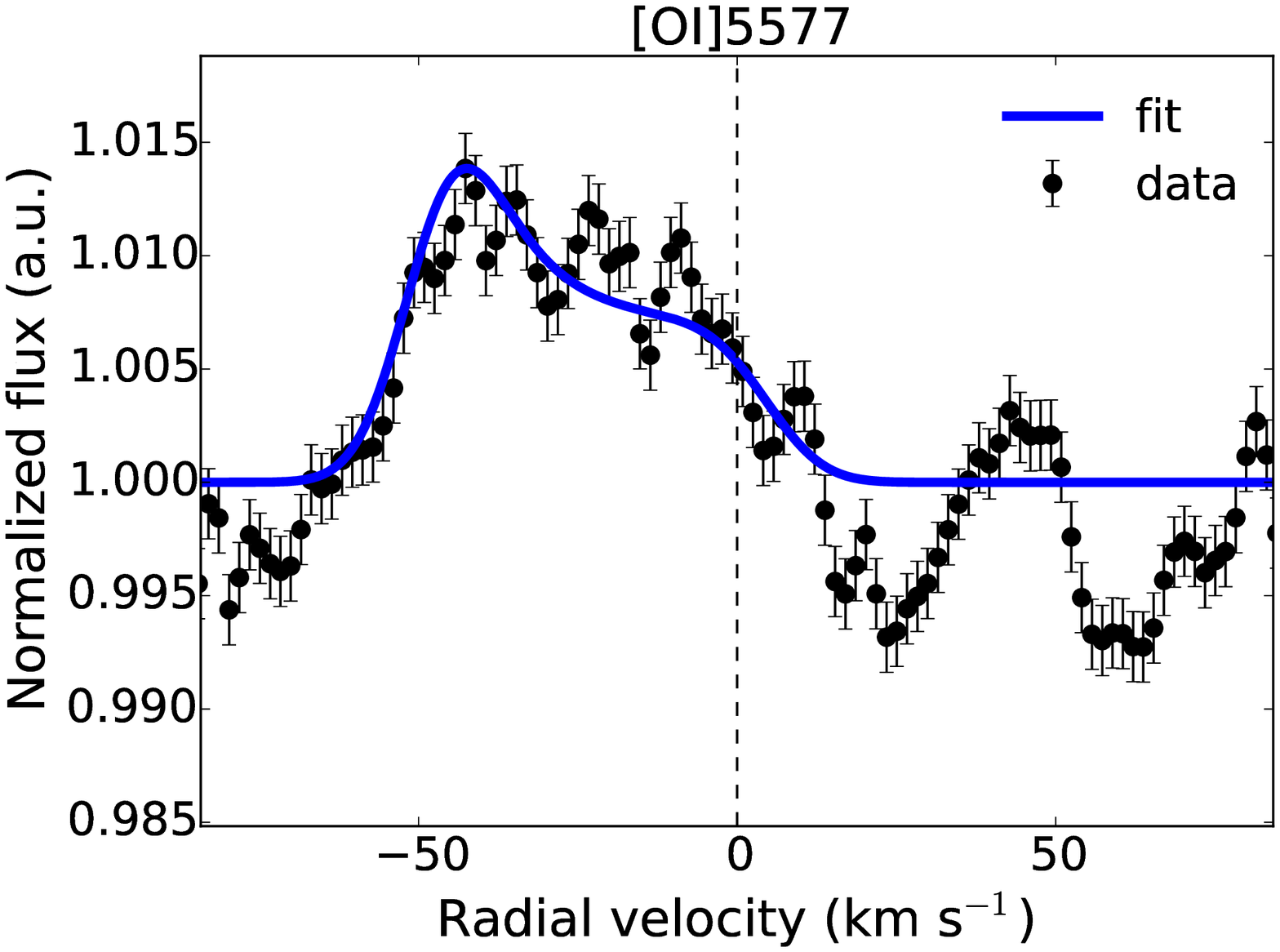} 
	\includegraphics[scale=0.3, trim=40 0 30 0,clip]{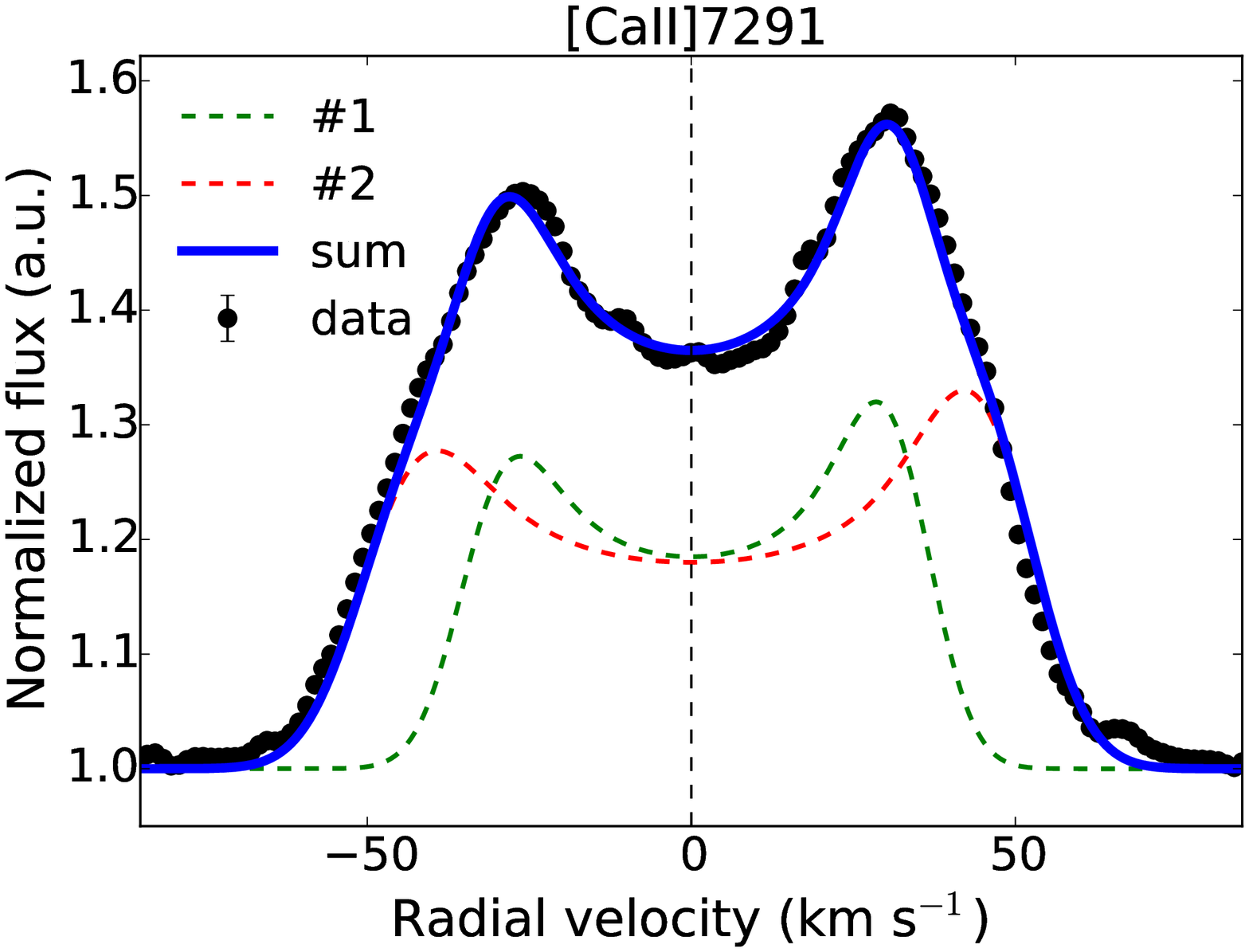}
	\includegraphics[scale=0.3, trim=42 0 10 0,clip]{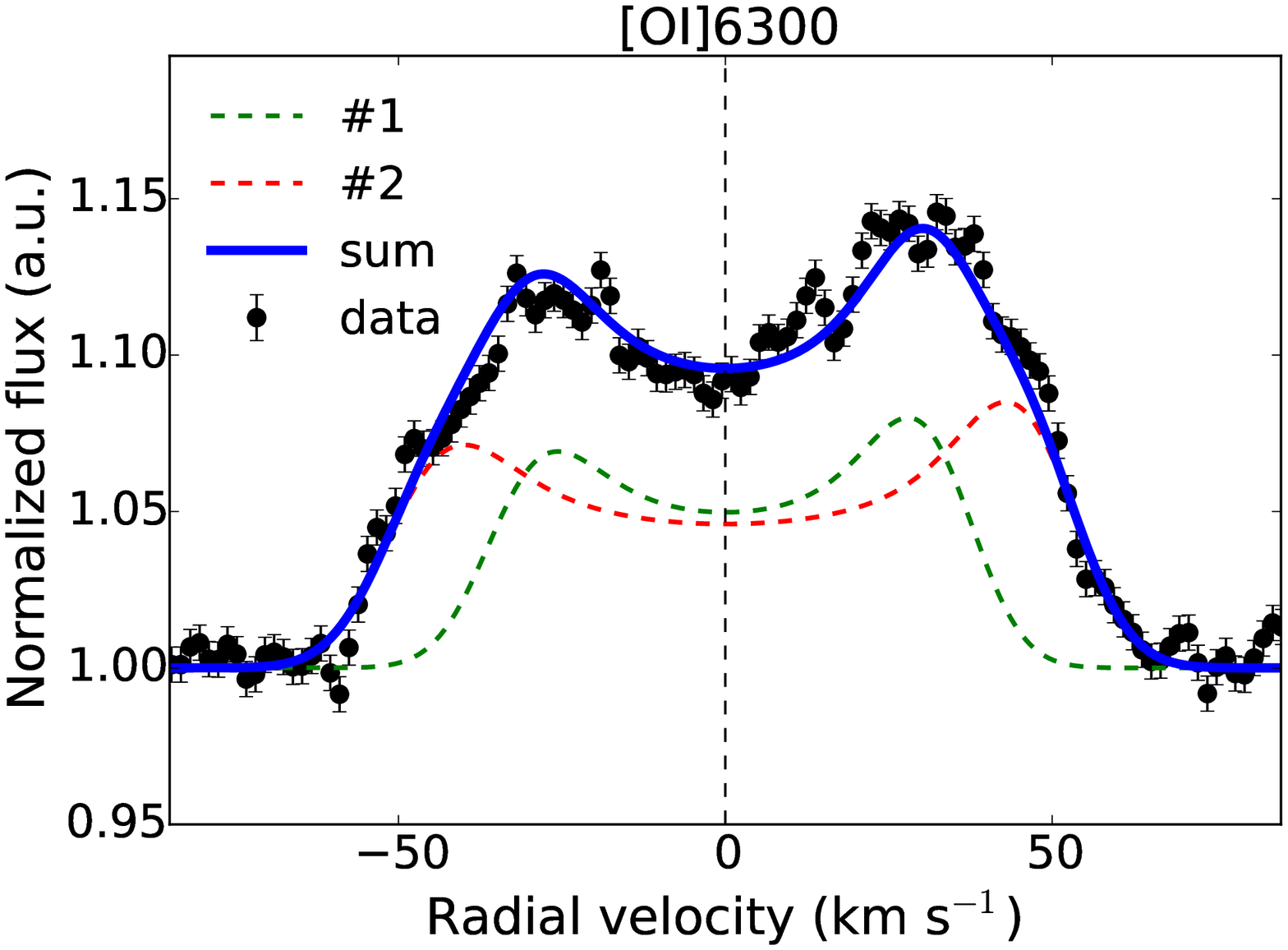} \\
	\caption{Examples of fits to the line profiles for our sample. Observations are shown as black dots, while the individual ring models as dashed lines and the reconstructed profile as solid blue lines. From top to bottom, \textit{CPD-52 9243:} the first two panels show double-peaked models for the \caii \lam7291 and the \oi \lam6300 lines from 2005-04-21, while the last one the asymmetric \oi \lam6300 line from 2015-10-11, \textit{CPD-57 2874:} the \caii and the \oi lines from 2015-05-13 along with three- and four-ring models, respectively, \textit{HD 327083:} the \caii line (from 2015-10-11 and 2016-04-13), and the \oi \lam6300 line (2015-10-11), respectively, with models of single rings with inhomogeneities, \textit{HD 62623:} the \oi \lam5577 (2010-05-03) line modeled with a partially filled ring, as well as the \caii \lam7291 (2015-05-10) and the \oi \lam6300 (2014-11-29) lines with their corresponding two-ring models with small inhomogeneities. [See text for more details for each source.] }
    \label{fig:fitlines-1}
\end{figure*}

In Fig. \ref{fig:variability-1} we present the line profiles of the optical emission lines from all available observations (from 1999 to 2016). We note the absence of \oi \lam5577 in all epochs. While the \caii doublet presents clearly double-peaked profiles the \oi displays  asymmetries (e.g. stronger blue peaks on 2016-04-13, 2015-05-13, and 2000-06-10)\footnote{The \oi \lam6363 line suffers from an absorption line on its red wing.}. Using the model described in Section \ref{sec:kinematicalmodel}, we fit each of these lines with a two-ring model (see Fig. \ref{fig:fitlines-1}). To account for the asymmetries of the \oi line we need to use partial rings (half-filled), and since these are present only on some epochs (e.g. not on 2016-08-02 or 2000-03-27) they may be indicative of a revolving inhomogeneity of the material. Nevertheless, the rotational velocities derived from the fitting process are similar in all epochs (see Table \ref{tab:kin-cpd-529243} for more details), and their averaged values are $30.4$ \kms and $48.9$ \kms for the \caii line and $32.1$ \kms and $51.4$ \kms for the \oi line. 

The identification of two rings for each line indicates that there are two emitting regions/rotational rings. Moreover, the similar values found for the \caii and \oi lines indicates that these gases coexist. Given the Keplerian rotation, the ring at $\sim31$ \kms is located further away from the star than the molecular ring (at $\sim36$ \kms). Since our typical ring-width is $\sim9$ \kms this ring may not be totally independent from the molecular one. The presence of another ring at a higher velocity (of $\sim50$ \kms) implies a region which is located closer to the star, and consists mainly of atomic gas (i.e. \caii and \oinos). Thus, in total we suggest that the circumstellar disk of CPD-52 9242 consists of two to three rings (as the outermost ring could be potentially overlapping with the molecular one). We used the stellar mass estimate by \cite{Cidale2012} of $17.4-18.6\,M_{\sun}$ to convert these velocities to distance radii from the star (see Table \ref{tab:results}). The derived radii at $\sim6.7$ AU (ring with atomic gas only: \caii and \oinos) and $\sim12-17$ AU (combined atomic and molecular gas) are consistent with the results of \cite{Cidale2012} who identified atomic gas at 5 and 12 AU.

Apart from the changes in the \oi profile, there are intensity variations which are systematic for all lines, with the strongest lines present during our latest observations (in 2015 and 2016). The H$\alpha$ line displays also variability on its blue peak, which is indicative of changes in the wind. It is interesting to note that the profiles of H$\alpha$ (up to 2005) and \oi \lam6300 (on dates with symmetric profiles) are similar to the ones obtained in 1988 by \cite{Zickgraf2003} (cf. his fig. 1 and 2, at similar resolution of $R\sim55000$), although the big gap between 1988 and 1999 does not allow for any strict conclusions about the variability of the lines.

Additionally, there is a small radial velocity offset for the \oi \lam6300 line present in different epochs (up to 5~\kms, larger than our typical calibration error of $\sim1$~\kms), and between this line and the \caii \lam7291 (up to 8 \kms) which may indicate either slight displacements of these rings with respect to each other or more elliptical rings than the circular ones assumed. This is similar to what we see for the elliptical binary GG Car  (\citealt{Marchiano2012, Kraus2013}; see \S\ref{sec:GGCar}), which may imply a binary nature for CPD-52 9243 also (as it has been suggested already by \citealt{Cidale2012}).

\begin{figure*}
	\includegraphics[scale=0.3]{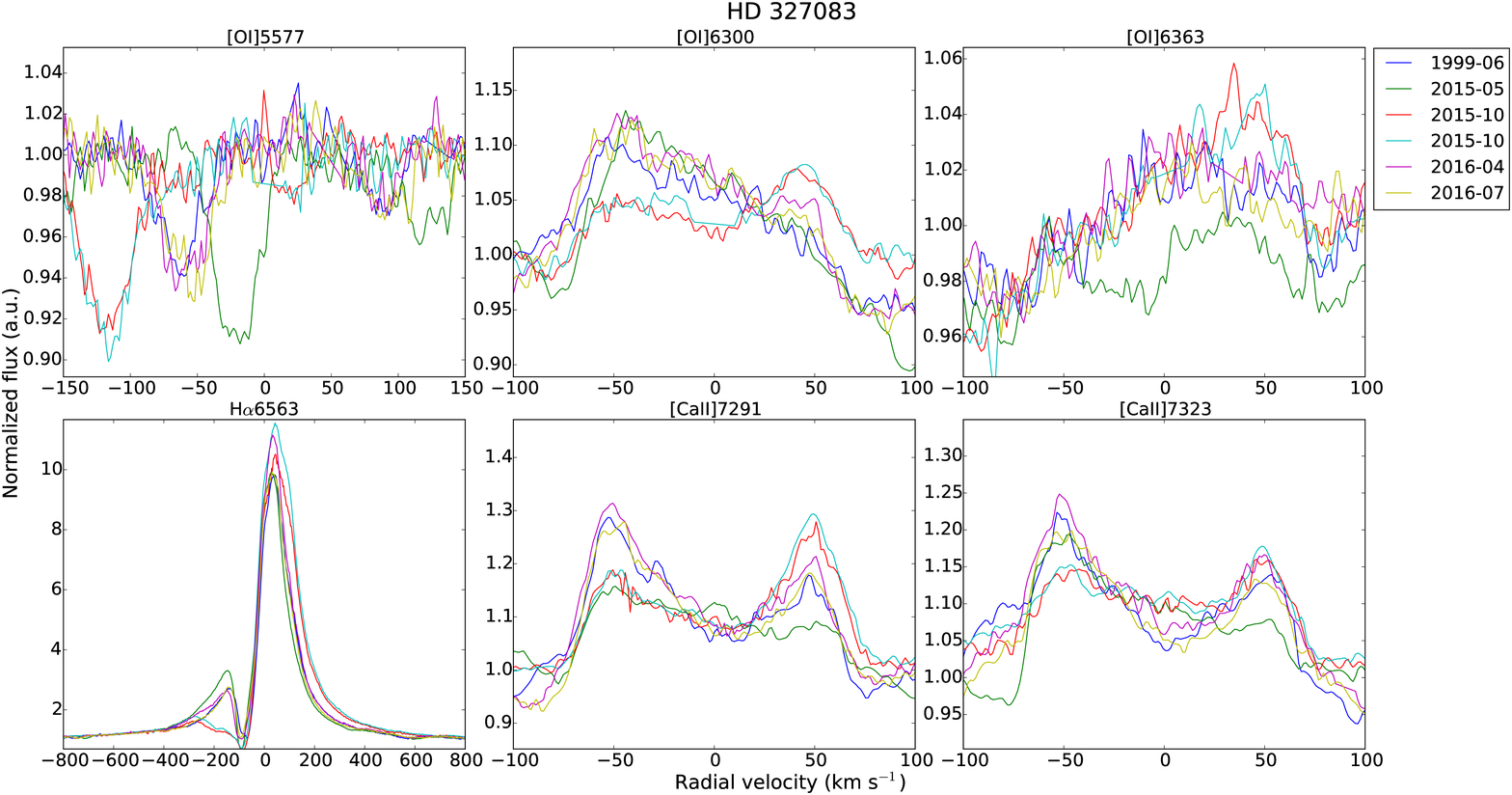}
	\includegraphics[scale=0.3]{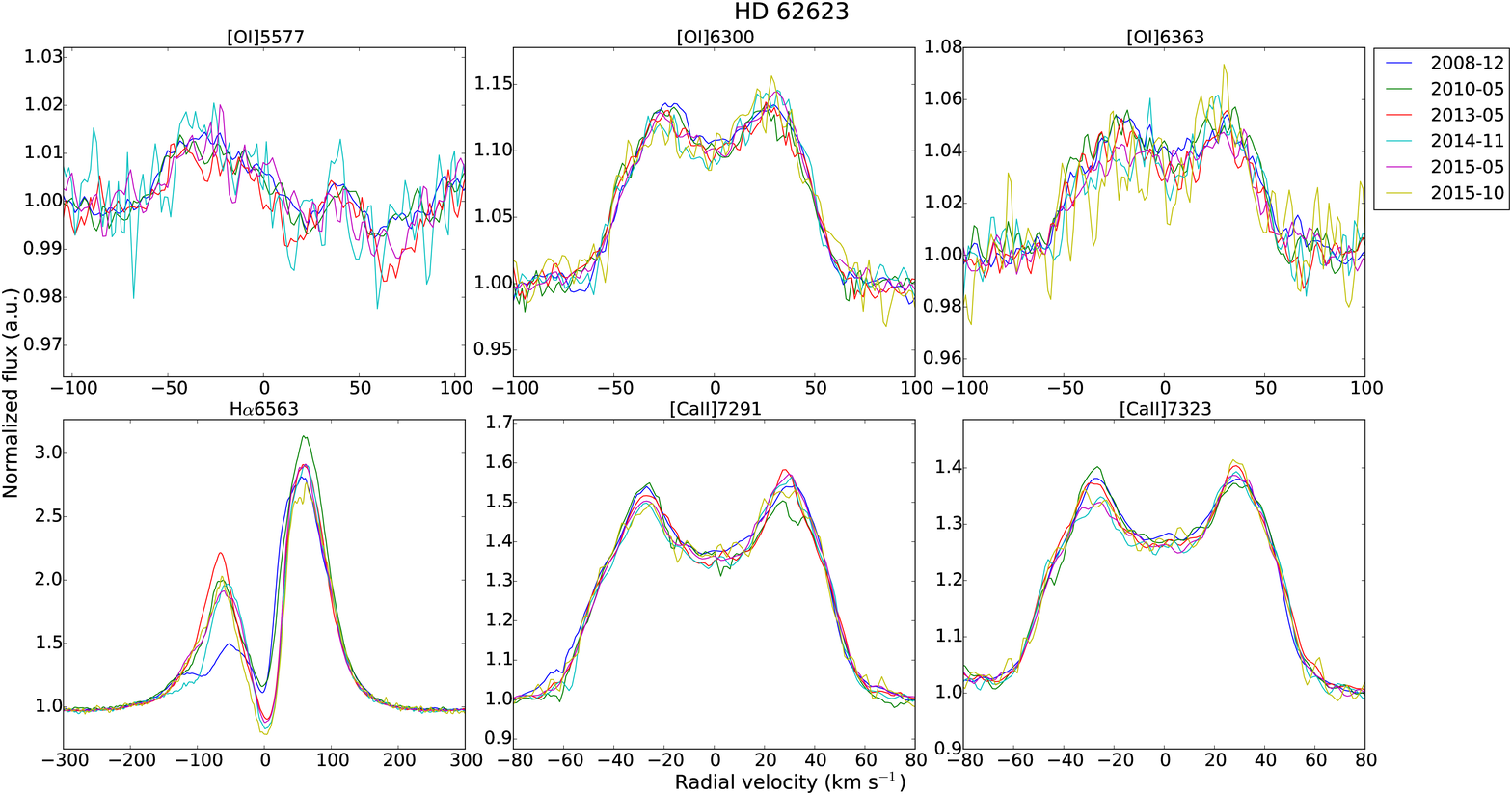}
    \caption{Similar to Fig. \ref{fig:variability-1} but for HD 327083 (1999-2016) and HD 62623 (2008-2015, excluding only the \oi \lam5577 line from 2015-10-12 because of the noise). }
    \label{fig:variability-2}
\end{figure*}

\begin{figure}
	\includegraphics[scale=0.42]{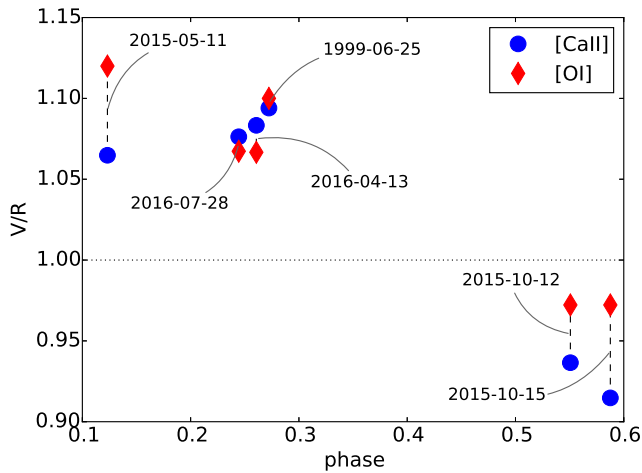}
    \caption{The \textit{V/R} variation for both the \caii \lam7291 and the \oi \lam6300 lines with respect to the phase for the binary HD~327083. The corresponding pair of spectral features  (connected with a dashed line) for each epoch (indicated with an arc) is also shown. }
    \label{fig:hd327083_vor}
\end{figure}

\begin{figure*}
	\includegraphics[scale=0.85]{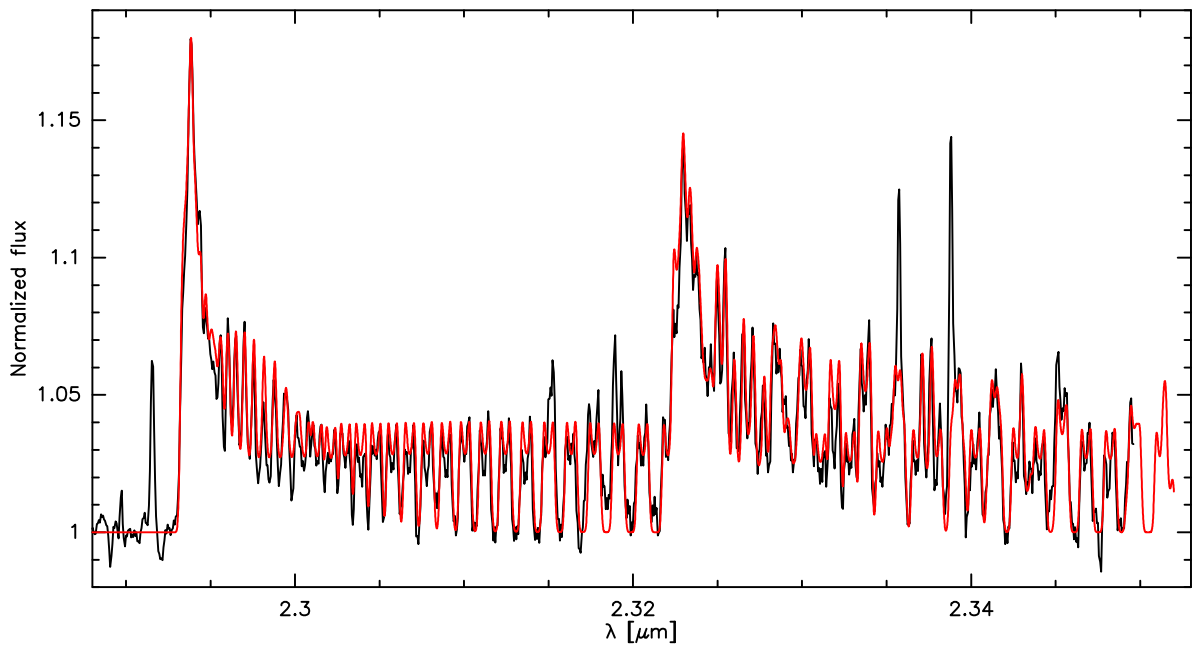}
    \caption{The GNIRS spectrum of HD 62623 (black line), showing the first and the second CO bandheads, and its corresponding model (red line). }
    \label{fig:hd62623_ir_gnirs}
\end{figure*}

\begin{figure*}
	\includegraphics[scale=0.3]{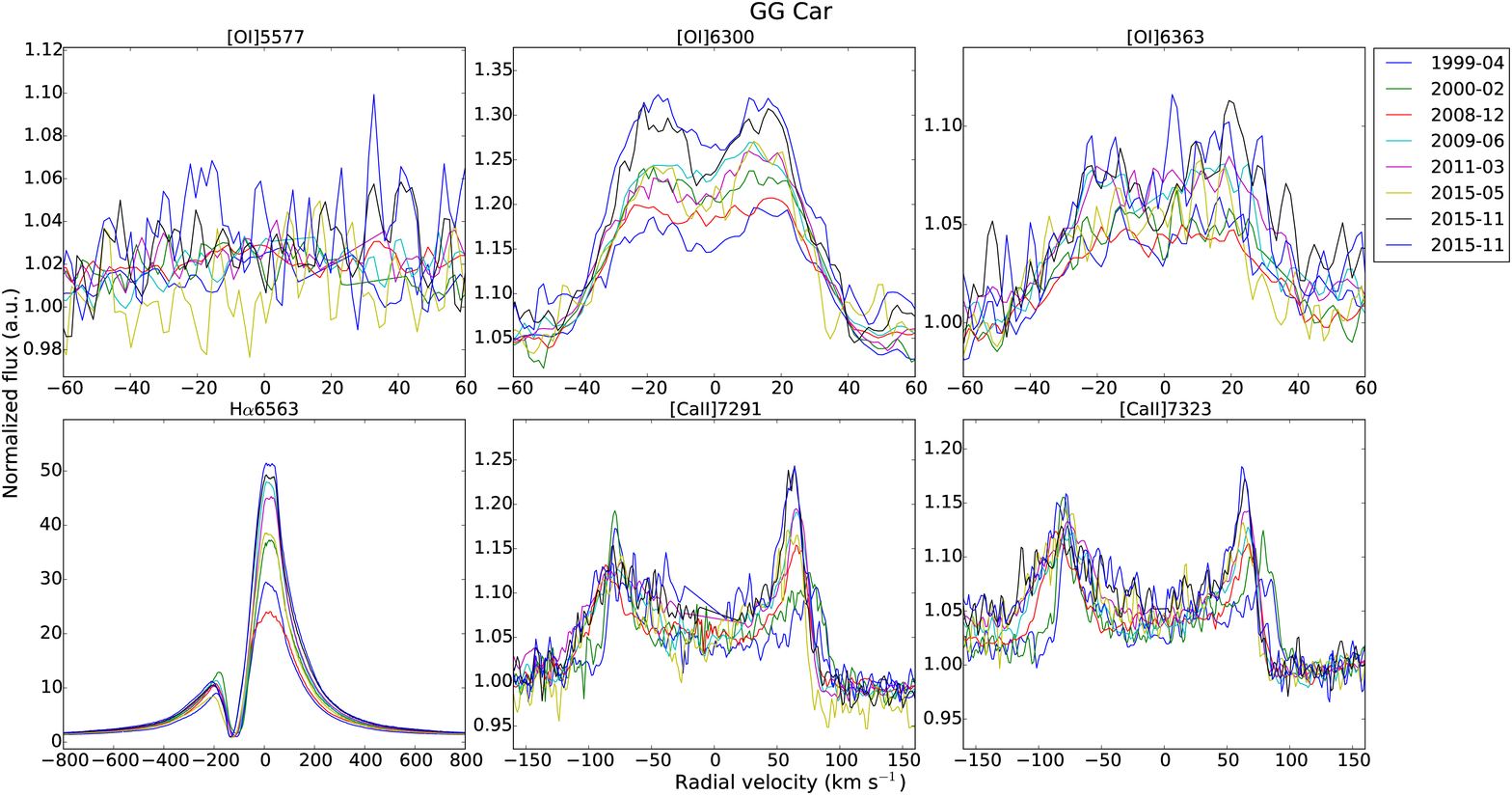}
	\includegraphics[scale=0.3]{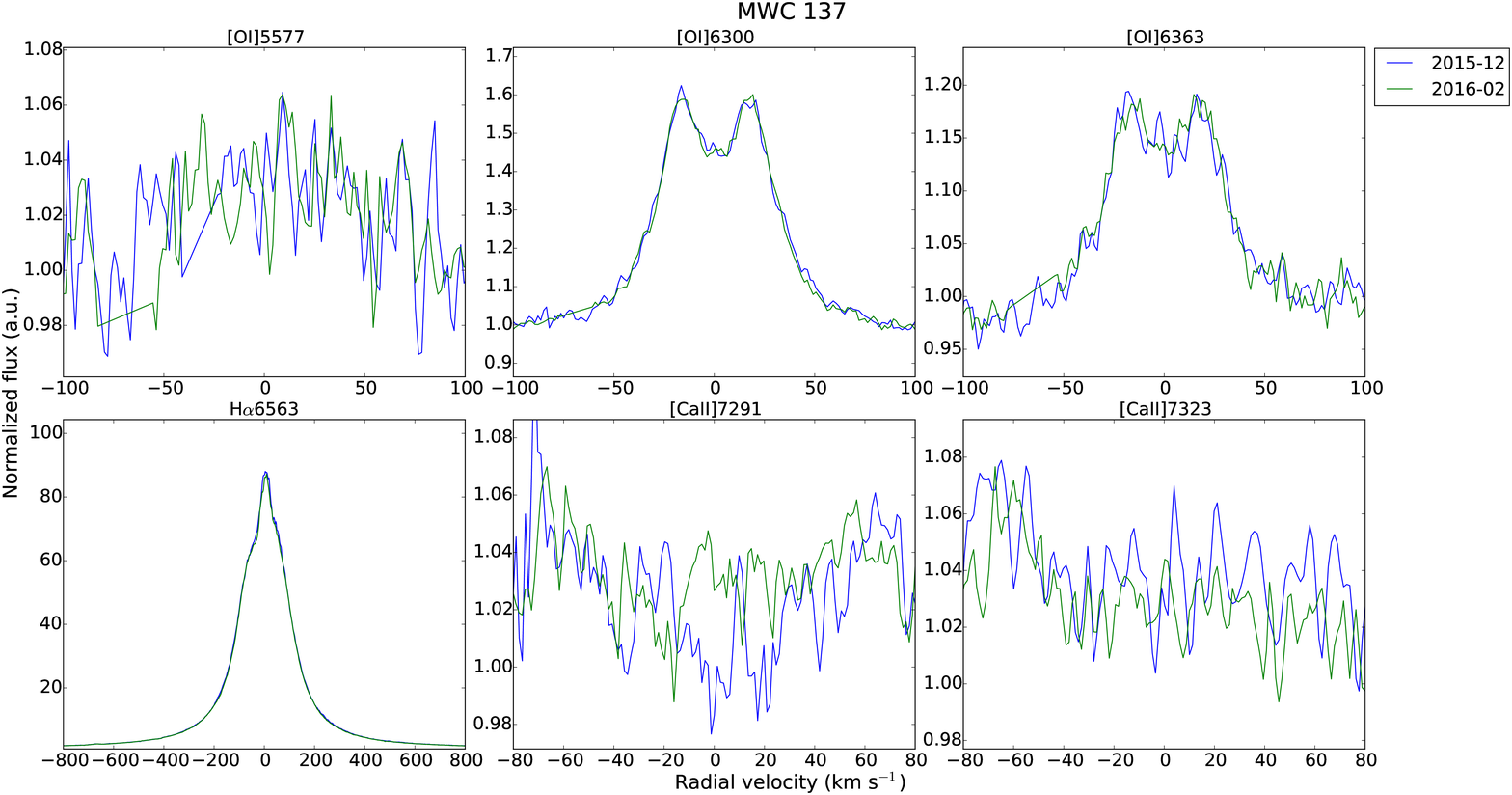}
    \caption{Similar to Fig. \ref{fig:variability-1} but for GG Car (1999-2015) and MWC 137 (2015-2016). }
    \label{fig:variability-3}
\end{figure*}

\begin{figure*}
	\textbf{GG Car}\\
	\includegraphics[scale=0.3, trim=10 0 20 0,clip]{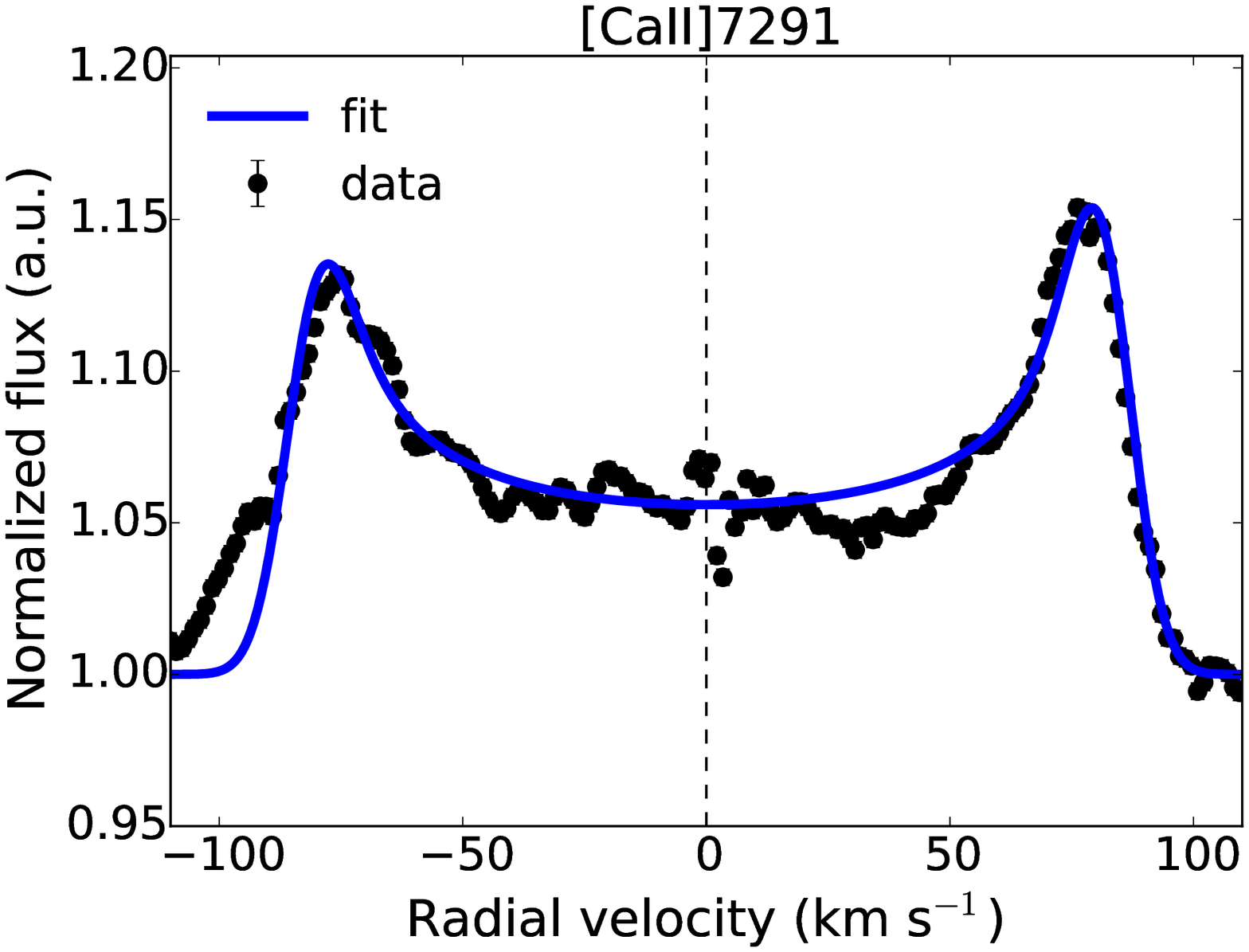}
	\includegraphics[scale=0.3, trim=30 0 10 0,clip]{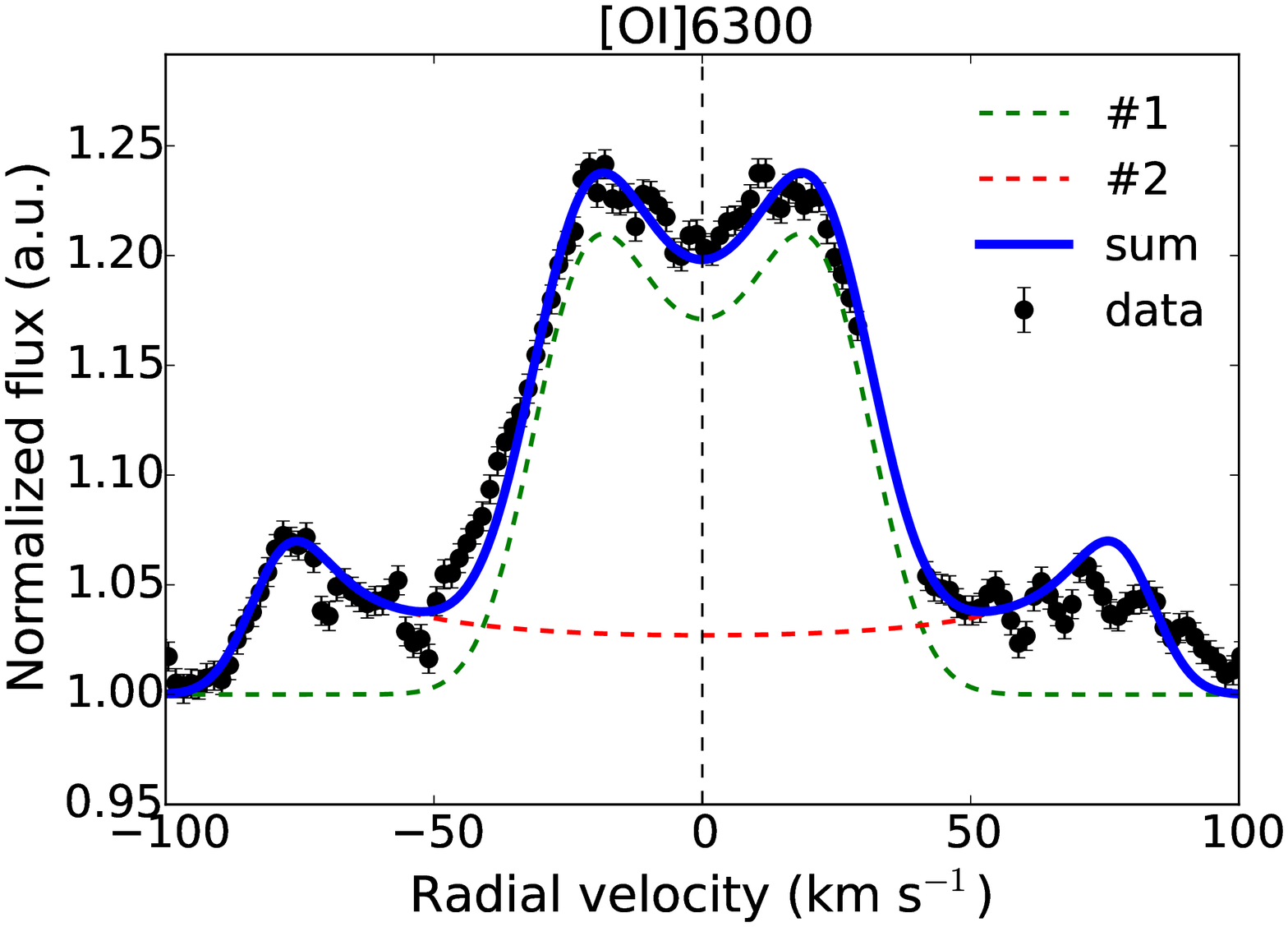} \\
	\textbf{MWC 137}\\
	\includegraphics[scale=0.3, trim=10 0 00 0,clip]{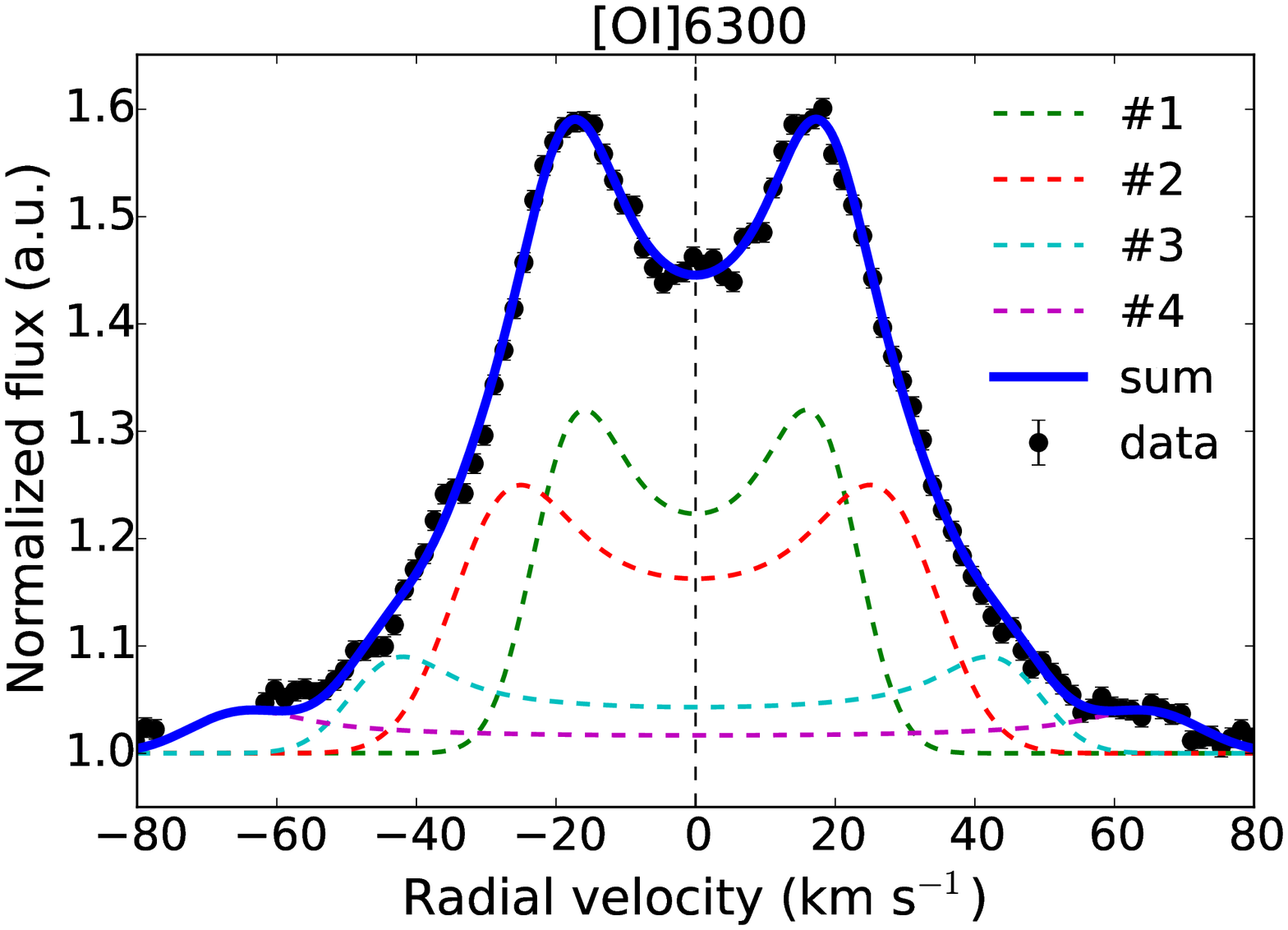} \\
	\textbf{HD 87643}\\
	\includegraphics[scale=0.3, trim=10 0 20 0,clip]{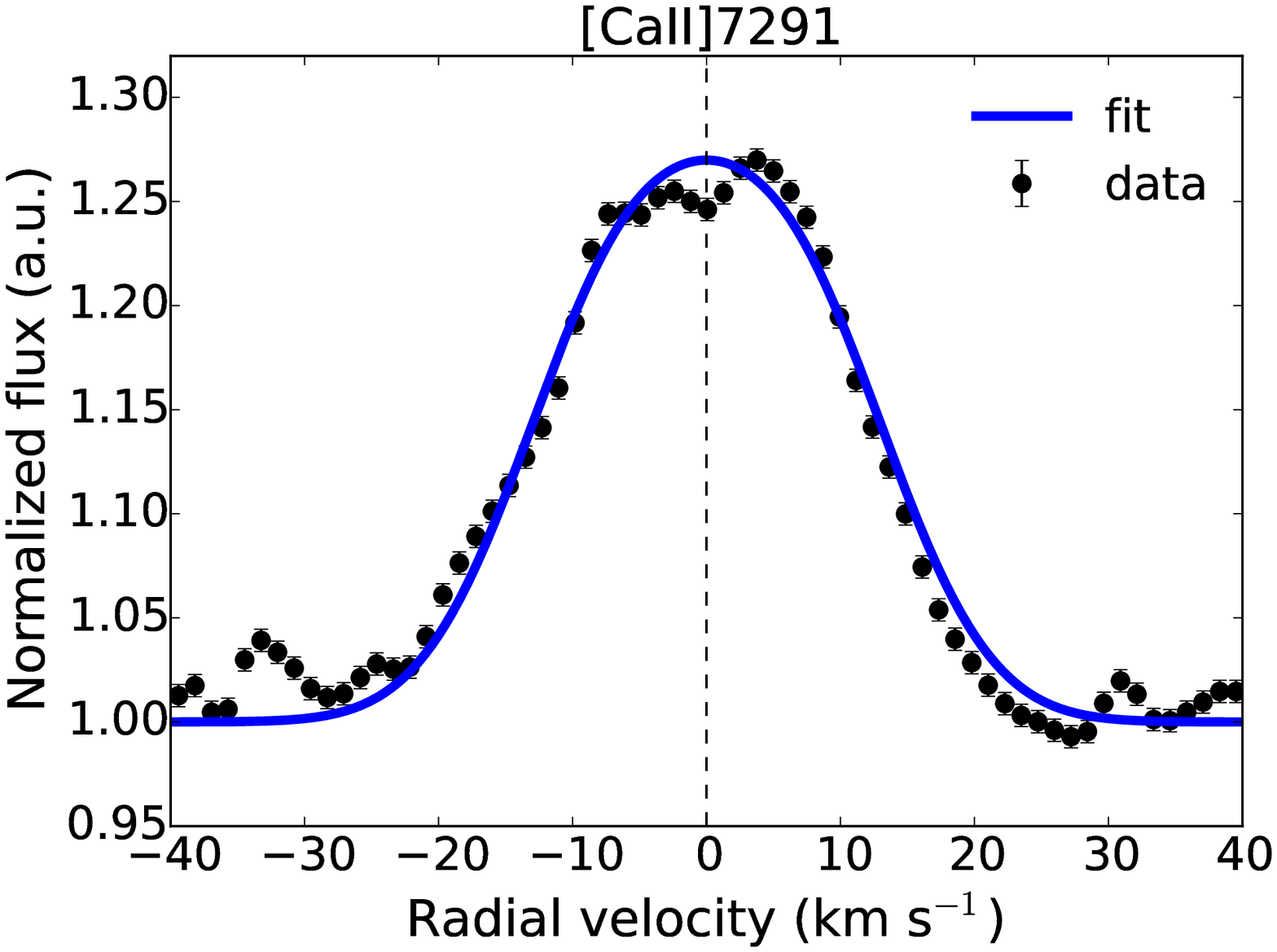}
	\includegraphics[scale=0.3, trim=35 0 20 0,clip]{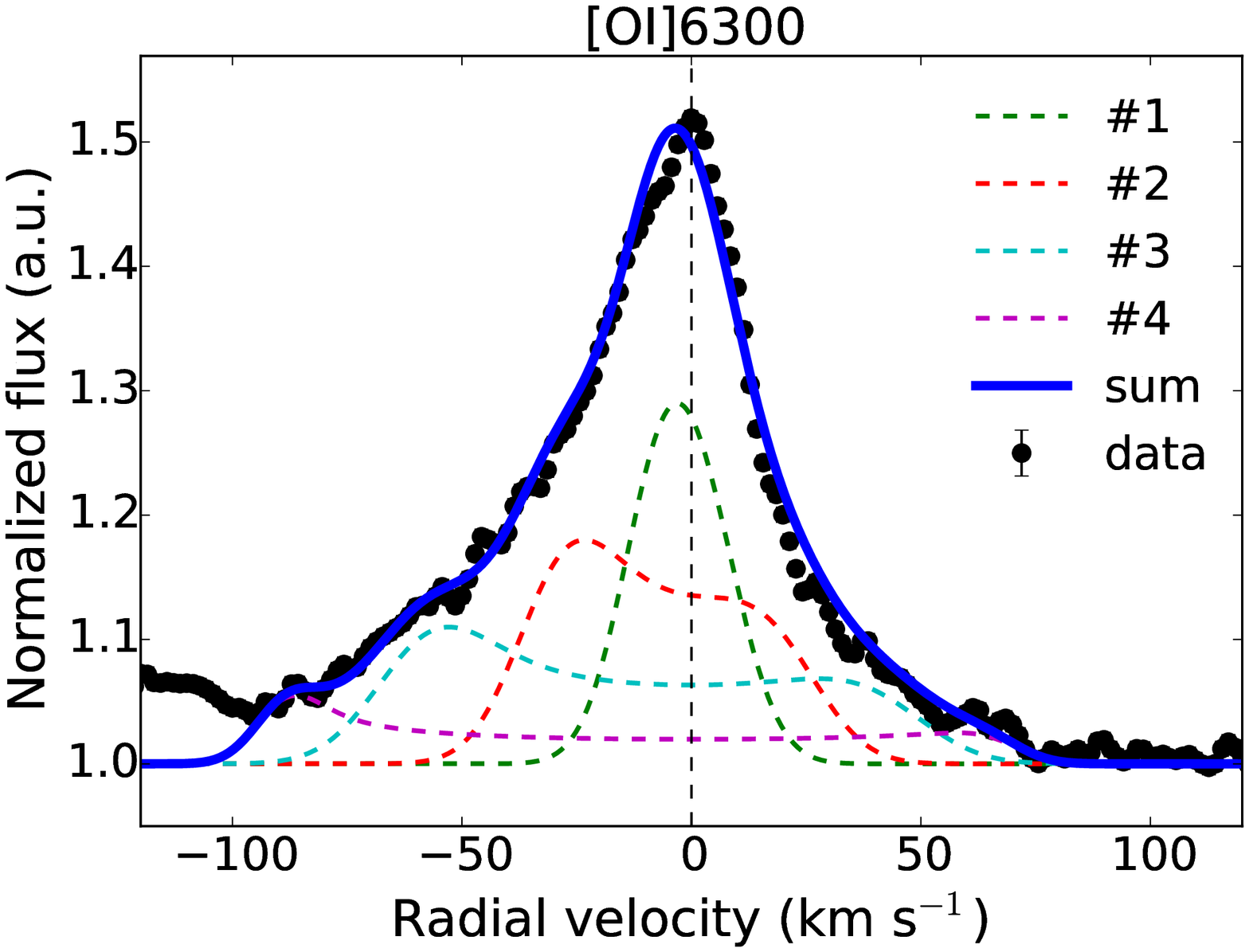} \\
	\textbf{Hen 3-298}\\
	\includegraphics[scale=0.3, trim=10 0 20 0,clip]{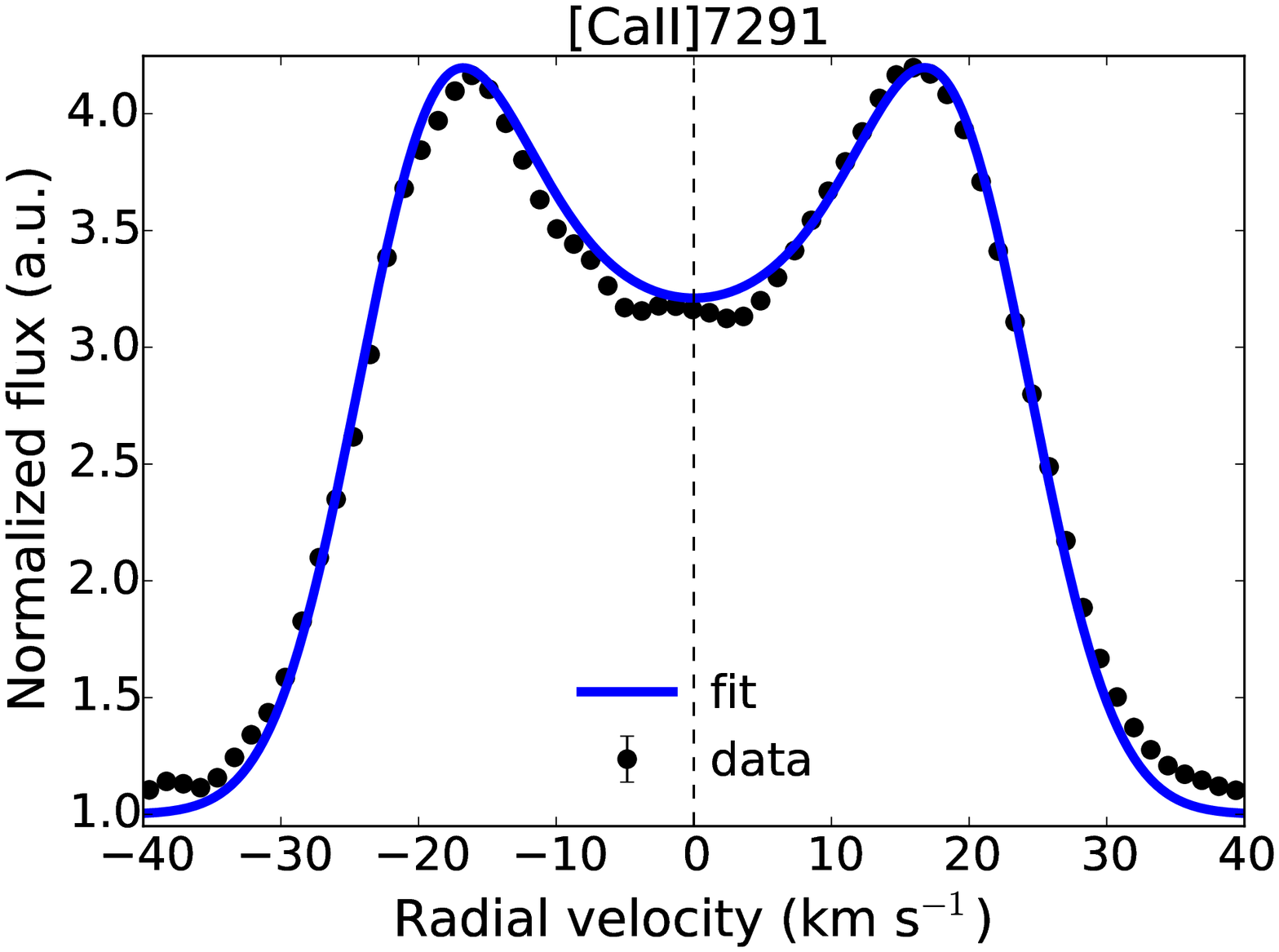}
	\includegraphics[scale=0.3, trim=33 0 18 0,clip]{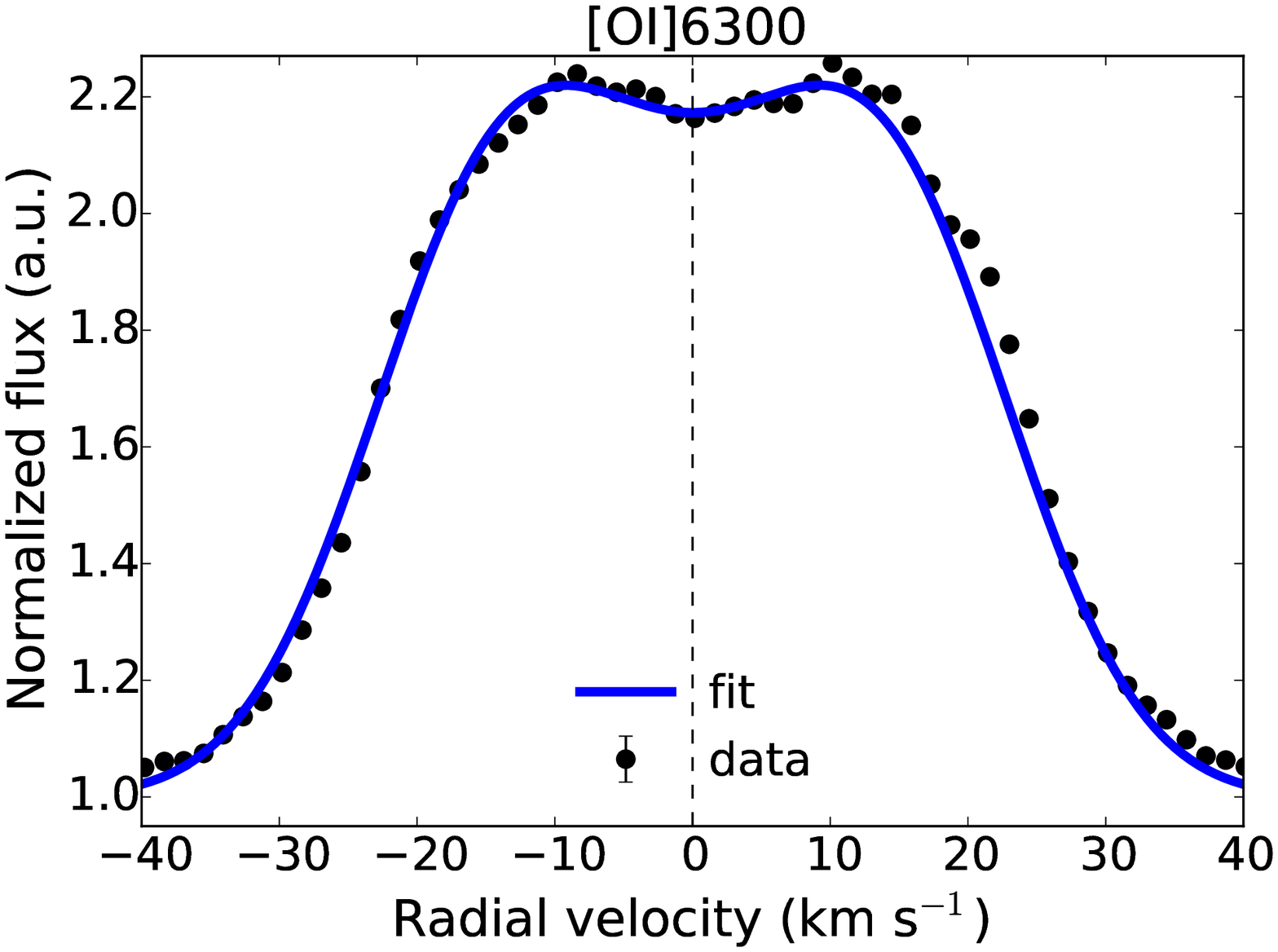} 
	\includegraphics[scale=0.3, trim=31 0 20 0,clip]{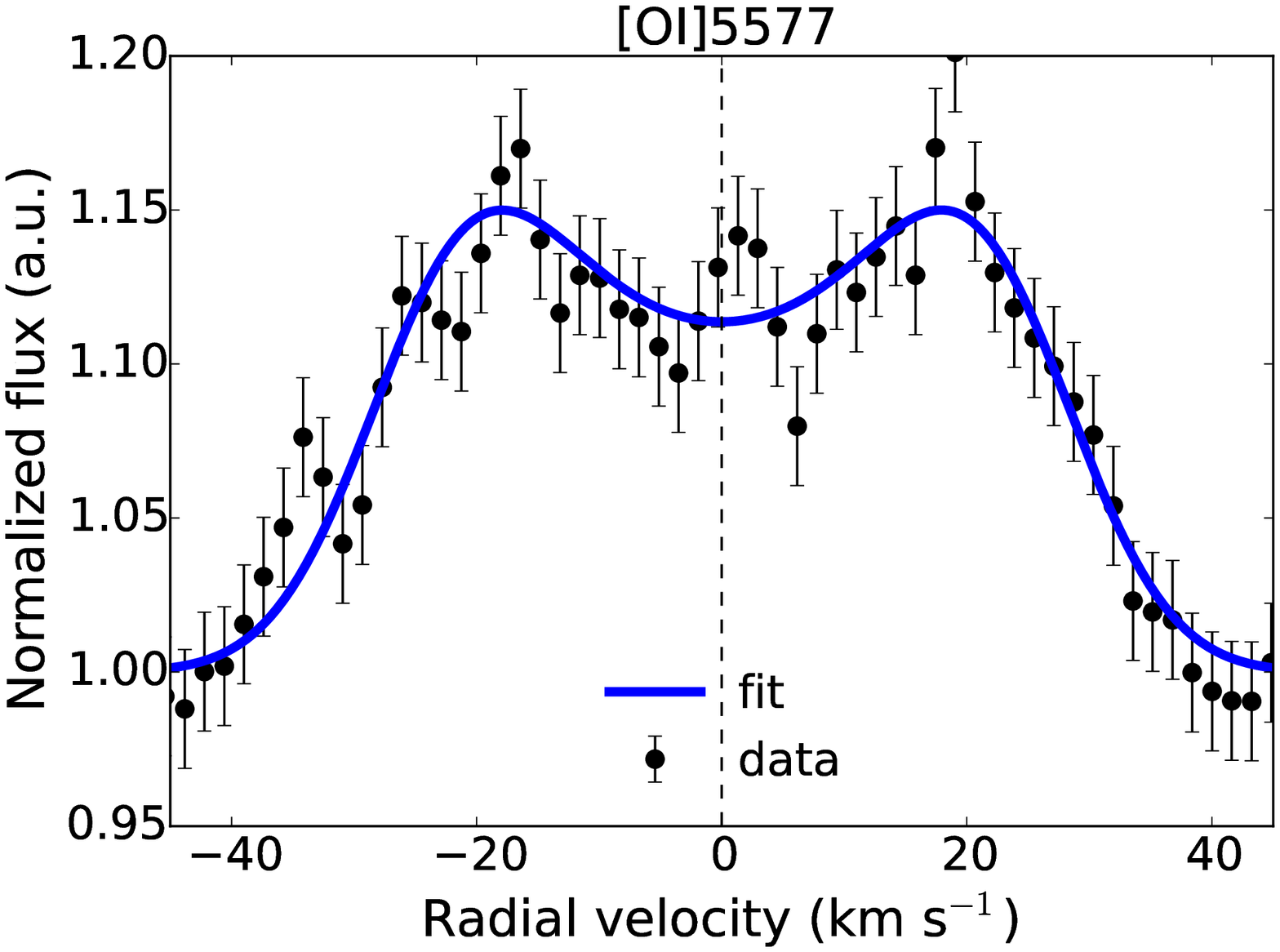} \\
    \caption{Similar to Fig. \ref{fig:fitlines-1}. From top to bottom, \textit{GG Car:} the models of a single partial ring and two complete rings for the \caii \lam7291 (2008-12-22) and the \oi \lam6300 (2000-02-23) lines, respectively, \textit{MWC 137:} the four-ring model for the \oi \lam6300 (2016-02-28) line, \textit{HD 87643:} the complete ring and the four partial-ring models for the \caii \lam7291 and the \oi \lam6300 lines (from 2015-10-13), \textit{Hen 3-298:} the \caii \lam7291 (2015-05-10), the \oi \lam6300 (2016-01-11), and the \oi \lam5577 (2016-01-11) lines with their corresponding models of single complete rings. }
    \label{fig:fitlines-2}
\end{figure*}

\begin{figure*}
	\includegraphics[scale=0.3]{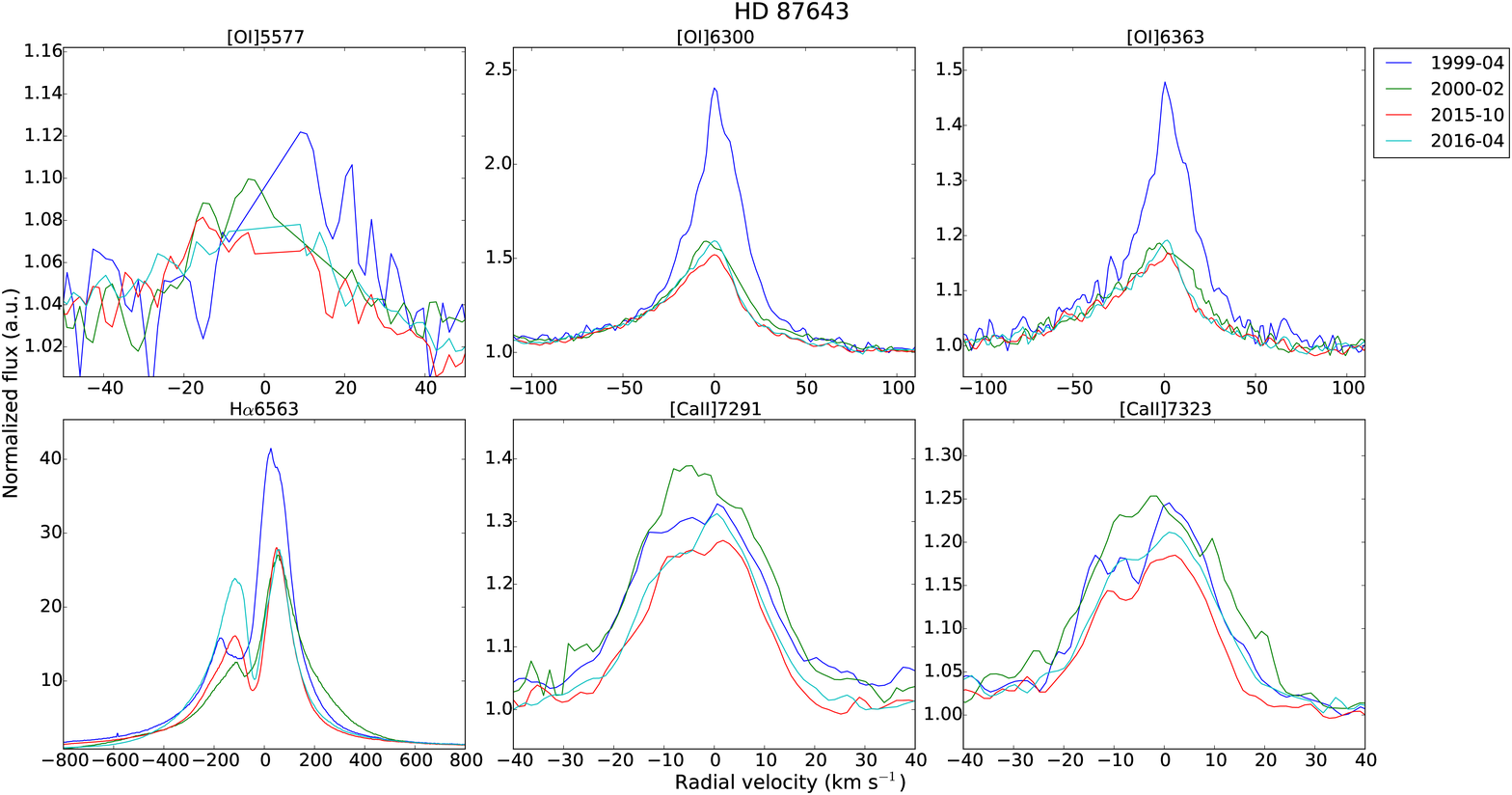}
	\includegraphics[scale=0.3]{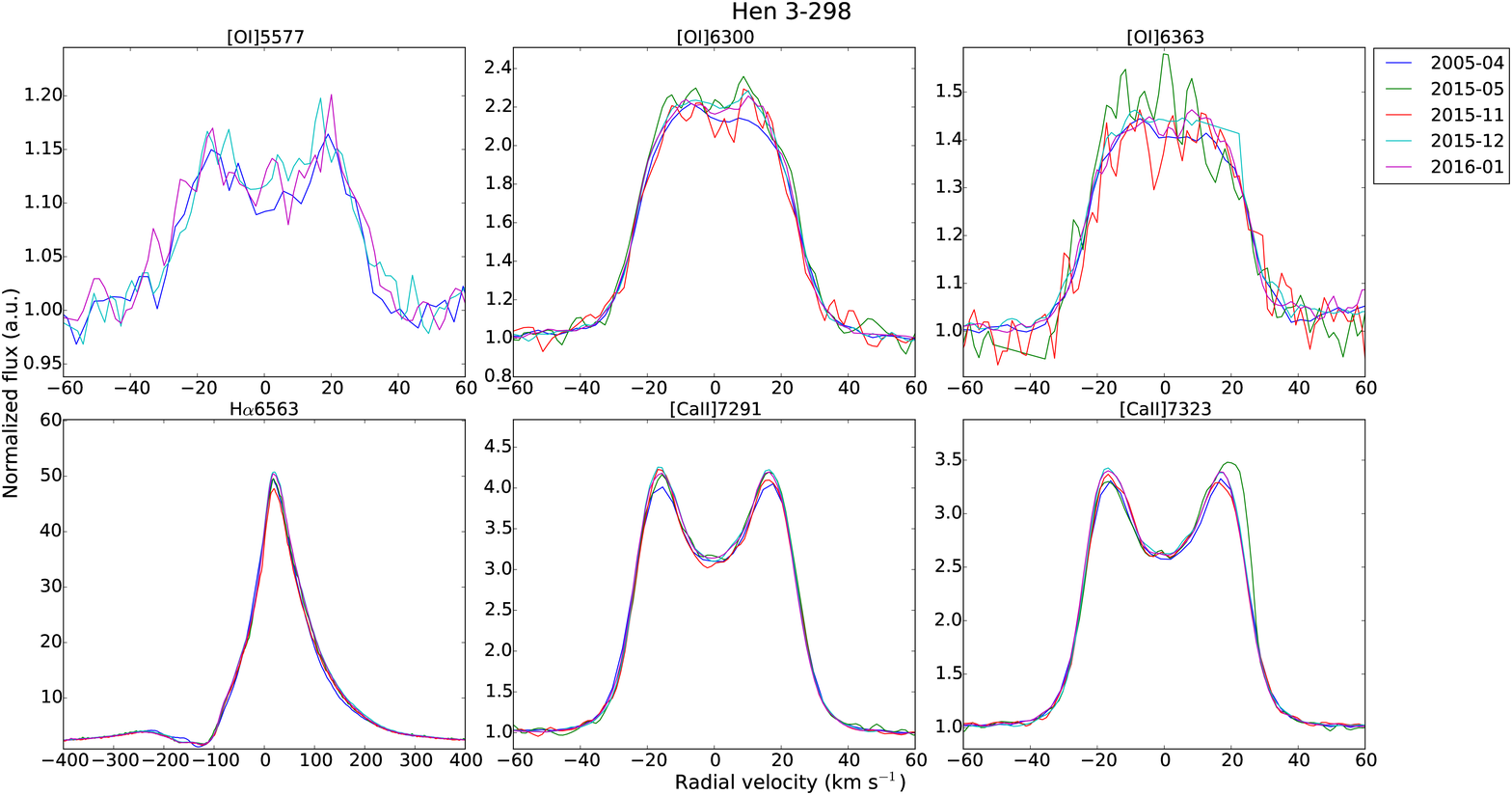}
    \caption{Similar to Fig. \ref{fig:variability-1} but for HD 87643 (1999-2016) and Hen 3-298 (2005-2016, excluding the \oi \lam5577 line from 2015-05-11 and 2015-11-26 because of the noise). }
    \label{fig:variability-4}
\end{figure*}

\subsection{CPD-57 2874}

The presence of the CO first overtone bandhead in a near-IR spectrum was first pointed out by \cite{McGregor1988}. \cite{Muratore2012} used high-resolution CRIRES spectra to identify the double-peaked features of CO and a preliminary fit of these data revealed a rotating ring at 65~\kms (line-of-sight velocity). In this work we reprocessed the original CRIRES data (from 2009-12-02) to derive a more robust conclusion regarding the kinematics of the CO. Using the first CO bandhead we find a de-projected rotational velocity of 130 \kms (see Fig. \ref{fig:ir_spectra}). Additionally to CO, \cite{Kraus2015} detected SiO emission, originating from a 110 \kms ring.

In Fig. \ref{fig:variability-1} we present the line profiles for the optical lines as derived from all available observations between 2008 and 2016. There is no sign of the \oi \lam5577 line. The \caii \lam7291 displays a deep central depression that demands a multi-ring model to properly fit its highly asymmetric profile. To account for the peaks we use two partial rings with similar rotational velocities but integrated over different velocity ranges (green and red dashed lines shown in Fig. \ref{fig:fitlines-1}; see Table \ref{tab:kin-cpd-572874} for more details). Then, we use another set of two (complete) rings that fit the extended wings of the \caii line (cyan and magenta dashed lines in Fig. \ref{fig:fitlines-1}). We interpret these results as three rings where \caii is forming (at 108.1, 158.3, and 210.5 \kms) of which the first one is a ring with inhomogeneities, due to either the absence of the gas or because the local conditions are not proper to produce detectable \caii emission. To model the \oi \lam6300 line we use four (complete) rings (at 44.5, 88.3, 120.3, and 166.3 \kms, shown as dashed lines in  Fig. \ref{fig:fitlines-1}). If we combine all these velocities then we see that we get a rather complex and alternate combination of emitting regions. By starting closer to the star (i.e. the largest rotational velocities) we get a \caii ring at 210.5 \kms, a \oi ring at 166.3 \kms, and a \caii ring at 158.3 \kms. Due to our typical ring width of $\sim11$ \kms the last two rings actually overlap. Further away from the star, we find the CO ring at 130 \kms, another \oi ring at 120.3 \kms (perhaps with some overlap with the CO and SiO rings), the SiO ring at 110 \kms, an non-homogeneous \caii ring at 108.1 \kms (overlapping with the SiO ring), and another two \oi rings at 88.3 and 44.5 \kms. It is possible though that the CSE in the case of CPD-57 2874 may not consist of individual rings, but is an inhomogeneous disk with alternate regions of molecular and atomic emission. 

Taking into account the stellar mass estimates for CPD-57 2874 of 15-20 $M_{\sun}$ by \cite{DomicianodeSouza2011} we can calculate the distances of these rings from the central star (see Table \ref{tab:results}). The derived ring radii show us that the gas emission extends up to $\sim9$ AU, which is consistent with the picture we have from the interferometry, as \cite{DomicianodeSouza2011} find a broad near-IR emission region at $\sim8.5$ AU and the bulk of the dust further away ($\sim11-14$~AU). 

In general, the lines do not show significant variability over the observed period of 8 years. There is only a small intensity increase in the last observation (2016-03-13) for the \oi and \caii lines, but with identical profiles. Likewise, the H$\alpha$ line displays a stronger red peak, while its blue peak shows some variability, possible due to changes in the wind. Observations from 1988 at similar resolution ($R\sim55000$) show that both the H$\alpha$ and the \oi \lam6300 lines appear similar to our observations \citep[][cf. fig. 1 and 2]{Zickgraf2003}. Nevertheless, it is hard to argue if there has been any change or not during the 1988-2008 period.

\subsection{HD 327083}

CO emission features have been detected in the near-IR spectra of HD 327083 since the works of \cite{Whitelock1983} and \cite{McGregor1988}. A preliminary derivation of the projected rotational velocity for the CO ring at $55\pm1$ \kms was given by \cite{Andruchow2012}, using the Phoenix/Gemini IR spectrometer in May 2010. From our CRIRES spectra in June 2010 we obtain a de-projected rotational velocity of $86\pm1$ \kms (Fig. \ref{fig:ir_spectra}). Additionally, \cite{Kraus2015} detected SiO, originating from a rotating ring at 78 \kms.

Fig. \ref{fig:variability-2} presents the line profiles for the 1999-2016 period, for which there is no sign of the \oi \lam5577 line. The \caii and \oi doublets display asymmetrical profiles. To properly model the strong central depression of the \caii line, we have to use two partial rings that have similar velocities but different integration ranges (see Fig. \ref{fig:fitlines-1} and Table \ref{tab:kin-hd327083} for details). We find a range of 70 -- 77 \kms throughout all epochs. As the typical ring-width is 10.5 \kms we conclude that the \caii emission line originates actually from a single ring region (with an averaged value of 75.4 \kms) with inhomogeneities. Alternate profiles on different dates are a possible indication of revolving inhomogeneities.

Regarding the \oi \lam6300 line we fit the first epoch (1999-06-25) with a single partial ring of 75 \kms. For the next epoch (2015-05-11) we find that a two partial-ring model (with velocities of 58 and 76 \kms) is necessary. In October 2015 the outermost ring (at 58 \kms) is not present (probably dissolved or not dense enough). However, we still need two partial rings of similar velocities (at 70 and 74 \kms) to fit the observed profiles, corresponding to a single ring with inhomogeneities. Starting with 2016-04-13 we have an asymmetric profile without any good constraint on its red part (in contrast to its blue part) that makes the fit quite loose. Nevertheless, we need to use a two partial-ring model with rotational velocities of 66 and 80 \kms. This is required since a single ring in between those values (e.g. $\sim72-74$ \kms) does not fit the observed profile. Similarly for 2016-07-28 we get two rings at 61 and 78.5 \kms. The innermost rings (at $\sim78-80$ \kms) identified in these two epochs of 2016 are consistent with previous ones. The outermost rings (at 66-61 \kms) though are more interesting. Alike the ring at 58 \kms observed in 2015-05-11, these rings may indicate material that left the single ring we see in October 2015 and dissolve further away with time (from 66 to 61 \kms). This would correspond to an extremely fast process, for which we do not have much evidence (such a ring exists only for GG Car). An alternative scenario is that of a relatively tenuous outer ring, which is fragmented to such a degree that only some parts become observable in certain epochs. 

Hence, we opt to use two rings for the emitting region of the \oinos, one with inhomogeneities (similar to \caiinos) at an averaged rotational velocity of 74.1 \kms and a tentative fragmented one at 61.7 \kms. The velocities found for the \caii and \oi forming regions are similar to the SiO ring, which imply a common location for these gases. CO forms (at slightly higher rotational velocity) another ring further closer to the star, although it is possible that due to the typical ring-width of 10.5 \kms it might not be totally separated from the other ring.

A number of stellar mass estimates have been derived in the literature, ranging from 60 $M_{\sun}$ (using non-LTE wind modeling of the Balmer lines; \citealt{Machado2003}) to 20 $M_{\sun}$ (detection of absorption lines from neutral metals and radial velocity variations; \citealt{Miroshnichenko2003}) and 25~$M_{\sun}$ (using interferometry; \citealt{Wheelwright2012b}). The CO ring, which is found closer to the star, is located at a distance of 7.2, 2.4, or 3.0 AU, considering the different masses respectively. In all cases, these radii are larger than the binary separation (at $\sim1.7$ AU, \citealt{Wheelwright2012}) that makes the whole structure circumbinary. We use the latest mass estimate, derived from the interferometric results, to calculate the ring radii in Table \ref{tab:results}. Given this mass, the rings (up to $\sim4$ AU) are located closer to the source than the dusty disk revealed by \cite{Wheelwright2012} at 5.1~AU. We do not detect any other emitting region closer to the source than the CO, although \cite{Wheelwright2012} observed a more compact region of Br$\gamma$ emission (possible originating from either of the stars in the binary). 

In Fig. \ref{fig:hd327083_vor} we present the ratio of the blue to red peak (\textit{V/R}) at each epoch. The \textit{V/R} of the \oi and the \caii lines vary in phase, with $V/R>1$ for phases $<0.5$ and $V/R<1$ for phases $>0.5$ (assuming an orbital period of $\sim107$ days; Cidale et al., in preparation). However, for the latter case only observations from October 2015 fall into this phase domain. Given the circumbinary nature of the structure around HD 327083 and the possibility that these rings are not necessarily circular or homogeneous, these asymmetries may be due to the excitation/heating of the gas (hence stronger emission) at phases when the hotter component is closer to one or the other ride of the rotating rings.

The H$\alpha$ line displays also profile changes, especially in October 2015 observations when the blue peak almost disappears. Except for these, H$\alpha$ displays smaller intensity changes. Even though we do not detect any sign of the \oi \lam5577 line, we do notice the presence of some absorption features (attributed to the cooler companion) that display radial velocity variations. At least partly, this variability could be due to the orbital modulations imposed by the binary inside the circumbinary structure.

\subsection{HD 62623}

HD 62623 is the only A[e] supergiant known in our Galaxy (A2.7Ib; \citealt{Chentsov2010}). Even though it is the brightest source in our sample ($V=3.93$ mag) and has been a target of many studies (see the historic overview in \citealt{Chentsov2010}), CO emission features in the near-IR have remained rather elusive. HD 62623 is the only object for which we have an additional, though slightly lower resolution, near-IR spectrum (from GEMINI/GNIRS; Fig. \ref{fig:hd62623_ir_gnirs}). Despite the time difference of 29 months between the individual observations, we obtained from both spectra the same rotational velocity of 53 \kms, indicating that the molecular ring is stable. In addition, since the GNIRS spectrum covers also the second band head, we could determine the temperature $(T_{\textrm{CO}}=1800\pm100\, \textrm{K})$ and the column density $(N_{\textrm{CO}}=(2.5\pm0.5)\times10^{20}\, \textrm{cm}^{-2})$ of the CO forming region. With the derived velocity, the CO ring resides slightly closer to the star than the SiO ring, for which a velocity of 48 \kms was found \citep{Kraus2015}. 

Optical observations cover a period of 7 years (2008-2015) and they are shown in Fig.  \ref{fig:variability-2}. In the case of HD 62623 a rather weak signal from \oi \lam5577 is apparent, displaying always a blue peak. We can model this profile as a single partial ring, which is less than half filled (see Table \ref{tab:kin-hd62623}), and with a rotational velocity of 57.1$\pm$6.3 \kms. Due to the large error and the weakness of the (purely blue-shifted) line we would consider the identification of this emission feature as \oi and hence this ring as rather tentative.

Regarding the \caii and \oi doublets, we can fit both with a two ring model at $\sim40.7$ \kms and $\sim61.3$ \kms (see Fig. \ref{fig:fitlines-1}). That means that the two atomic gases coexist in two clearly separated forming regions (as the typical ring-width for HD 62623 is $\sim9$ \kms). It is interesting to note though observations up to 2013 can be fit with complete rings, while after 2014 we need to insert some inhomogeneity to properly model their slightly asymmetrical profiles. This could be interpreted as the development of inhomogeneities within both rings, however, its simultaneous appearance in both rings may indicate a change in the geometry along the line-of-sight. In between the rings of atomic gas we find the molecular rings (CO and SiO). In the innermost rings we have a common forming region of \caii and \oi  \lam6300 lines (at $\sim61.3$ \kms) and the presence of the (tentative) \oi \lam5577 line (at 57.1 \kms). 

According to the interferometric results by \cite{Millour2011}, the equatorial CSE consists of a hot ionized disk close to the star (up to 1.3 AU) and a dusty disk further outside (at 4 AU). Using the mass range of 9-10.5 $M_{\sun}$ \citep{Aret2016}, we can estimate the ring radii (see Table \ref{tab:results}) within the range 2.3-5.0 AU, fairly close to the location of the dusty ring. Although not directly detected, a binary scenario has been used to explain the CSE of HD 62623 \citep{Millour2011}. In this case, the maximum possible binary separation (1.2-1.7 AU) is smaller than the smallest ring-radius estimate (at 2.3 AU), which makes this structure circumbinary (similar to HD 327083). 

In general, there are only minor intensity variations in the peaks of \caii lines, with the \oi lines being more stable. The most striking change is in the blue peak of H$\alpha$, which is characterized by a sharp increase from 2008 to 2010. From then on, the ratio between the blue and red peak remains almost stable, with a small increase in 2014 and returning back to the previous state in 2015. Given the relative stability of the other lines this variation in H$\alpha$ may be the result of changes in the wind of the star, which do not affect the circumstellar rings.

\subsection{GG Car}
\label{sec:GGCar}

\begin{figure}
	\includegraphics[scale=0.42]{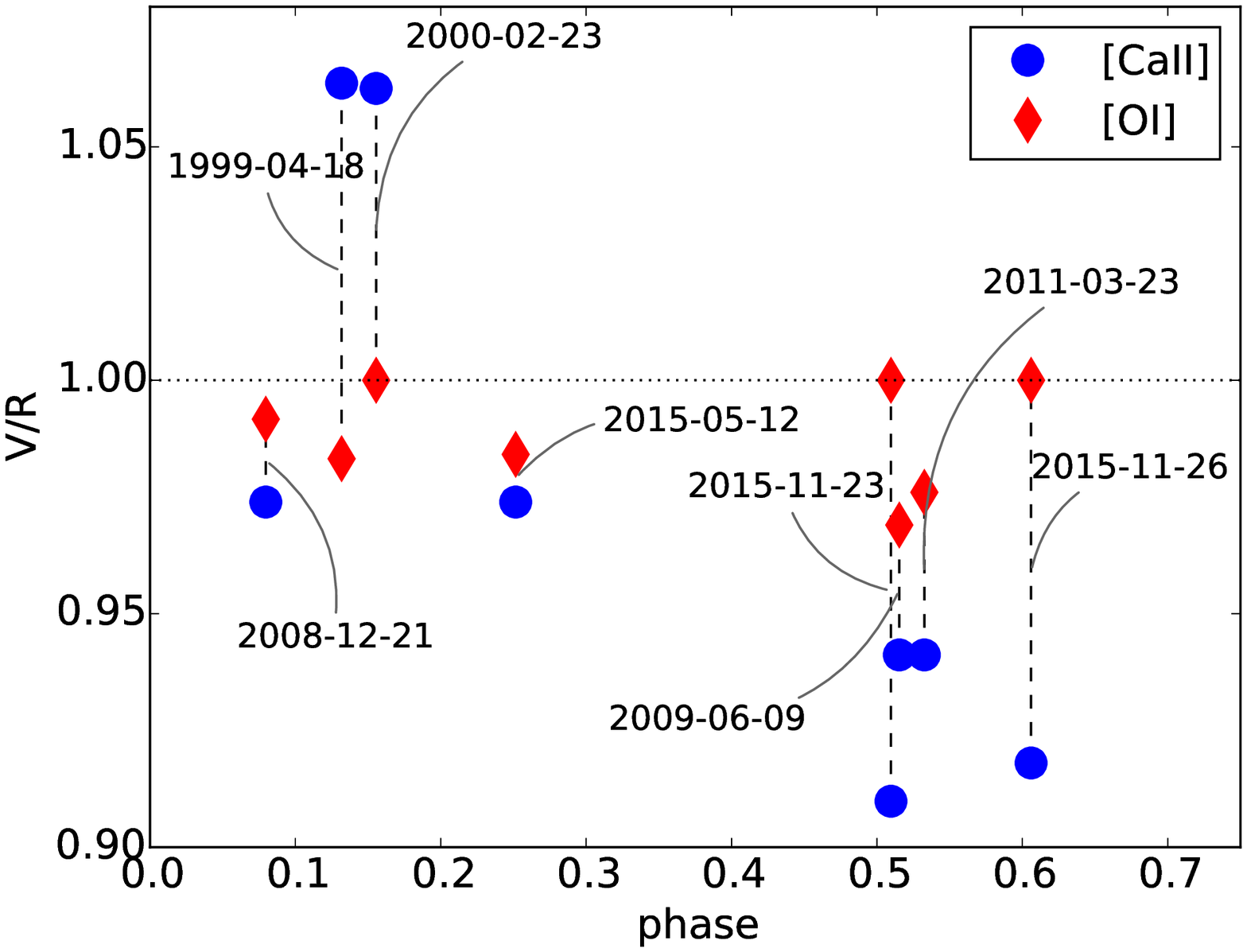}
    \caption{Same as Fig. \ref{fig:hd327083_vor}, but for GG Car. The \caii \lam7291 line displays stronger variation of \textit{V/R} than the \oi \lam6300 line. }
    \label{fig:ggcar_vor}
\end{figure}

\begin{figure}
	\includegraphics[scale=0.42]{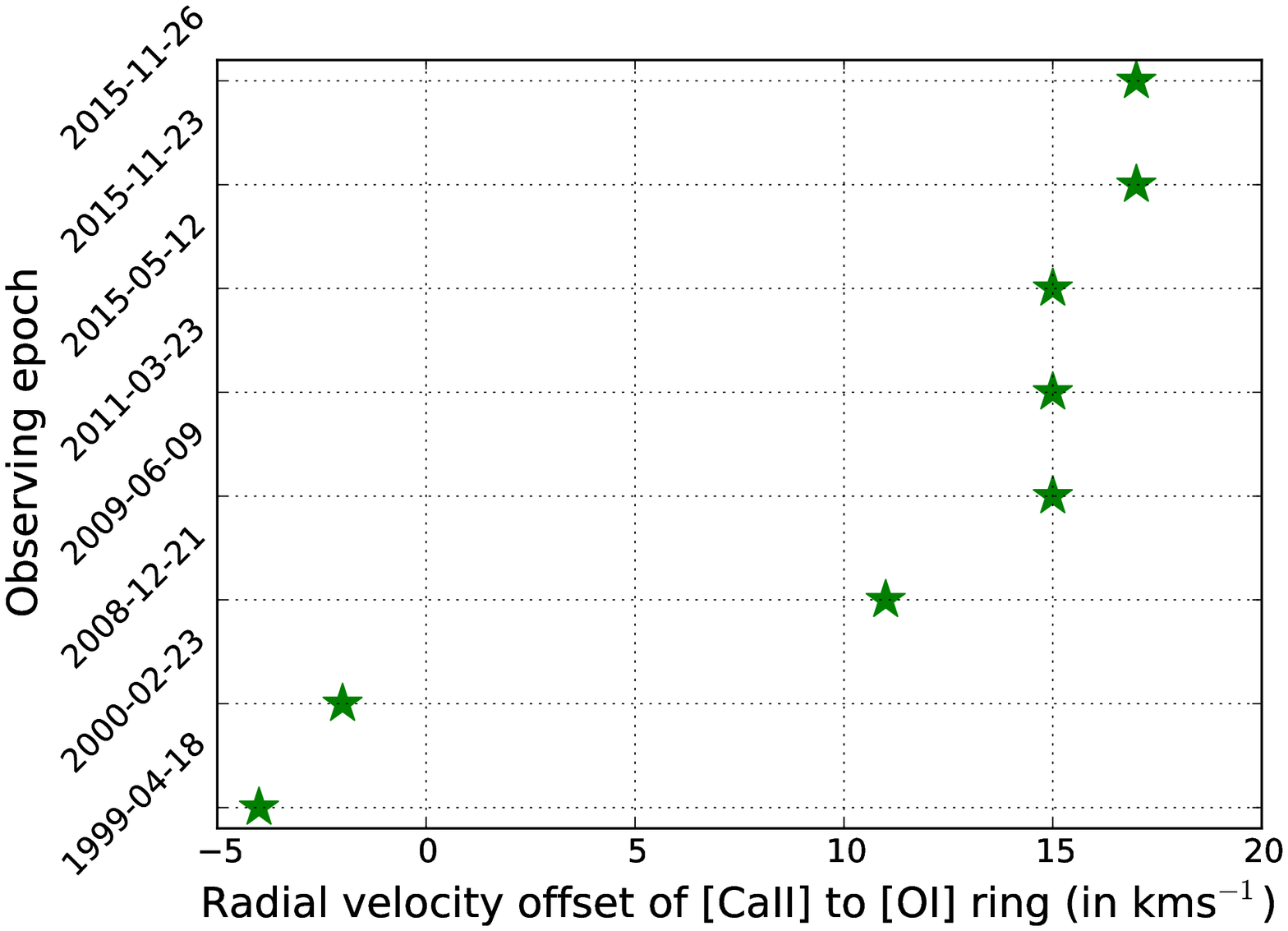}
    \caption{The time evolution of the radial velocity offset of the \caii \lam7291 line to the \oi \lam6300 line, for the case of GG Car. }
    \label{fig:ggcar_rvoffset}
\end{figure}

The IR spectrum of GG Car has been described extensively in previous works \citep[e.g.][]{McGregor1988, Morris1996}. However, the kinematical properties of the CO emission features were investigated by \cite{Kraus2013}, who identified a detached rotating ring at 91.5 \kms (see in Fig. \ref{fig:ir_spectra} this model on top the CRIRES spectrum). 

All observations (1999-2015) of GG Car are presented in Fig. \ref{fig:variability-3}. There is no sign of the \oi \lam5577 line, contrary to the doublets of \caii and \oinos. We model the \caii \lam7291 line with a single ring at 95.8 \kms, which implies a stable forming region throughout all epochs (see Fig. \ref{fig:fitlines-2}) although the profile asymmetries point to inhomogeneous rings. The \oi \lam6300 line can be modeled with two complete rings throughout all epochs, with velocities ranging from 91.5-77 to 35.5-28 \kms. However, for the 2008-12-22 line profile we need to add another (third) ring at 51 \kms located in between the other two. This might imply some movement of gas from a region of high rotational velocity to a low one as it dissolves, although we note that we do not find this (or any similar) ring later on (even though the next observation is only half a year later, on 2009-06-09; see Table \ref{tab:kin-ggcar}). Moreover, it is interesting to point out that after 2009 the velocity found for the innermost ring decreases from $\sim91$ \kms to $\sim80$ \kms, and we can only speculate that this may be associated with the formation of the third ring observed in 2008. In contrast, the outermost one has remained stable. Therefore, we opt to describe the forming regions for \oi with two rings at averaged velocities of 84.3 and 29.9~\kms.

Given our typical ring-width of 9 \kms, we suggest the existence of two distinct rings around GG Car, one close to the star where the atomic gas coexists with CO, and another one further outside where only \oi emission is excited. Using the stellar mass estimate of 38 $M_{\sun}$ \citep{Kraus2013} we find ring radii of 3.7-4.7 AU and 37.7 AU. GG Car is an eccentric binary system (with a maximum separation of 0.83 AU) and the CO emission originates from a circumbinary ring \citep{Kraus2013,Marchiano2012}. Since the atomic gas of the innermost ring follows the CO behavior we conclude that the whole structure is circumbinary. This is in contrast to the findings of \cite{Marchiano2012} who used a spectral energy distribution fitting to calculate a gaseous envelope located at $\sim0.5$ AU, lying in between the two components of the binary.

With respect to the intensity variability of the lines, the \oi line displays the weakest lines in 1999 and 2008, and the strongest ones in November 2015. The \caii line though does not show significant intensity variation, its \textit{V/R} varies considerably with respect to the \oi line (Fig. \ref{fig:ggcar_vor}). Furthermore, we find that the \caii line displays an offset with respect to values measured for the \oi line ranging from -4 to +17 \kms (on top of the systemic velocity of -22 \kms). In Fig. \ref{fig:ggcar_rvoffset} we show the evolution of this offset with time, which may be indicative of displacements of these rings/lines with respect to each other.  

We note also the strong variation in the H$\alpha$ line. There are significant changes in the intensity of its red peak with a \textit{V/R} range of $\sim0.2-0.45$ (in 2015 and 2008, respectively). There is also an evolution of the central absorption feature from $\sim-111$ \kms (1999) to $\sim-135$ \kms (2015-Nov) which may indicate an increase in the expansion of the ionized material, as the H$\alpha$ line includes much broader regions (e.g. polar wind) than the equatorial disk/rings that give rise to the other observed lines.

\subsection{MWC 137}

MWC 137 is embedded into a rich CSE \citep{Kraus2017b}. Although its nature has been debated,  observations favor a post-main sequence scenario \citep{Mehner2016}, especially with the detection of both $^{12}$CO and $^{13}$CO emission features \citep{Oksala2013, Muratore2015}. In Fig.~\ref{fig:ir_spectra} we present the, rather noisy, CRIRES spectrum of MWC~137 with a model for the CO originating from a rotating ring at 84 \kms \citep{Muratore2015}. Due to the unknown inclination angle, this velocity (as well as the ones derived from the forbidden optical lines) corresponds to the line-of-sight velocities, i.e. they are lower limits of the real rotational velocities.

Fig. \ref{fig:variability-3} shows the optical line profiles from two observations in 2015 and 2016. We do not see any sign of the \oi \lam5577 line, but most striking is the absence of the \caii doublet. We get a strong signal from the \oi doublet, with a clear symmetrical profile that requires a four-ring model (at 20.3, 31.0, 46.8, and 68.0 \kms; see Fig. \ref{fig:fitlines-2} and Table \ref{tab:kin-mwc137}). Considering the typical ring-width of 7 \kms these four rings form distinct regions. As these velocities are smaller than that of the CO ring, their corresponding rings are located further away from the CO ring (with no other emitting region closer to the star). Using a stellar mass estimate of 10-15~$M_{\sun}$ \citep{Mehner2016} we can calculate the ring radii for the CO and the four [OI] rings (see Table \ref{tab:results}).

Due to the very short time difference between the two observations (only a few months) there is hardly any difference in the lines. \cite{Zickgraf2003} was able to resolve the \oi \lam6300 line in 1988 (cf. his fig. 2) and, although we cannot really be certain about its wings, the peak separation he found is $\sim30$ \kms comparable to our observations. Regarding H$\alpha$, there are additional spectra from 1986/1988 \citep{Zickgraf2003} and from 2011/2013 \citep{Kraus2017b} that do not show significant changes. It is interesting to note that \cite{Mehner2016} discovered a jet (better traced in [N\,{\sc ii}]\,\lam6583). They argue that the central position of MWC 137 in the nebula and the jet suggests that it is the origin of both, which is not confirmed by a more detailed analysis of the kinematics of the nebula around MWC 137 \citep{Kraus2017b}. In any case, the presence of such a jet is quite puzzling, and if it is actually connected with MWC 137 then it might (most likely) originate from an accretion disk around a compact object, i.e. a hint for a possible binary system like CI Cam \citep{Clark2006}.

\subsection{HD 87643}

HD 87643 is another source embedded in a reflection nebula \citep[e.g.][]{Crampton1971, vandenBergh1972, Surdej1983}. It has been studied extensively \citep[e.g.][]{Oudmaijer1998, Zickgraf2003, Millour2009}, but CO emission features have been elusive. In the near-IR survey by \cite{McGregor1988} no clear identification of CO was made (for observations obtained in Jan 1985). In the current work we report for the first time the unequivocal detection of CO band emission, modeled with a rotating ring at $11\pm1$~\kms (Fig. \ref{fig:ir_spectra}), corresponding to the line-of-sight velocity due to the unknown inclination angle.

In Fig. \ref{fig:variability-4} we plot the profiles from all observations in the 1999-2015 period. We have excluded the 2015-05-12 spectrum because the exposure has been either compromised or of too low SNR to be useful. There is no emission from \oi \lam5577 line and the small visible peaks are the residuals of the sky emission line at \lam5577.3 due to the imperfect sky subtraction process. We clearly detect though the \caii and the \oi doublets. To fit the observed profiles of the \caii \lam7291 line we use a single complete ring (see Fig. \ref{fig:fitlines-2}). It is interesting to point out that we see a decrease of the \caii rotational velocity with time from $\sim12$ to $\sim10$ \kms in 1999/2000 and 2015/2016, respectively, which is also obvious from the change in profile widths. This could indicate a movement of gas away from the central source. Nevertheless, we opt to describe the forming region of \caii with a single ring of an averaged rotational velocity of 10.6 \kms. The \caii velocities are similar to the CO one, which implies that the two gases coexist. 

Regarding the \oi \lam6300 line (the intensity changes are discussed further below) the observed profiles can be fit with a model of four half-filled rings. The first two (closer to the star) seem to be stable structures as their corresponding velocities do not change much with time (within the $\sim86-91$~\kms and $\sim54-33$ \kms range; see Table \ref{tab:kin-hd87643}). However, the other two display some differences: (i) from $\sim24$ to $\sim33$ \kms for the 1999/2000 and 2015/2016 epochs, (ii) from $<3$ \kms in 1999 to $\sim9-10$ \kms in 2000/2015/2016. Within the proposed scenario of Keplerian rotation, these changes would be interpreted as formation of new emitting \oi regions (further away from the star), possible after the dissolution of the previous rings. We opt again to describe the HD 87643's disk structure with four rings at (averaged) velocities of 9.7, 28.1, 61.0, and 89.1 \kms. Of these only one (at 9.7 \kms) coincides with the \caii and the CO gases, while the other three form closer to the star. Considering our typical ring-width is $\sim11$ \kms we conclude that the derived rings for the \oi form distinct regions. The need of partially-filled rings shows also a strong asymmetry on the equatorial distribution between the \oi regions and the complete \caii/CO rings.

Currently the stellar mass of HD 87643 is not well constrained. \cite{Oudmaijer1998} have used a mass of 25~$M_{\sun}$ for calculations regarding its stellar wind. Using this estimate, our closest (to the star) \oi ring and the CO ring would have a radius of $\sim3$ AU and $\sim200$ AU, respectively. \cite{Millour2009} have resolved HD 87643 to identify a binary system that consists of a primary giant/supergiant hot star with a dusty circumprimary disk and a companion which is much fainter and embedded in its own dusty envelope (with a separation of 51 AU, at 1.5 kpc), as well as a cooler circumbinary envelope. They find the dusty circumprimary disk at $\sim6$ AU and they estimate a radius of $\sim2.5-3.0$ AU for the inner gaseous disk. This is fairly consistent with the location of our first \oi ring, but not with the rest of the structure, as they are found in between the binary and/or further away (coexisting with much cooler dust). Even if we assume that the stellar mass estimate is wrong and we use the smallest mass definition for massive stars (8 $M_{\sun}$) the CO ring is found at $\sim59$ AU, still an order of magnitude difference and circumbinary. In any case, if the position of the CO ring was really somewhere in between these locations it would have been easily resolved with interferometry. The fact that this is not what we observe may actually provide a constraint regarding the inclination anle. Assuming that CO originates from the inner rim of the dusty circumprimary disk as found from interferometry (3 AU at 1.5 kpc), the mass estimate of 25 $M_{\sun}$, and the line-of-sight rotational velocity for CO at 11 \kms, we estimate an inclination angle of 7.4\degr, which corresponds to a pole-on system. Then, our results are compatible with interferometry as all the \oi rings are found within the gaseous disk ($<3$ AU). In Table \ref{tab:results} we present the ring radii derived using both the line-of-sight velocities and the de-projected ones using this inclination angle\footnote{Considering the \textit{GAIA} results to solve the distance uncertainty does not help to put more constraints in HD 87643's nature. The measured parallax of $6.92\pm0.27$ mas (from DR1) and $0.01\pm0.12$ mas (from DR2) corresponds approximately to $\sim0.14$ kpc and $\sim100$ kpc, respectively. These values are obviously inconsistent, while even the DR2 value alone is highly unreliable.} The proposed approach has admittedly some issues to explain for example how such a nearly pole-on system displays high level of polarization and possibly the photometric variability. Nevertheless, it is a valid attempt to construct a common picture derived from both near-IR and optical data. Certainly, higher spectral and angular resolution observations could help to better resolve the CSE.

There is a striking change in the intensity of the \oi doublet and the H$\alpha$ line between the first epoch (1999-04-18) and others. This is a quite dramatic change especially for the \oi lines when comparing with the second epoch (2000-02-23) which is only 10 months later. This line has been observed in 1988 by \cite{Zickgraf2003}, who shows a profile similar to our observations after 1999 but with a less extended blue wing (cf. his fig. 2). The H$\alpha$ line exhibits significant changes too. The \textit{V/R} is increasing steadily since 1999, from $\sim0.4$ to $\sim0.8$ in 2016. There are radial velocity changes in the blue peak and the absorption component, but without any significant changes for the red peak. When comparing with the observations by \cite{Zickgraf2003} in 1986 and 1988 (cf. his fig. 1) we see that there are small differences between the two epochs. More importantly though, the \textit{V/R} ratio remains low, similar perhaps to our 1999/2000 observations. Unfortunately there are no data for the 1988-1999 period to show us if this ratio has been systematically constant or not over this period, hiding any possible periodicity.

\subsection{Hen 3-298}

CO emission features have been reported previously by a number of works \citep{Miroshnichenko2005,Muratore2012,Oksala2013}. From our CRIRES data (2009-12-02) we model the CO emission from a rotating ring at $19\pm1$ \kms (Fig. \ref{fig:ir_spectra}). This value is in agreement with the $17.8\pm0.4$ \kms found by \cite{Miroshnichenko2005}, by fitting the second CO bandhead (from observations in Dec 2002 and May 2004). We note that for Hen 3-298 also the derived rotational velocities are lower limits, corresponding to the line-of-sight velocities.

In Fig. \ref{fig:variability-4} we show the line profiles from observations from 2005 to 2016. We detect all the optical forbidden emission with clear double-peaked profiles. \cite{Miroshnichenko2005} detected the \caii and \oi doublets (without any reference on the \oi \lam5577 line) as single-peaked, mainly because of the lower resolution ($R\sim15000$) of their observations.  We fit each of the lines using a single and complete emitting ring, with rotating velocities of 23.2, 21.5, and 18.7~\kms for the \oi \lam5577, the \caii \lam7291, and the \oi \lam6300 lines, respectively (see Fig. \ref{fig:fitlines-2} and Table \ref{tab:kin-hen3-298}). From these velocities we find the \oi \lam5577 closer to the star, the \caii ring further out, and the outermost \oi (doublet) ring further away, coexisting with the CO emission region. Given the  typical ring-width of 8 \kms for Hen 3-298, it is probable that these regions overlap. To calculate the ring radii we estimate a stellar mass of 20 $M_{\sun}$, by using the evolutionary track that best fits its position in the Hertzsprung-Russell diagram (see fig. 12 in \citealt{Oksala2013}). 

Over an 11-year period Hen 3-298 displays remarkable stability, with only minor changes in the profiles of the \oi and \caii doublets. The observed H$\alpha$ line profiles are similar to the profile presented by \cite{Miroshnichenko2005}. From their observations in 2002 (cf. their fig. 1d) H$\alpha$ shows a P-Cygni profile with a strong red peak ($\sim55$ \kms) and a central absorption at $\sim-120$ \kms, similar to our data. Thus, we can conclude that H$\alpha$ has remained stable over the 2002-2016 period. However, we cannot claim any differences for the metal lines between the 2002 observations by \cite{Miroshnichenko2005} and our first dataset (2005) because of the lower resolution of the former work.

\section{Discussion}
\label{sec:discussion}

\begin{figure*}
	\includegraphics[scale=0.45]{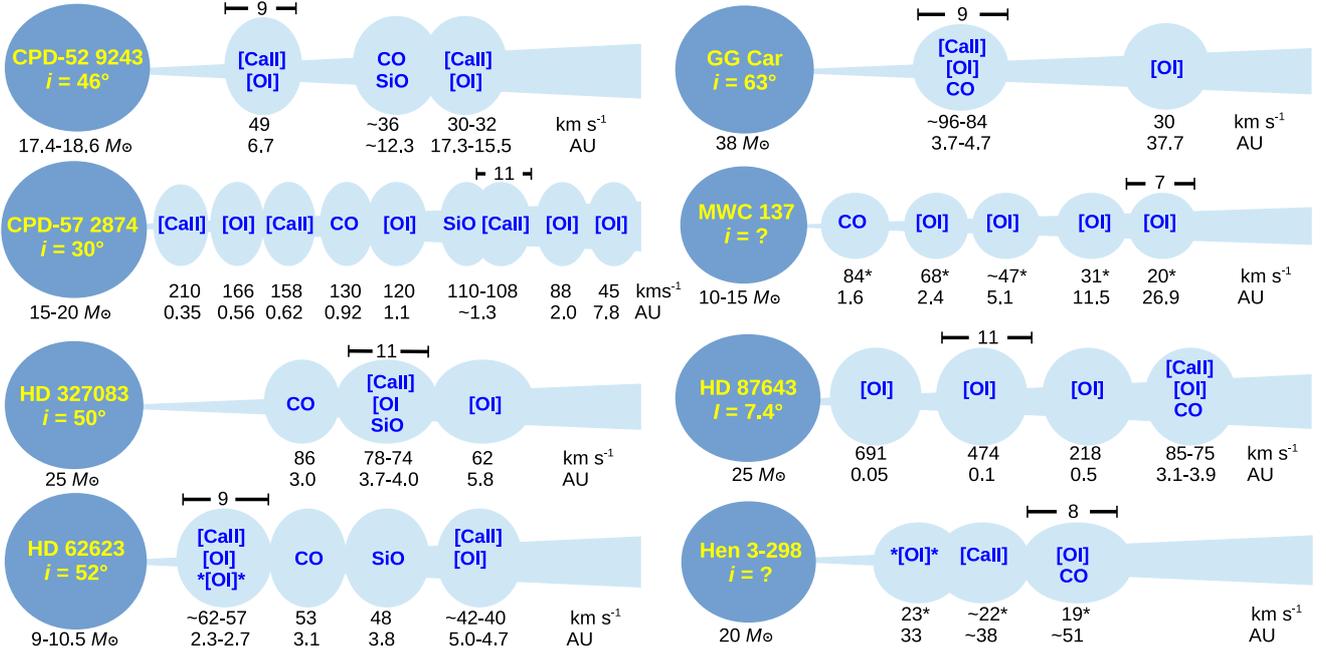}
    \caption{A cartoon illustration of the disk-structures as derived from our analysis. We represent the \oi \lam5577 line as *[OI]*, the \oi \lam\lam6300, 6363 doublet as [OI], and the \caii \lam\lam7291, 7323 as [CaII]. The arrows above the rings symbolize the typical ring-widths and are given in \kms. For more details on the data used and references see Table \ref{tab:results}. Note that the relative structures and sizes are not in scale. }
    \label{fig:physical_views}
\end{figure*}

\subsection{A common picture?}

Our main motivation for this work is to discuss the CSE properties of B[e]SGs with a consistent and homogeneous approach (both during the observations and the analysis) in an attempt to investigate whether there is a common description. To better illustrate our results (from Table \ref{tab:results}) we have created a cartoon presentation of the disk structures in Fig. \ref{fig:physical_views}. We present a section along the disk of each object and on top of this we show the individual ring structures identified.  

Regarding our disk tracers, the \oi \lam5577 line is found only in two objects, in the binary HD 62623 and in Hen~3-298. Its position is always located in a single region and very close to the star as expected \citep{Kraus2007, Kraus2010, Aret2012}. Its presence in a confirmed binary system and a (currently considered) single one does not provide any further constraints regarding its appearance. On the other hand, the \caii \lam\lam7291,7323 and the \oi \lam\lam6300,6363 doublets are detected in all objects, with the exception of MWC~137 in which no \caii is detected. In many cases we find that the two doublets are formed in the same or near-by regions, and sometimes coexist with the molecular gases (CO and SiO). Our results show that the circumstellar disks around B[e]SGs are far different from the classic picture of homogeneous outflowing structures \citep{Zickgraf1985}. Given the Keplerian rotation of the gas we find that these disks are a composition of atomic and molecular gases that form local enhancements with the appropriate conditions to give rise to the corresponding lines, a result that seems to be common among sources in the Galaxy and the Magellanic Clouds \citep{Aret2012, Kraus2016,Kraus2017b,Torres2018}. In all stars, we find distinct, but not necessarily homogeneous, rings, except for CPD-57 2874 where we see a continuous set of rings that points perhaps to an inhomogeneous disk. A visual inspection of our results, as illustrated in Fig. \ref{fig:physical_views}, does not provide a common picture regarding the distribution of the atomic and molecular rings in the CSE. It seems that each source, regardless if it is a binary system with an orbital period of a few tens of days (GG Car, HD 327083) or much longer (HD 87643), a single star (Hen 3-298), or embedded in a nebula (MWC 137, HD 87643), displays a unique combination. 

It is appropriate though to include also information about the position of the dust as derived from the inteferometric studies. Dust is found to coexist with the rings of atomic (\oi \lam6300 and \caii \lam7291 lines) and molecular gas (CO and SiO). In particular we have:

\begin{itemize}

\item In CPD-52 9243 dust is found at $\sim15$ AU \citep{Cidale2012} in between the molecular (CO and SiO) and atomic (\oi and \caiinos) rings at 12 and 17 AU, respectively. 

\item In HD 327083 the dust's position at $\sim5$ AU \citep{Wheelwright2012b} is very close to the position of the CO ring at 3 AU and the \oi, \caii, and SiO ring at $\sim4$ AU. 

\item In HD 62623 the dusty disk is located at $\sim4$ AU \citep{Millour2011} where we find the SiO ring, in between the CO ring at $\sim3$ AU and the \oi and \caii ring at $\sim5$~AU.

\item In CPD-57 2874 dust is located approximately at 11-14~AU \citep{DomicianodeSouza2011} from mid-IR observations, but they state that they are not able to resolve the near-IR emission (located at $\sim8$ AU), which is a combination of ionized material and includes the tail of the dust identified in the mid-IR. Our structure extends up to $\sim8$~AU, so possible we do have some presence of dust coexisting with parts of the inhomogeneous disk, which consists of alternate regions of atomic and molecular gas.  

\item In order to better correlate our results with interferometry for HD 87643, we have used the position of the CO ring to match the inner rim of the dusty ring (3 AU) to derive the possible inclination angle for this system. Nevertheless, we do find in the same region CO and atomic gas, with dust. 

\item For GG Car an estimate of a dusty envelope is made through SED fitting \citep{Marchiano2012} starting at approximately 34 AU. This location coincides with the position of a \oi (at $\sim35$ AU), but is further apart than the other ring which combines CO, \oi, and \caii (at $\sim4$ AU). 

\item There are no interferometric observations for MWC 137 and Hen 3-298, so no solid conclusions regarding the position of the dust with respect to the identified rings can be made for these objects. 

\end{itemize}

In total, we see that in all objects for which we do have a resolved CSE, excluding MWC 137, Hen 3-298, and possible GG Car, we see that the presence of the dust correlates directly with the presence of atomic and molecular gas.  Due to the different conditions of temperature and density for the emission forming regions, it is possible that we see the contribution of different layers in a disk displaying different optical depths for the stellar radiation\footnote{The coexistence of the atomic gases is not unphysical since the corresponding excitation energies are similar. In particular, for the \oi \lam5577 line, the \oi \lam\lam6300,6363 doublet, and the \caii \lam\lam7291,7323 doublet these are 2.23, 1.97, and 1.70 eV, respectively (data derived from the NIST Atomic Spectra Database Levels Form, at \url{https://physics.nist.gov/PhysRefData/ASD/levels_form.html}, accessed on February, 7, 2018)}.

As the B[e] phenomenon occurs in different evolution stages \citep{Lamers1998}, including pre-main sequence stars that are known to have gaps in their circumstellar disks \citep[e.g.][]{Menu2015}, it would be tempting to claim that the observed differences may be due to the different nature of the objects. However, our sample consists of stars with strong evidence in favor of their supergiant nature. An exception might be HD 87643 which is still considered a pre-main sequence source, although it does not fully comply with this group (see e.g. \citealt{Carmona2011} regarding the lack of H$_2$ emission lines). Nevertheless, the uniqueness of each CSE shows how the formation mechanisms work in each system. For example, in binary systems the circumbinary ring formation is connected with the phases of interaction. The length and violence of these phases critically depends on the closeness of the components, their stellar parameters, the eccentricity of the system, etc., resulting in unique sets of rings. In single stars, the situation is more tricky as the degeneracy of their evolutionary state (pre- vs post-RSGs) means that they have experienced different mass-loss episodes. In addition, physical processes like rotation and pulsations may be involved, but to investigate these in more detail is beyond the scope of the current paper.

\subsection{On the CSE formation}

Binary interaction is considered as one major channel to form these structures \citep[e.g.][]{Miroshnichenko2007, Millour2011, Wheelwright2012b}. Half of our sample is actually confirmed binaries: HD 327083, HD 62623, GG Car, and HD 87643 (excluding CPD-52 9243 which has only been suggested to be a binary; \citealt{Cidale2012}), but this fraction does not correspond to their total population ($\sim20\%$, see \citealt{Kraus2017}). For both HD 327083 and GG Car we do not find any other emission forming region closer to the star than the CO ring, while the opposite is true for HD 62623 and HD 87643. This follows our previous result on the uniqueness of the CSE around each B[e]SG. In the cases of HD 327083, HD 62623, and GG Car, the innermost ring is always larger than the binary separation. The fact that the rings around these binary B[e]SGs are found to be circumbinary can fit well in the binary interaction scenario. Nonetheless, in HD 87643 we actually find a circumprimary structure.  

In the absence of any confirmation for binarity in the rest of our sample, we should consider a different mechanism that forms these structures in single stars. It is possible that they are the results of mass loss triggered by pulsations and/or other instabilities. B[e]SGs may be the successors of Yellow Hypergiants, after passing the Yellow Void \citep{Davies2007, Aret2017}. Since the mass-loss in these hypergiants is believed to originate from pulsations \citep{deJager1998} it is possible that the CSE in these two phases are the result of a common/similar pulsation mechanism. Indications for such stellar pulsations have been found for the LHA 120-S 73 \citep{Kraus2016} and LHA 120-S 35 \citep{Torres2018}. Alternative scenarios include either the presence of objects that can clear their paths and stabilize these ring structures similar to shepherd moons \citep{Kraus2016}. 

Most observational evidence favor a Keplerian rotation for the disks \citep[e.g.][]{Marchiano2012, Kraus2015}. Nevertheless, it is possible that these multi-ring structures that we see may be due to a different distribution of the CSE. In an hourglass-like or in a spiral-arm structure these rings could correspond to density enhancements as the projection along the line-of-sight. Such a formation could result from wind-wind interactions. For example, \cite{Chita2008} simulate the interaction of an asymmetric wind of a post-RSG B[e]SG which interacts with the material shed spherically during its RSG phase. To date it is not known whether B[e]SGs are post- or pre-RSG objects. An indication of their age can be obtained from the ratio $^{12}{\rm CO}/^{13}{\rm CO}$ (which decreases as the star evolves; \citealt{Kraus2009}), but yet this method is applicable only to objects with detected CO emission and with known rotation speed.

\subsection{On the variability}

\begin{figure*}
	\includegraphics[scale=0.6,trim=80 40 100 90,clip]{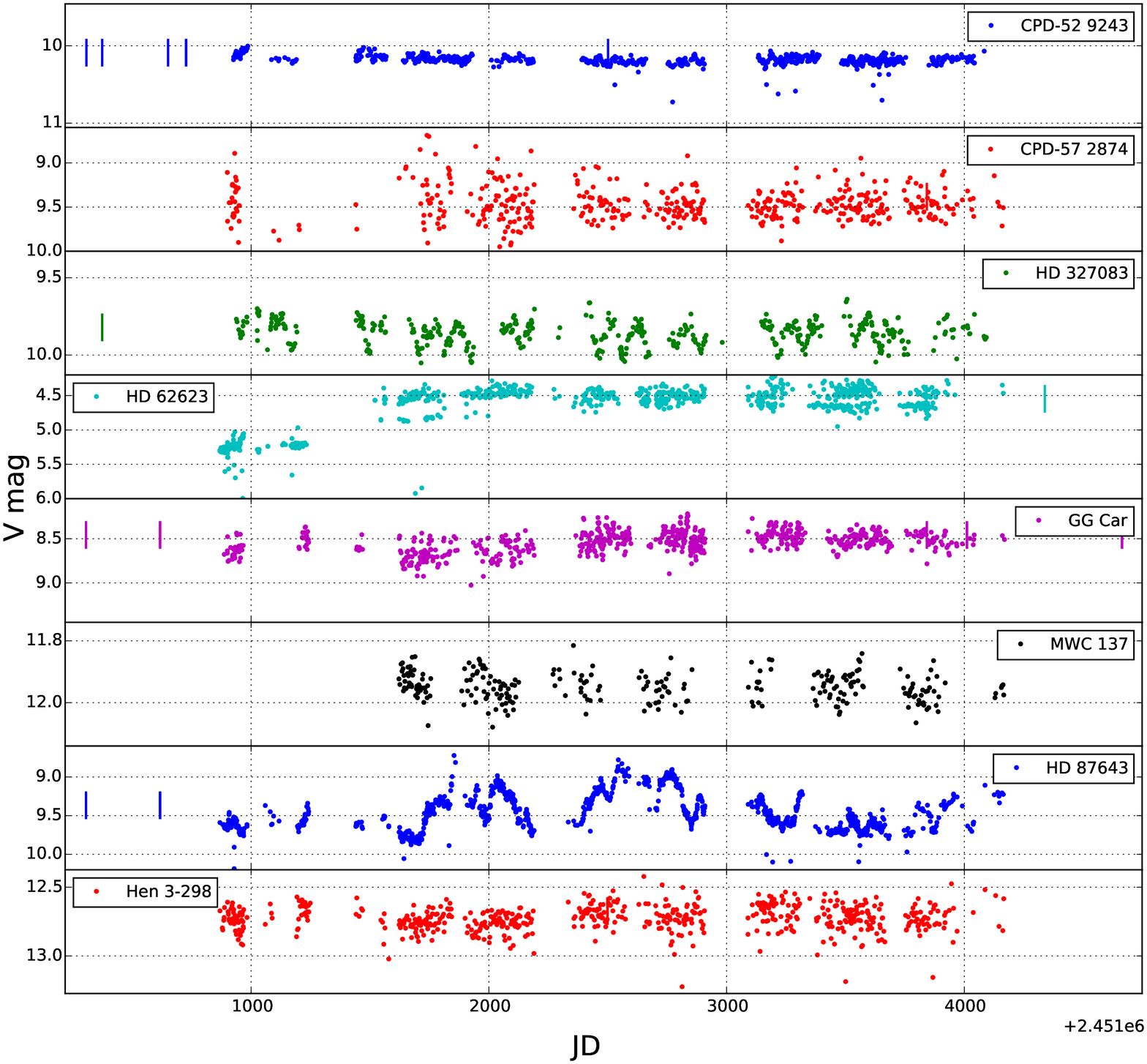}
    \caption{The ASAS \textit{V}-band light curves for our objects, covering approximately the period November 2000 to 2009 (dots). We also indicate the dates of our optical spectroscopic observations as vertical lines. (See text for a discussion on the variability and the derived periods for each source.)}
    \label{fig:photometric_var}
\end{figure*}

We have described the observed spectral variability for each source individually in Section \ref{sec:results}, so in this section we discuss all sources globally.

Among all sources, the line profiles of Hen 3-298 and CPD-57 2874 remain almost constant throughout all epochs, spanning 11 years (2005-2016) and 8 years (2008-2016), respectively. The \oi and \caii lines of the binary HD 62623 are stable, in contrast to the variable H$\alpha$ line (over the 2008-2015 period). CPD-52 9243 and the binary HD 87643 display more evident variability in their line intensities (with these changes being much stronger in the second source), over a similar time period (1999-2016). The binaries GG Car and HD 327083 exhibit the most drastic changes regarding both their line intensities and their profiles (over the 1999-2016 period). Assuming that the observed variability could be attributed at least partly to the binary interaction, then we could argue that binaries display a more variable CSE (for GG Car, HD 327083, HD 87643, and HD 62623 - albeit only in H$\alpha$). Interestingly, CPD-52 9243 exhibits some variability and following that trend we could suggest potential binarity, further supporting previous indications \citep{Cidale2012}. We cannot comment on the variability of MWC 137 for which we have obtained only two datasets very close in time ($\sim3$ months) and there is hardly any difference.  

To further explore the correlation of binaries with variability we examine the light curves obtained from the ASAS survey\footnote{\url{http://www.astrouw.edu.pl/asas/} \citep{Pojmanski1997}.}, since it is more suitable for the relatively bright sources of our sample (which saturate in OGLE). In Fig. \ref{fig:photometric_var} we show the \textit{V}-band photometry covering approximately the November 2000 to November 2009 period (9 years). We used the most reliable measurements (flagged with a quality grade of A or B) for which we calculated an average magnitude from all five apertures accompanied by an error estimated as the square root of all errors. Along with the photometric points we also show the dates of our optical spectroscopic observations (vertical lines). Unfortunately, the overlap between the photometric and the spectroscopic observations is scarce, so we cannot derive any conclusion with respect to this correlation. From Fig. \ref{fig:photometric_var} the variability of the confirmed binaries HD 327083 and HD 87643 is prominent, while for GG Car and HD 62623 is much less evident (the latter one displays also a brightening event after the start of the observations at $V\sim5.5$ mag to $V\sim4.5$ mag, but care should be taken in total since it may well be saturated even for ASAS). For the rest, the variability may be stochastic or not. To further investigate this we obtained the Lomb--Scargle periodograms \citep{Lomb1976,Scargle1982} for the entire light curves as well as for  a number of individual datasets (e.g. one or more consequent years) of continuous coverage. The last step was performed in order to exclude periods that do not persist throughout the data and may arise from data sampling of the full light curves. Then we identify all potential periods above the 99 per cent confidence level, which was estimated by simulating light curves based on the noise characteristics of the data and repeating the analysis for each simulated light curve. The results are presented in Table \ref{tab:periods}. 

We can identify the previously derived periods for GG Car and HD 62623, as well as the period of $\sim107$ d for HD 327083 which is also found by Cidale et al. (in preparation). We note here the identification of some periods for HD 87643, although its orbital period is considered to be of several decades \citep{Millour2009}, well beyond the baseline of this dataset. Similar with the spectral variability we see that all the four confirmed binaries do display numerous periods. CPD-52 9243's behavior is again similar to the other binaries. For the least variable objects CPD-57 2874 and MWC 137 we have identified only one (and rather weak) period. For Hen 3-298 we cannot detect any significant period (even at the 90\% confidence level). We point though that the current, not exhausting, analysis of these light curves does solve for the nature of these periods, as some may not be related to the binaries but to other causes (e.g. disk activity, stellar pulsations). 

\begin{table}
\centering
\caption{Detected periods at 99\% confidence level} 
\label{tab:periods}
 \begin{tabular}{ccc}
 \hline \hline
 Object & \multicolumn{2}{c}{Periods (days)}  \\
        & this work  &  previous  \\
 \hline
 CPD-52 9243  & 24.7, 66.3, 174.2 & -- \\

 CPD-57 2874  & 21.5 & -- \\

 HD 327083    & 20.0, 23.9, 53.8, 82.3, 93.4, & \\ 
              & 107.9, 118.4, 152.6, 171.6 & 107.687 $^1$ \\

 HD 62623     & 136.9, 162.0, 237.1, 276.6 & 138.5, 161.3 $^2$\\
              & 308.8, 349.4, 414.8 & \\

 GG Car       & 31.0, 33.8 & 31.033 $^3$\\

 MWC 137      & 332.4 & -- \\

 HD 87643     & 104.8, 117.1, 147.1,  & -- \\
              & 172.7, 192.7 & \\
              
 Hen 3-298    & none* & --\\

 \hline
 \end{tabular}
 
 \flushleft{
 \textit{Notes:}
 * No period detection even at the 90\% confidence level.\\
 $^1$ Cidale et al. , in preparation; 
 $^2$ \cite{Plets1995};
 $^3$ \cite{Marchiano2012} 
 }

\end{table}

\section{Conclusion}
\label{sec:conclusion}

With this work we present our results obtained from high-resolution optical (FEROS) and IR (CRIRES) spectroscopy for 8 Galactic B[e]SGs with CO emission. Our goal is to understand the structure of the disks around the B[e]SGs. Assuming that CO emission originates from a Keplerian-rotating disk and that the atomic gas of the circmustellar environment follows this distribution, we modeled the \oi \lam5577, \lam6300, and \caii7291 lines as emission from rotating rings. Their derived kinematics correspond to certain distances that allow us to probe the structure of their disks. We find that all B[e]SGs of our sample are surrounded by multiple ring-like structures. The distribution of these gas rings is unique for each object, without any particular preference or dependance on binarity. This interpretation is based on the Keplerian rotation of the disk but alternative scenarios, e.g. an hourglass or spiral-arm distribution, can result in local enhancements similar to the observed structures.

Multiple epochs of spectroscopic observations help us to investigate the variability and the stability of these structures over time. We have identified the trend that binaries display the largest intensity and profile variability of their optical lines. This is further supported by a preliminary analysis of ASAS optical light curves, from which we have detected a number of periods for binaries. However, to date there is no consensus on whether binarity causes the formation of the ring structures, in particular since not all studied objects have a companion. Alternative scenarios such as (quasi-)periodic mass ejections triggered by pulsations or other instabilities are worth being studied in more detail.


\section*{Acknowledgments}

The authors would like to thank the anonymous referee for their constructive feedback. GM acknowledges support from CONICYT, Programa de Astronom\'ia/PCI, FONDO ALMA 2014, Proyecto No 31140024, and GA\,\v{C}R under grant number 14-21373S. He also thanks Theodoros Bitsakis and Alexandros Maragkoudakis for fruitful discussions during the project. MK acknowledges financial support from GA\v{C}R (grant number 17-02337S). The Astronomical Institute Ond\v{r}ejov is supported by the project RVO:67985815. This research was also supported by the European Union European Regional Development Fund, project ``Benefits for Estonian Society from Space Research and Application'' (KOMEET, 2014\,-\,2020.\,4.\,01.\,16\,-\,0029) and by the institutional research funding IUT40-1 of the Estonian Ministry of Education and Research. Parts of the observations obtained with the MPG 2.2m telescope were supported by the Ministry of Education, Youth and Sports project - LG14013 (Tycho Brahe: Supporting Ground-based Astronomical Observations). We would like to thank the observers (S. Ehlerova, P. Kabath, A. Kawka, S. Vennes) for obtaining the data. LSC thanks financial support from the Agencia de Promoci\'on Cient\'ifica y Tecnol\'ogica (Pr\'estamo BID PICT 2016/1971), CONICET (PIP 0177), and the Universidad Nacional de La Plata (Programa de Incentivos G11/137), Argentina. LSC and MC thank the support from the project CONICYT and PAI/Atracci\'on de capital humano avanzado del extranjero (folio PAI80160057). MC acknowledges support from Centro de Astrof\'isica de Valpara\'iso.

This research has made use of NASA's Astrophysics Data System, and of the SIMBAD database, operated at CDS, Strasbourg, France. It has used matplotlib, a Python library for publication quality graphics \citep{matplotlib}, Astropy, a community-developed core Python package for Astronomy \citep{astropy}, and NumPy \citep{numpy}. IRAF is distributed by the National Optical Astronomy Observatory, which is operated by the Association of Universities for Research in Astronomy (AURA) under cooperative agreement with the National Science Foundation \citep{IRAF}. PyRAF is a product of the Space Telescope Science Institute, which is operated by AURA for NASA. This research has made use of the services of the ESO Science Archive Facility. This publication makes use of data products from the Two Micron All Sky Survey, which is a joint project of the University of Massachusetts and the Infrared Processing and Analysis Center/California Institute of Technology, funded by the National Aeronautics and Space Administration and the National Science Foundation. This work has made use of data from the European Space Agency (ESA) mission \textit{Gaia} (\url{https://www.cosmos.esa.int/gaia}), processed by the \textit{Gaia} Data Processing and Analysis Consortium (DPAC, \url{https://www.cosmos.esa.int/web/gaia/dpac/consortium}). Funding for the DPAC has been provided by national institutions, in particular the institutions participating in the \textit{Gaia} Multilateral Agreement.

 This study is based on observations under the ESO proposals: 384.D-0613(A) and 385.D-0513(A) for CRIRES,  075.D-0177(A), 082.A-9209(A), 085.D-0185(A), 094.A-9029(D), 095.A-9032(A), 096.A-9024(A), 096.A-9030(A), 096.A-9039(A), 097.A-9024(A), 097.A-9039(C) for FEROS. Based on observations done with the 1.52-m and 2.2-m telescopes at the European Southern Observatory (La Silla, Chile), under agreements ESO-Observat\'orio Nacional/MCTIC and MPI-Observat\'orio Nacional/MCTIC. Based on data obtained from the ESO Science Archive Facility under request numbers: 301274, 302259, 302651 (from ESO proposals: 091.D-0221(A), 096.A-9014(A)). Based, in part, on observations obtained at the Gemini Observatory (GN-2012A-Q-21 for GNIRS), which is operated by the Association of Universities for Research in Astronomy, Inc., under a cooperative agreement with the NSF on behalf of the Gemini partnership: the National Science Foundation (United States), the National Research Council (Canada), CONICYT (Chile), Ministerio de Ciencia, Tecnolog\'{i}a e Innovaci\'{o}n Productiva (Argentina), and Minist\'{e}rio da Ci\^{e}ncia, Tecnologia e Inova\c{c}\~{a}o (Brazil).

\bibliographystyle{mnras}
\bibliography{/home/grigoris/Resources/Papers-Library/00_bibs_00/supergiants,/home/grigoris/Resources/Papers-Library/00_bibs_00/extrainfo}

\begin{thebibliography}{}
\makeatletter
\relax
\def\mn@urlcharsother{\let\do\@makeother \do\$\do\&\do\#\do\^\do\_\do\%\do\~}
\def\mn@doi{\begingroup\mn@urlcharsother \@ifnextchar [ {\mn@doi@}
  {\mn@doi@[]}}
\def\mn@doi@[#1]#2{\def\@tempa{#1}\ifx\@tempa\@empty \href
  {http://dx.doi.org/#2} {doi:#2}\else \href {http://dx.doi.org/#2} {#1}\fi
  \endgroup}
\def\mn@eprint#1#2{\mn@eprint@#1:#2::\@nil}
\def\mn@eprint@arXiv#1{\href {http://arxiv.org/abs/#1} {{\tt arXiv:#1}}}
\def\mn@eprint@dblp#1{\href {http://dblp.uni-trier.de/rec/bibtex/#1.xml}
  {dblp:#1}}
\def\mn@eprint@#1:#2:#3:#4\@nil{\def\@tempa {#1}\def\@tempb {#2}\def\@tempc
  {#3}\ifx \@tempc \@empty \let \@tempc \@tempb \let \@tempb \@tempa \fi \ifx
  \@tempb \@empty \def\@tempb {arXiv}\fi \@ifundefined
  {mn@eprint@\@tempb}{\@tempb:\@tempc}{\expandafter \expandafter \csname
  mn@eprint@\@tempb\endcsname \expandafter{\@tempc}}}

\bibitem[\protect\citeauthoryear{{Andruchow} et~al.,}{{Andruchow}
  et~al.}{2012}]{Andruchow2012}
{Andruchow} I.,  et~al., 2012, in {Carciofi} A.~C.,  {Rivinius} T.,  eds,
  Astronomical Society of the Pacific Conference Series Vol. 464, Circumstellar
  Dynamics at High Resolution. p.~319

\bibitem[\protect\citeauthoryear{{Aret}, {Kraus}, {Muratore}  \& {Borges
  Fernandes}}{{Aret} et~al.}{2012}]{Aret2012}
{Aret} A.,  {Kraus} M.,  {Muratore} M.~F.,   {Borges Fernandes} M.,  2012,
  \mn@doi [\mnras] {10.1111/j.1365-2966.2012.20871.x}, \href
  {http://adsabs.harvard.edu/abs/2012MNRAS.423..284A} {423, 284}

\bibitem[\protect\citeauthoryear{{Aret}, {Kraus}  \& {{\v S}lechta}}{{Aret}
  et~al.}{2016}]{Aret2016}
{Aret} A.,  {Kraus} M.,   {{\v S}lechta} M.,  2016, \mn@doi [\mnras]
  {10.1093/mnras/stv2758}, \href
  {http://adsabs.harvard.edu/abs/2016MNRAS.456.1424A} {456, 1424}

\bibitem[\protect\citeauthoryear{{Aret}, {Kolka}, {Kraus}  \&
  {Maravelias}}{{Aret} et~al.}{2017}]{Aret2017}
{Aret} A.,  {Kolka} I.,  {Kraus} M.,   {Maravelias} G.,  2017, in
  {Miroshnichenko} A.,  {Zharikov} S.,  {Kor{\v c}{\'a}kov{\'a}} D.,   {Wolf}
  M.,  eds,  Astronomical Society of the Pacific Conference Series Vol. 508,
  The B[e] Phenomenon: Forty Years of Studies. p.~239 (\mn@eprint {arXiv}
  {1611.04490})

\bibitem[\protect\citeauthoryear{{Astropy Collaboration} et~al.,}{{Astropy
  Collaboration} et~al.}{2013}]{astropy}
{Astropy Collaboration} et~al., 2013, \mn@doi [\aap]
  {10.1051/0004-6361/201322068}, \href
  {http://adsabs.harvard.edu/abs/2013A%26A...558A..33A} {558, A33}

\bibitem[\protect\citeauthoryear{{Carmona}, {van der Plas}, {van den Ancker},
  {Audard}, {Waters}, {Fedele}, {Acke}  \& {Pantin}}{{Carmona}
  et~al.}{2011}]{Carmona2011}
{Carmona} A.,  {van der Plas} G.,  {van den Ancker} M.~E.,  {Audard} M.,
  {Waters} L.~B.~F.~M.,  {Fedele} D.,  {Acke} B.,   {Pantin} E.,  2011, \mn@doi
  [\aap] {10.1051/0004-6361/201116561}, \href
  {http://adsabs.harvard.edu/abs/2011A%26A...533A..39C} {533, A39}

\bibitem[\protect\citeauthoryear{{Chentsov}, {Klochkova}  \&
  {Miroshnichenko}}{{Chentsov} et~al.}{2010}]{Chentsov2010}
{Chentsov} E.~L.,  {Klochkova} V.~G.,   {Miroshnichenko} A.~S.,  2010, \mn@doi
  [Astrophysical Bulletin] {10.1134/S1990341310020057}, \href
  {http://adsabs.harvard.edu/abs/2010AstBu..65..150C} {65, 150}

\bibitem[\protect\citeauthoryear{{Chita}, {Langer}, {van Marle},
  {Garc{\'{\i}}a-Segura}  \& {Heger}}{{Chita} et~al.}{2008}]{Chita2008}
{Chita} S.~M.,  {Langer} N.,  {van Marle} A.~J.,  {Garc{\'{\i}}a-Segura} G.,
  {Heger} A.,  2008, \mn@doi [\aap] {10.1051/0004-6361:200810087}, \href
  {http://adsabs.harvard.edu/abs/2008A%26A...488L..37C} {488, L37}

\bibitem[\protect\citeauthoryear{{Cidale} et~al.,}{{Cidale}
  et~al.}{2012}]{Cidale2012}
{Cidale} L.~S.,  et~al., 2012, \mn@doi [\aap] {10.1051/0004-6361/201220120},
  \href {http://adsabs.harvard.edu/abs/2012A%26A...548A..72C} {548, A72}

\bibitem[\protect\citeauthoryear{{Clark}}{{Clark}}{2006}]{Clark2006}
{Clark} J.~S.,  2006, in {Kraus} M.,  {Miroshnichenko} A.~S.,  eds,
  Astronomical Society of the Pacific Conference Series Vol. 355, Stars with
  the B[e] Phenomenon. p.~269

\bibitem[\protect\citeauthoryear{{Crampton}}{{Crampton}}{1971}]{Crampton1971}
{Crampton} D.,  1971, \mn@doi [\aj] {10.1086/111114}, \href
  {http://adsabs.harvard.edu/abs/1971AJ.....76..260C} {76, 260}

\bibitem[\protect\citeauthoryear{{Cur{\'e}}}{{Cur{\'e}}}{2004}]{Cure2004}
{Cur{\'e}} M.,  2004, \mn@doi [\apj] {10.1086/423776}, \href
  {http://adsabs.harvard.edu/abs/2004ApJ...614..929C} {614, 929}

\bibitem[\protect\citeauthoryear{{Cur{\'e}}, {Rial}  \& {Cidale}}{{Cur{\'e}}
  et~al.}{2005}]{Cure2005}
{Cur{\'e}} M.,  {Rial} D.~F.,   {Cidale} L.,  2005, \mn@doi [\aap]
  {10.1051/0004-6361:20052686}, \href
  {http://adsabs.harvard.edu/abs/2005A%26A...437..929C} {437, 929}

\bibitem[\protect\citeauthoryear{{Cutri} et~al.,}{{Cutri}
  et~al.}{2003}]{Cutri2003}
{Cutri} R.~M.,  et~al., 2003, VizieR Online Data Catalog, \href
  {http://cdsads.u-strasbg.fr/abs/2003yCat.2246....0C} {2246}

\bibitem[\protect\citeauthoryear{{Davies}, {Oudmaijer}  \& {Sahu}}{{Davies}
  et~al.}{2007}]{Davies2007}
{Davies} B.,  {Oudmaijer} R.~D.,   {Sahu} K.~C.,  2007, \mn@doi [\apj]
  {10.1086/523692}, \href {http://adsabs.harvard.edu/abs/2007ApJ...671.2059D}
  {671, 2059}

\bibitem[\protect\citeauthoryear{{Domiciano de Souza} et~al.,}{{Domiciano de
  Souza} et~al.}{2011}]{DomicianodeSouza2011}
{Domiciano de Souza} A.,  et~al., 2011, \mn@doi [\aap]
  {10.1051/0004-6361/201015194}, \href
  {http://adsabs.harvard.edu/abs/2011A%26A...525A..22D} {525, A22}

\bibitem[\protect\citeauthoryear{{Dunstall} et~al.,}{{Dunstall}
  et~al.}{2015}]{Dunstall2015}
{Dunstall} P.~R.,  et~al., 2015, \mn@doi [\aap] {10.1051/0004-6361/201526192},
  \href {http://adsabs.harvard.edu/abs/2015A%26A...580A..93D} {580, A93}

\bibitem[\protect\citeauthoryear{Hunter}{Hunter}{2007}]{matplotlib}
Hunter J.~D.,  2007, Computing In Science \& Engineering, 9, 90

\bibitem[\protect\citeauthoryear{{Kaeufl} et~al.,}{{Kaeufl}
  et~al.}{2004}]{Kaeufl2004}
{Kaeufl} H.-U.,  et~al., 2004, in {Moorwood} A.~F.~M.,  {Iye} M.,  eds,
  \procspie Vol. 5492, Ground-based Instrumentation for Astronomy. pp
  1218--1227, \mn@doi{10.1117/12.551480}

\bibitem[\protect\citeauthoryear{{Kaufer}, {Stahl}, {Tubbesing},
  {N{\o}rregaard}, {Avila}, {Francois}, {Pasquini}  \& {Pizzella}}{{Kaufer}
  et~al.}{1999}]{Kaufer1999}
{Kaufer} A.,  {Stahl} O.,  {Tubbesing} S.,  {N{\o}rregaard} P.,  {Avila} G.,
  {Francois} P.,  {Pasquini} L.,   {Pizzella} A.,  1999, The Messenger, \href
  {http://adsabs.harvard.edu/abs/1999Msngr..95....8K} {95, 8}

\bibitem[\protect\citeauthoryear{{Kraus}}{{Kraus}}{2006}]{Kraus2006}
{Kraus} M.,  2006, \mn@doi [\aap] {10.1051/0004-6361:20065204}, \href
  {http://adsabs.harvard.edu/abs/2006A%26A...456..151K} {456, 151}

\bibitem[\protect\citeauthoryear{{Kraus}}{{Kraus}}{2009}]{Kraus2009}
{Kraus} M.,  2009, \mn@doi [\aap] {10.1051/0004-6361:200811020}, \href
  {http://adsabs.harvard.edu/abs/2009A%26A...494..253K} {494, 253}

\bibitem[\protect\citeauthoryear{{Kraus}}{{Kraus}}{2017}]{Kraus2017}
{Kraus} M.,  2017, in {Miroshnichenko} A.,  {Zharikov} S.,  {Kor{\v
  c}{\'a}kov{\'a}} D.,   {Wolf} M.,  eds,  Astronomical Society of the Pacific
  Conference Series Vol. 508, The B[e] Phenomenon: Forty Years of Studies.
  p.~219 (\mn@eprint {arXiv} {1610.05000})

\bibitem[\protect\citeauthoryear{{Kraus}, {Kr{\"u}gel}, {Thum}  \&
  {Geballe}}{{Kraus} et~al.}{2000}]{Kraus2000}
{Kraus} M.,  {Kr{\"u}gel} E.,  {Thum} C.,   {Geballe} T.~R.,  2000, \aap, \href
  {http://adsabs.harvard.edu/abs/2000A%26A...362..158K} {362, 158}

\bibitem[\protect\citeauthoryear{{Kraus}, {Borges Fernandes}  \& {de
  Ara{\'u}jo}}{{Kraus} et~al.}{2007}]{Kraus2007}
{Kraus} M.,  {Borges Fernandes} M.,   {de Ara{\'u}jo} F.~X.,  2007, \mn@doi
  [\aap] {10.1051/0004-6361:20066325}, \href
  {http://adsabs.harvard.edu/abs/2007A%26A...463..627K} {463, 627}

\bibitem[\protect\citeauthoryear{{Kraus}, {Borges Fernandes}  \& {de
  Ara{\'u}jo}}{{Kraus} et~al.}{2010}]{Kraus2010}
{Kraus} M.,  {Borges Fernandes} M.,   {de Ara{\'u}jo} F.~X.,  2010, \mn@doi
  [\aap] {10.1051/0004-6361/200913964}, \href
  {http://adsabs.harvard.edu/abs/2010A%26A...517A..30K} {517, A30}

\bibitem[\protect\citeauthoryear{{Kraus}, {Oksala}, {Nickeler}, {Muratore},
  {Borges Fernandes}, {Aret}, {Cidale}  \& {de Wit}}{{Kraus}
  et~al.}{2013}]{Kraus2013}
{Kraus} M.,  {Oksala} M.~E.,  {Nickeler} D.~H.,  {Muratore} M.~F.,  {Borges
  Fernandes} M.,  {Aret} A.,  {Cidale} L.~S.,   {de Wit} W.~J.,  2013, \mn@doi
  [\aap] {10.1051/0004-6361/201220442}, \href
  {http://adsabs.harvard.edu/abs/2013A%26A...549A..28K} {549, A28}

\bibitem[\protect\citeauthoryear{{Kraus}, {Oksala}, {Cidale}, {Arias}, {Torres}
   \& {Borges Fernandes}}{{Kraus} et~al.}{2015}]{Kraus2015}
{Kraus} M.,  {Oksala} M.~E.,  {Cidale} L.~S.,  {Arias} M.~L.,  {Torres} A.~F.,
   {Borges Fernandes} M.,  2015, \mn@doi [\apjl] {10.1088/2041-8205/800/2/L20},
  \href {http://adsabs.harvard.edu/abs/2015ApJ...800L..20K} {800, L20}

\bibitem[\protect\citeauthoryear{{Kraus} et~al.,}{{Kraus}
  et~al.}{2016}]{Kraus2016}
{Kraus} M.,  et~al., 2016, \mn@doi [\aap] {10.1051/0004-6361/201628493}, \href
  {http://adsabs.harvard.edu/abs/2016A%26A...593A.112K} {593, A112}

\bibitem[\protect\citeauthoryear{{Kraus} et~al.,}{{Kraus}
  et~al.}{2017}]{Kraus2017b}
{Kraus} M.,  et~al., 2017, \mn@doi [\aj] {10.3847/1538-3881/aa8df6}, \href
  {http://adsabs.harvard.edu/abs/2017AJ....154..186K} {154, 186}

\bibitem[\protect\citeauthoryear{{Krti{\v c}ka}, {Kurf{\"u}rst}  \& {Krti{\v
  c}kov{\'a}}}{{Krti{\v c}ka} et~al.}{2015}]{Krticka2015}
{Krti{\v c}ka} J.,  {Kurf{\"u}rst} P.,   {Krti{\v c}kov{\'a}} I.,  2015,
  \mn@doi [\aap] {10.1051/0004-6361/201424867}, \href
  {http://adsabs.harvard.edu/abs/2015A%26A...573A..20K} {573, A20}

\bibitem[\protect\citeauthoryear{{Kurf{\"u}rst}, {Feldmeier}  \& {Krti{\v
  c}ka}}{{Kurf{\"u}rst} et~al.}{2014}]{Kurfurst2014}
{Kurf{\"u}rst} P.,  {Feldmeier} A.,   {Krti{\v c}ka} J.,  2014, \mn@doi [\aap]
  {10.1051/0004-6361/201424272}, \href
  {http://adsabs.harvard.edu/abs/2014A%26A...569A..23K} {569, A23}

\bibitem[\protect\citeauthoryear{{Lamers}, {Zickgraf}, {de Winter}, {Houziaux}
  \& {Zorec}}{{Lamers} et~al.}{1998}]{Lamers1998}
{Lamers} H.~J.~G.~L.~M.,  {Zickgraf} F.-J.,  {de Winter} D.,  {Houziaux} L.,
  {Zorec} J.,  1998, \aap, \href
  {http://adsabs.harvard.edu/abs/1998A%26A...340..117L} {340, 117}

\bibitem[\protect\citeauthoryear{{Liermann}, {Kraus}, {Schnurr}  \&
  {Fernandes}}{{Liermann} et~al.}{2010}]{Liermann2010}
{Liermann} A.,  {Kraus} M.,  {Schnurr} O.,   {Fernandes} M.~B.,  2010, \mn@doi
  [\mnras] {10.1111/j.1745-3933.2010.00915.x}, \href
  {http://adsabs.harvard.edu/abs/2010MNRAS.408L...6L} {408, L6}

\bibitem[\protect\citeauthoryear{{Lomb}}{{Lomb}}{1976}]{Lomb1976}
{Lomb} N.~R.,  1976, \mn@doi [\apss] {10.1007/BF00648343}, \href
  {http://adsabs.harvard.edu/abs/1976Ap%26SS..39..447L} {39, 447}

\bibitem[\protect\citeauthoryear{{Machado} \& {de Ara{\'u}jo}}{{Machado} \& {de
  Ara{\'u}jo}}{2003}]{Machado2003}
{Machado} M.~A.~D.,  {de Ara{\'u}jo} F.~X.,  2003, \mn@doi [\aap]
  {10.1051/0004-6361:20031159}, \href
  {http://adsabs.harvard.edu/abs/2003A%26A...409..665M} {409, 665}

\bibitem[\protect\citeauthoryear{{Maeder} \& {Desjacques}}{{Maeder} \&
  {Desjacques}}{2001}]{Maeder2001}
{Maeder} A.,  {Desjacques} V.,  2001, \mn@doi [\aap]
  {10.1051/0004-6361:20010539}, \href
  {http://adsabs.harvard.edu/abs/2001A%26A...372L...9M} {372, L9}

\bibitem[\protect\citeauthoryear{{Maravelias}}{{Maravelias}}{2014}]{Maravelias2014}
{Maravelias} G.,  2014, PhD thesis, University of Crete, Heraklion, Greece

\bibitem[\protect\citeauthoryear{{Maravelias}, {Kraus}, {Aret}, {Cidale},
  {Arias}  \& {Borges Fernandes}}{{Maravelias} et~al.}{2017}]{Maravelias2017}
{Maravelias} G.,  {Kraus} M.,  {Aret} A.,  {Cidale} L.,  {Arias} M.~L.,
  {Borges Fernandes} M.,  2017, in {Miroshnichenko} A.,  {Zharikov} S.,
  {Kor{\v c}{\'a}kov{\'a}} D.,   {Wolf} M.,  eds,  Astronomical Society of the
  Pacific Conference Series Vol. 508, The B[e] Phenomenon: Forty Years of
  Studies. p.~213 (\mn@eprint {arXiv} {1610.00607})

\bibitem[\protect\citeauthoryear{{Marchiano}, {Brandi}, {Muratore}, {Quiroga},
  {Ferrer}  \& {Garc{\'{\i}}a}}{{Marchiano} et~al.}{2012}]{Marchiano2012}
{Marchiano} P.,  {Brandi} E.,  {Muratore} M.~F.,  {Quiroga} C.,  {Ferrer}
  O.~E.,   {Garc{\'{\i}}a} L.~G.,  2012, \mn@doi [\aap]
  {10.1051/0004-6361/201117715}, \href
  {http://adsabs.harvard.edu/abs/2012A%26A...540A..91M} {540, A91}

\bibitem[\protect\citeauthoryear{{McGregor}, {Hyland}  \& {Hillier}}{{McGregor}
  et~al.}{1988}]{McGregor1988}
{McGregor} P.~J.,  {Hyland} A.~R.,   {Hillier} D.~J.,  1988, \mn@doi [\apj]
  {10.1086/165964}, \href {http://adsabs.harvard.edu/abs/1988ApJ...324.1071M}
  {324, 1071}

\bibitem[\protect\citeauthoryear{{Mehner} et~al.,}{{Mehner}
  et~al.}{2016}]{Mehner2016}
{Mehner} A.,  et~al., 2016, \mn@doi [\aap] {10.1051/0004-6361/201527180}, \href
  {http://adsabs.harvard.edu/abs/2016A%26A...585A..81M} {585, A81}

\bibitem[\protect\citeauthoryear{{Menu}, {van Boekel}, {Henning}, {Leinert},
  {Waelkens}  \& {Waters}}{{Menu} et~al.}{2015}]{Menu2015}
{Menu} J.,  {van Boekel} R.,  {Henning} T.,  {Leinert} C.,  {Waelkens} C.,
  {Waters} L.~B.~F.~M.,  2015, \mn@doi [\aap] {10.1051/0004-6361/201525654},
  \href {http://adsabs.harvard.edu/abs/2015A%26A...581A.107M} {581, A107}

\bibitem[\protect\citeauthoryear{{Millour} et~al.,}{{Millour}
  et~al.}{2009}]{Millour2009}
{Millour} F.,  et~al., 2009, \mn@doi [\aap] {10.1051/0004-6361/200811592},
  \href {http://adsabs.harvard.edu/abs/2009A%26A...507..317M} {507, 317}

\bibitem[\protect\citeauthoryear{{Millour}, {Meilland}, {Chesneau}, {Stee},
  {Kanaan}, {Petrov}, {Mourard}  \& {Kraus}}{{Millour}
  et~al.}{2011}]{Millour2011}
{Millour} F.,  {Meilland} A.,  {Chesneau} O.,  {Stee} P.,  {Kanaan} S.,
  {Petrov} R.,  {Mourard} D.,   {Kraus} S.,  2011, \mn@doi [\aap]
  {10.1051/0004-6361/201016193}, \href
  {http://adsabs.harvard.edu/abs/2011A%26A...526A.107M} {526, A107}

\bibitem[\protect\citeauthoryear{{Miroshnichenko}}{{Miroshnichenko}}{2007}]{Miroshnichenko2007}
{Miroshnichenko} A.~S.,  2007, \mn@doi [\apj] {10.1086/520798}, \href
  {http://adsabs.harvard.edu/abs/2007ApJ...667..497M} {667, 497}

\bibitem[\protect\citeauthoryear{{Miroshnichenko}, {Levato}, {Bjorkman}  \&
  {Grosso}}{{Miroshnichenko} et~al.}{2003}]{Miroshnichenko2003}
{Miroshnichenko} A.~S.,  {Levato} H.,  {Bjorkman} K.~S.,   {Grosso} M.,  2003,
  \mn@doi [\aap] {10.1051/0004-6361:20030824}, \href
  {http://adsabs.harvard.edu/abs/2003A%26A...406..673M} {406, 673}

\bibitem[\protect\citeauthoryear{{Miroshnichenko}, {Bjorkman}, {Grosso},
  {Hinkle}, {Levato}  \& {Marang}}{{Miroshnichenko}
  et~al.}{2005}]{Miroshnichenko2005}
{Miroshnichenko} A.~S.,  {Bjorkman} K.~S.,  {Grosso} M.,  {Hinkle} K.,
  {Levato} H.,   {Marang} F.,  2005, \mn@doi [\aap]
  {10.1051/0004-6361:20052726}, \href
  {http://adsabs.harvard.edu/abs/2005A%26A...436..653M} {436, 653}

\bibitem[\protect\citeauthoryear{{Morris}, {Eenens}, {Hanson}, {Conti}  \&
  {Blum}}{{Morris} et~al.}{1996}]{Morris1996}
{Morris} P.~W.,  {Eenens} P.~R.~J.,  {Hanson} M.~M.,  {Conti} P.~S.,   {Blum}
  R.~D.,  1996, \mn@doi [\apj] {10.1086/177892}, \href
  {http://adsabs.harvard.edu/abs/1996ApJ...470..597M} {470, 597}

\bibitem[\protect\citeauthoryear{{Muratore}, {de Wit}, {Kraus}, {Aret},
  {Cidale}, {Borges Fernandes}, {Oudmaijer}  \& {Wheelwright}}{{Muratore}
  et~al.}{2012}]{Muratore2012}
{Muratore} M.~F.,  {de Wit} W.~J.,  {Kraus} M.,  {Aret} A.,  {Cidale} L.~S.,
  {Borges Fernandes} M.,  {Oudmaijer} R.~D.,   {Wheelwright} H.~E.,  2012, in
  {Carciofi} A.~C.,  {Rivinius} T.,  eds,  Astronomical Society of the Pacific
  Conference Series Vol. 464, Circumstellar Dynamics at High Resolution. p.~67
  (\mn@eprint {arXiv} {1212.4798})

\bibitem[\protect\citeauthoryear{{Muratore}, {Kraus}, {Oksala}, {Arias},
  {Cidale}, {Borges Fernandes}  \& {Liermann}}{{Muratore}
  et~al.}{2015}]{Muratore2015}
{Muratore} M.~F.,  {Kraus} M.,  {Oksala} M.~E.,  {Arias} M.~L.,  {Cidale} L.,
  {Borges Fernandes} M.,   {Liermann} A.,  2015, \mn@doi [\aj]
  {10.1088/0004-6256/149/1/13}, \href
  {http://adsabs.harvard.edu/abs/2015AJ....149...13M} {149, 13}

\bibitem[\protect\citeauthoryear{{Oksala}, {Kraus}, {Cidale}, {Muratore}  \&
  {Borges Fernandes}}{{Oksala} et~al.}{2013}]{Oksala2013}
{Oksala} M.~E.,  {Kraus} M.,  {Cidale} L.~S.,  {Muratore} M.~F.,   {Borges
  Fernandes} M.,  2013, \mn@doi [\aap] {10.1051/0004-6361/201321568}, \href
  {http://adsabs.harvard.edu/abs/2013A%26A...558A..17O} {558, A17}

\bibitem[\protect\citeauthoryear{{Oudmaijer}, {Proga}, {Drew}  \& {de
  Winter}}{{Oudmaijer} et~al.}{1998}]{Oudmaijer1998}
{Oudmaijer} R.~D.,  {Proga} D.,  {Drew} J.~E.,   {de Winter} D.,  1998, \mn@doi
  [\mnras] {10.1046/j.1365-8711.1998.01875.x}, \href
  {http://adsabs.harvard.edu/abs/1998MNRAS.300..170O} {300, 170}

\bibitem[\protect\citeauthoryear{{Pelupessy}, {Lamers}  \& {Vink}}{{Pelupessy}
  et~al.}{2000}]{Pelupessy2000}
{Pelupessy} I.,  {Lamers} H.~J.~G.~L.~M.,   {Vink} J.~S.,  2000, \aap, \href
  {http://adsabs.harvard.edu/abs/2000A%26A...359..695P} {359, 695}

\bibitem[\protect\citeauthoryear{{Petrov}, {Vink}  \& {Gr{\"a}fener}}{{Petrov}
  et~al.}{2016}]{Petrov2016}
{Petrov} B.,  {Vink} J.~S.,   {Gr{\"a}fener} G.,  2016, \mn@doi [\mnras]
  {10.1093/mnras/stw382}, \href
  {http://adsabs.harvard.edu/abs/2016MNRAS.458.1999P} {458, 1999}

\bibitem[\protect\citeauthoryear{{Plets}, {Waelkens}  \& {Trams}}{{Plets}
  et~al.}{1995}]{Plets1995}
{Plets} H.,  {Waelkens} C.,   {Trams} N.~R.,  1995, \aap, \href
  {http://adsabs.harvard.edu/abs/1995A%26A...293..363P} {293, 363}

\bibitem[\protect\citeauthoryear{{Podsiadlowski}, {Morris}  \&
  {Ivanova}}{{Podsiadlowski} et~al.}{2006}]{Podsiadlowski2006}
{Podsiadlowski} P.,  {Morris} T.~S.,   {Ivanova} N.,  2006, in {Kraus} M.,
  {Miroshnichenko} A.~S.,  eds,  Astronomical Society of the Pacific Conference
  Series Vol. 355, Stars with the B[e] Phenomenon. p.~259

\bibitem[\protect\citeauthoryear{{Pojmanski}}{{Pojmanski}}{1997}]{Pojmanski1997}
{Pojmanski} G.,  1997, \actaa, \href
  {http://adsabs.harvard.edu/abs/1997AcA....47..467P} {47, 467}

\bibitem[\protect\citeauthoryear{{Porter}}{{Porter}}{2003}]{Porter2003}
{Porter} J.~M.,  2003, \mn@doi [\aap] {10.1051/0004-6361:20021698}, \href
  {http://adsabs.harvard.edu/abs/2003A%26A...398..631P} {398, 631}

\bibitem[\protect\citeauthoryear{{Sana} et~al.,}{{Sana}
  et~al.}{2012}]{Sana2012}
{Sana} H.,  et~al., 2012, \mn@doi [Science] {10.1126/science.1223344}, \href
  {http://adsabs.harvard.edu/abs/2012Sci...337..444S} {337, 444}

\bibitem[\protect\citeauthoryear{{Sana} et~al.,}{{Sana}
  et~al.}{2013}]{Sana2013}
{Sana} H.,  et~al., 2013, \mn@doi [\aap] {10.1051/0004-6361/201219621}, \href
  {http://adsabs.harvard.edu/abs/2013A%26A...550A.107S} {550, A107}

\bibitem[\protect\citeauthoryear{{Scargle}}{{Scargle}}{1982}]{Scargle1982}
{Scargle} J.~D.,  1982, \mn@doi [\apj] {10.1086/160554}, \href
  {http://adsabs.harvard.edu/abs/1982ApJ...263..835S} {263, 835}

\bibitem[\protect\citeauthoryear{{Surdej} \& {Swings}}{{Surdej} \&
  {Swings}}{1983}]{Surdej1983}
{Surdej} J.,  {Swings} J.~P.,  1983, \aap, \href
  {http://adsabs.harvard.edu/abs/1983A%26A...117..359S} {117, 359}

\bibitem[\protect\citeauthoryear{{Tody}}{{Tody}}{1993}]{IRAF}
{Tody} D.,  1993, in {Hanisch} R.~J.,  {Brissenden} R.~J.~V.,   {Barnes} J.,
  eds,  Astronomical Society of the Pacific Conference Series Vol. 52,
  Astronomical Data Analysis Software and Systems II. p.~173

\bibitem[\protect\citeauthoryear{{Torres}, {Cidale}, {Kraus}, {Arias},
  {Barb{\'a}}, {Maravelias}  \& {Borges Fernandes}}{{Torres}
  et~al.}{2018}]{Torres2018}
{Torres} A.~F.,  {Cidale} L.~S.,  {Kraus} M.,  {Arias} M.~L.,  {Barb{\'a}}
  R.~H.,  {Maravelias} G.,   {Borges Fernandes} M.,  2018, \mn@doi [\aap]
  {10.1051/0004-6361/201731723}, \href
  {http://adsabs.harvard.edu/abs/2018A%26A...612A.113T} {612, A113}

\bibitem[\protect\citeauthoryear{Van Der~Walt, Colbert  \& Varoquaux}{Van
  Der~Walt et~al.}{2011}]{numpy}
Van Der~Walt S.,  Colbert S.~C.,   Varoquaux G.,  2011, Computing in Science \&
  Engineering, 13, 22

\bibitem[\protect\citeauthoryear{{Wheelwright}, {de Wit}, {Oudmaijer}  \&
  {Vink}}{{Wheelwright} et~al.}{2012a}]{Wheelwright2012}
{Wheelwright} H.~E.,  {de Wit} W.~J.,  {Oudmaijer} R.~D.,   {Vink} J.~S.,
  2012a, \mn@doi [\aap] {10.1051/0004-6361/201117766}, \href
  {http://adsabs.harvard.edu/abs/2012A%26A...538A...6W} {538, A6}

\bibitem[\protect\citeauthoryear{{Wheelwright}, {de Wit}, {Weigelt},
  {Oudmaijer}  \& {Ilee}}{{Wheelwright} et~al.}{2012b}]{Wheelwright2012b}
{Wheelwright} H.~E.,  {de Wit} W.~J.,  {Weigelt} G.,  {Oudmaijer} R.~D.,
  {Ilee} J.~D.,  2012b, \mn@doi [\aap] {10.1051/0004-6361/201219325}, \href
  {http://adsabs.harvard.edu/abs/2012A%26A...543A..77W} {543, A77}

\bibitem[\protect\citeauthoryear{{Whitelock}, {Feast}, {Roberts}, {Carter}  \&
  {Catchpole}}{{Whitelock} et~al.}{1983}]{Whitelock1983}
{Whitelock} P.~A.,  {Feast} M.~W.,  {Roberts} G.,  {Carter} B.~S.,
  {Catchpole} R.~M.,  1983, \mn@doi [\mnras] {10.1093/mnras/205.4.1207}, \href
  {http://adsabs.harvard.edu/abs/1983MNRAS.205.1207W} {205, 1207}

\bibitem[\protect\citeauthoryear{{Zickgraf}}{{Zickgraf}}{2003}]{Zickgraf2003}
{Zickgraf} F.-J.,  2003, \mn@doi [\aap] {10.1051/0004-6361:20030999}, \href
  {http://adsabs.harvard.edu/abs/2003A%26A...408..257Z} {408, 257}

\bibitem[\protect\citeauthoryear{{Zickgraf}, {Wolf}, {Stahl}, {Leitherer}  \&
  {Klare}}{{Zickgraf} et~al.}{1985}]{Zickgraf1985}
{Zickgraf} F.-J.,  {Wolf} B.,  {Stahl} O.,  {Leitherer} C.,   {Klare} G.,
  1985, \aap, \href {http://adsabs.harvard.edu/abs/1985A%26A...143..421Z} {143,
  421}

\bibitem[\protect\citeauthoryear{{de Jager}}{{de Jager}}{1998}]{deJager1998}
{de Jager} C.,  1998, \mn@doi [\aapr] {10.1007/s001590050009}, \href
  {http://adsabs.harvard.edu/abs/1998A%26ARv...8..145D} {8, 145}

\bibitem[\protect\citeauthoryear{{de Mink}, {Sana}, {Langer}, {Izzard}  \&
  {Schneider}}{{de Mink} et~al.}{2014}]{deMink2014}
{de Mink} S.~E.,  {Sana} H.,  {Langer} N.,  {Izzard} R.~G.,   {Schneider}
  F.~R.~N.,  2014, \mn@doi [\apj] {10.1088/0004-637X/782/1/7}, \href
  {http://adsabs.harvard.edu/abs/2014ApJ...782....7D} {782, 7}

\bibitem[\protect\citeauthoryear{{de Wit}, {Oudmaijer}  \& {Vink}}{{de Wit}
  et~al.}{2014}]{deWit2014}
{de Wit} W.~J.,  {Oudmaijer} R.~D.,   {Vink} J.~S.,  2014, \mn@doi [Advances in
  Astronomy] {10.1155/2014/270848}, \href
  {http://adsabs.harvard.edu/abs/2014AdAst2014E..10D} {2014, 270848}

\bibitem[\protect\citeauthoryear{{van den Bergh}}{{van den
  Bergh}}{1972}]{vandenBergh1972}
{van den Bergh} S.,  1972, \mn@doi [\pasp] {10.1086/129339}, \href
  {http://adsabs.harvard.edu/abs/1972PASP...84..594V} {84, 594}

\makeatother
\end{thebibliography}



\appendix

\section{Details on the individual fits}
\label{app:individualkinematics}

In this Appendix we give the detailed information for each object regarding our fits for each line per epoch. For each fit we provide the corresponding velocities, i.e. the rotational velocity ($v_{\rm rot}$) and the Gaussian component ($v_g$), as well as the range of integrating angles and possible rotational velocities. 

During the fit process we integrate the line flux over a range of angles which are arbitrary selected. A starting angle of $0\degr$ is set at the right (red) part of the circle along the line-of-sight, and then the integration continues by going behind the star, coming in front of it from the left (blue) part to reach the initial position again (corresponding to a complete ring - in case of partial rings we adjust these angles accordingly). This introduces a direction of the ring rotation which is arbitrary. To overcome this we convert these angles to a range of possible velocities. For complete rings the range is given as $[-v_{\rm rot},+v_{\rm rot}]$, while for partial rings it is defined as $[ |\cos(angle_{{\rm min}}) \times v_{\rm rot}|, |\cos(angle_{{\rm max}}) \times v_{\rm rot}|  ]$.


\begin{table*}
\caption{Fitting the kinematics of CPD-52 9243.}
\label{tab:kin-cpd-529243}
\begin{tabular}{l cccc cccc}
\hline
Date (UT)    &  \multicolumn{4}{c}{[CaII] \lam7291} & \multicolumn{4}{c}{[OI] \lam6300} \\
             &   $v_{\rm rot}$  &  $v_g$  & ring & range & $v_{\rm rot}$  &  $v_g$  & ring & range  \\
             & (\kms) & (\kms) & & (\kms) & (\kms) & (\kms) & & (\kms) \\  
\hline
1999-04-19   & 30.0$\pm$1.0 & 14.0$\pm$1.0 & complete & [-30.0,30.0] & 33.0$\pm$1.0 & 12.0$\pm$1.0 & complete & [-33.0,33.0] \\
             & 49.0$\pm$1.0 & 9.0$\pm$1.0 & complete & [-49.0,49.0] & 55.0$\pm$1.0 & 10.0$\pm$1.0 & complete & [-55.0,55.0] \\
1999-06-25   & 30.0$\pm$1.0 & 12.0$\pm$1.0 & complete & [-30.0,30.0] & 34.0$\pm$1.0 & 12.0$\pm$1.0 & complete & [-34.0,34.0] \\
             & 49.0$\pm$1.0 & 10.0$\pm$1.0 & complete & [-49.0,49.0] & 55.0$\pm$1.0 & 10.0$\pm$1.0 & complete & [-55.0,55.0]  \\
2000-03-28   & 32.0$\pm$1.0 & 9.0$\pm$0.5 & complete & [-32.0,32.0] & 35.0$\pm$3.0 & 11.5$\pm$2.5 & complete & [-35.0,35.0] \\
             & 52.0$\pm$1.0 & 10.0$\pm$1.0 & complete & [-52.0,52.0] & 56.0$\pm$3.0 & 10.5$\pm$1.5 & complete & [-56.0,56.0] \\
2000-06-11   & 31.0$\pm$1.0 & 10.0$\pm$1.0 & 0\degr--160\degr & [|31.0|,|-29.1|] & 31.0$\pm$2.0$^b$ & 11.0$\pm$1.0 & 0\degr--125\degr & [|31.0|,|-17.8|] \\
             & 48.0$\pm$1.0 & 10.0$\pm$1.0 & complete & [-48.0,48.0] & 48.0$\pm$1.0 & 11.0$\pm$1.0& complete & [-48.0,48.0] \\
2005-04-21   & 29.0$\pm$1.0 & 13.0$\pm$0.5 & complete & [-29.0,29.0] & 29.5$\pm$1.5 & 13.0$\pm$1.0 & complete & [-29.5,29.5]\\
             & 48.0$\pm$1.0 & 12.0$\pm$1.0 & complete & [-48.0,48.0] & 50.0$\pm$2.0 & 12.0$\pm$1.0 & complete & [-50.0,50.0] \\
2015-05-13   & 30.0$\pm$1.0 & 9.0$\pm$0.5 & complete & [-30.0,30.0] & 36.0$\pm$1.0 & 11.0$\pm$1.0 & 0\degr--145\degr & [|36.0|,|-29.5|] \\
             & 48.0$\pm$1.0 & 9.0$\pm$1.0 & complete & [-48.0,48.0] & 50.0$\pm$2.0 & 12.0$\pm$1.0 & complete & [-50.0,50.0]  \\
2015-10-11   & 30.0$\pm$1.0 & 8.0$\pm$0.5 & complete & [-30.0,30.0] & 28.0$\pm$1.0 & 10.0$\pm$1.0 & 0\degr--160\degr & [|28.0|,|-26.3|]  \\
             & 48.0$\pm$1.0 & 10.0$\pm$1.0 & complete & [-48.0,48.0] & 48.0$\pm$1.0 & 10.0$\pm$1.0 & complete & [-48.0,48.0] \\
2016-04-13   & 31.0$\pm$0.5 & 8.5$\pm$0.5 & 0\degr--165\degr & [|31.0|,|-29.9|] & 33.0$\pm$1.0 & 11.0$\pm$1.0 & 0\degr--140\degr & [|33.0|,|-25.3|]\\
             & 48.0$\pm$1.0 & 10.0$\pm$1.0 & complete & [-48.0,48.0] & 50.0$\pm$1.0 & 11.0$\pm$1.0 & complete & [-50.0,50.0] \\
2016-08-02   & 31.0$\pm$1.0$^g$ & 9.0$\pm$0.5 & 30\degr--360\degr & [|26.8|,|31.0|] & 29.0$\pm$1.0 & 11.0$\pm$1.0 & complete & [-29.0,29.0] \\
             & 50.0$\pm$1.0 & 11.0$\pm$1.0 & complete & [-50.0,50.0] & 51.0$\pm$1.0 & 11.5$\pm$0.5& complete & [-51.0,51.0] \\

\hline
\end{tabular}
\end{table*}

\begin{table*}
\caption{Fitting the kinematics of CPD-57 2874.}
\label{tab:kin-cpd-572874}
\begin{tabular}{l cccc cccc}
\hline
Date (UT)    &  \multicolumn{4}{c}{[CaII] \lam7291} & \multicolumn{4}{c}{[OI] \lam6300} \\
             &   $v_{\rm rot}$  &  $v_g$  & ring & range & $v_{\rm rot}$  &  $v_g$  & ring & range  \\
             & (\kms) & (\kms) & & (\kms) & (\kms) & (\kms) & & (\kms) \\  
\hline
2008-12-22   & 110.0$\pm$3.0 & 14.0$\pm$1.0 & 100\degr--260\degr & [|-19.1|,|-19.1|] & 42.0$\pm$2.0 & 12.0$\pm$1.0 & complete & [-42.0,42.0] \\
             & 109.0$\pm$3.0 & 14.0$\pm$1.0 & 0\degr--90\degr & [|109|,0] & 88.0$\pm$3.0 & 11.0$\pm$1.0 & complete & [-88.0,88.0] \\
             & 161.0$\pm$3.0 & 13.0$\pm$1.0 & complete & [-161.0,161.0] & 120.0$\pm$2.0 & 12.0$\pm$1.0 & complete & [-120.0,120.0] \\
             & 208.0$\pm$8.0 & 10.0$\pm$2.0 & complete & [-208.0,208.0] & 165.0$\pm$7.0 & 12.0$\pm$2.0 & complete & [-165.0,165.0] \\

2015-05-13   & 105.0$\pm$3.0 & 13.0$\pm$1.0 & 100\degr--260\degr & [|-18.2|,|-18.2|] & 44.5$\pm$2.5 & 11.0$\pm$1.0 & complete & [-44.5,44.5] \\
             & 105.0$\pm$3.0 & 13.0$\pm$1.0 & 0\degr--85\degr & [|105|,|9.2|] & 88.0$\pm$4.0 & 11.0$\pm$2.0 & complete & [-88.0,88.0] \\
             & 157.0$\pm$3.0 & 13.0$\pm$1.0 & complete & [-157.0,157.0] & 121.0$\pm$3.0 & 13.0$\pm$2.0 & complete & [-121.0,121.0] \\
             & 220.0$\pm$5.0 & 10.0$\pm$2.0 & complete & [-220.0,220.0] & 167.0$\pm$6.0 & 12.0$\pm$2.0 & complete & [-167.0,167.0] \\

2016-01-13   & 107.0$\pm$2.0 & 13.0$\pm$1.0 & 100\degr--260\degr & [|-18.6|,|-18.6|] & 48.0$\pm$2.0 & 12.0$\pm$1.0 & complete & [-48.0,48.0] \\
             & 108.0$\pm$1.0 & 14.0$\pm$1.0 & 0\degr--88\degr & [|108|,|3.8|] & 90.0$\pm$3.0 & 11.0$\pm$1.0 & complete & [-90.0,90.0] \\
             & 160.0$\pm$2.0 & 13.0$\pm$1.0 & complete & [-160.0,160.0] & 123.0$\pm$4.0 & 13.0$\pm$2.0 & complete & [-123.0,123.0] \\
             & 213.0$\pm$5.0 & 12.0$\pm$2.0 & complete & [-213.0,213.0] & 166.0$\pm$6.0 & 12.0$\pm$2.0 & complete & [-166.0,166.0] \\

2016-03-13   & 111.0$\pm$4.0 & 14.0$\pm$1.0 & 100\degr--260\degr & [|-19.3|,|-19.3|] & 44.5$\pm$2.5 & 12.0$\pm$1.0 & complete & [-44.5,44.5] \\
             & 110.0$\pm$3.0$^c$ & 13.0$\pm$1.0 & 0\degr--88\degr & [|110|,|3.8|] & 87.0$\pm$4.0 & 11.0$\pm$2.0 & complete & [-87.0,87.0] \\
             & 155.0$\pm$4.0 & 12.0$\pm$1.0 & complete & [-155.0,155.0] & 117.0$\pm$3.0 & 13.0$\pm$1.0 & complete & [-117.0,117.0] \\
             & 201.0$\pm$11.0 & 12.0$\pm$2.0 & complete & [-201.0,201.0] & 167.0$\pm$3.0 & 12.0$\pm$2.0 & complete & [-167.0,167.0] \\

\hline
\end{tabular}

 
\end{table*}

\begin{table*}
\caption{Fitting the kinematics of HD 327083.}
\label{tab:kin-hd327083}
\begin{tabular}{l cccc cccc}
\hline
Date (UT)    &  \multicolumn{4}{c}{[CaII] \lam7291} & \multicolumn{4}{c}{[OI] \lam6300} \\
             &   $v_{\rm rot}$  &  $v_g$  & ring & range & $v_{\rm rot}$  &  $v_g$  & ring & range  \\
             & (\kms) & (\kms) & & (\kms) & (\kms) & (\kms) & & (\kms) \\  
\hline
1999-06-25   & 76.0$\pm$1.0 & 11.0$\pm$1.0 & 0\degr--80\degr  &  [|76.0|,|13.2|] & 75.0$\pm$1.0 & 14.0$\pm$1.0  &  0\degr--122\degr & [|75.0|,|13.0|] \\
             & 75.0$\pm$1.0 & 8.0$\pm$1.0 & 90\degr--270\degr & [0,0] &  &  & \\

2015-05-11   & 76.0$\pm$1.0 & 12.0$\pm$1.0 &  0\degr--110\degr  & [|76.0|,|-26.0|] & 58.0$\pm$2.0 & 11.0$\pm$1.0 &  0\degr--110\degr & [|58.0|,|-19.8|]\\
             & 77.0$\pm$1.0 & 12.0$\pm$2.0 & 40\degr--270\degr & [|59.0|,0] &  76.0$\pm$2.0 & 9.0$\pm$1.0  &  0\degr--140\degr & [|76.0|,|-58.2|] \\

2015-10-12   & 76.0$\pm$1.0 & 12.5$\pm$0.5 &  0\degr--90\degr & [|76.0|,0] &  70.0$\pm$2.0 &  14.0$\pm$1.0 &  110\degr--270\degr & [|-23.9|,0]\\
             & 76.0$\pm$1.0 & 13.5$\pm$0.5 & 120\degr--270\degr  & [|-38.0|,0] & 74.0$\pm$2.0 & 12.0$\pm$1.0 &  0\degr--85\degr  & [|74.0|,|6.4|]\\

2015-10-15   & 76.0$\pm$1.0 & 13.0$\pm$1.0 & 0\degr--90\degr & [|76.0|,0] &  70.0$\pm$1.0 &  14.0$\pm$1.0 &  110\degr--270\degr & [|-23.9|,0] \\
             & 76.5$\pm$1.0 & 12.5$\pm$0.5 & 120\degr--270\degr & [|-38.5|,0] & 74.0$\pm$1.0 & 14.0$\pm$1.0 &  0\degr--90\degr & [|74.0|,0] \\

2016-04-13   & 77.0$\pm$1.0 & 12.0$\pm$0.5 & 0\degr--85\degr & [|77.0|,|6.7|] &  66.0$\pm$2.0 &  8.0$\pm$1.0 &  90\degr--270\degr & [0,0]\\
             & 71.0$\pm$1.0 & 11.0$\pm$1.0 & 110\degr--270\degr & [|-24.3|,0] & 80.0$\pm$1.0 & 13.0$\pm$1.0  &  0\degr--105\degr & [|80|,|-20.7|]\\

2016-07-28   & 76.0$\pm$1.0 & 12.0$\pm$1.0 & 0\degr--85\degr & [|77.0|,|6.7|] &  61.0$\pm$3.0 &  9.0$\pm$1.0 &  90\degr--270\degr & [0,0]\\
             & 72.0$\pm$1.0 & 10.5$\pm$1.5 & 110\degr--270\degr & [|-24.6|,0] & 78.5$\pm$1.5 & 13.0$\pm$1.0  &  0\degr--106\degr & [|78.5|,|-21.6|]\\

\hline 
\end{tabular}


\flushleft{ 
Note: The fit of the \oi \lam6300 line for the 2016-04-13 epoch is not well-constrained due to the extended red wing it displays. 
}
\end{table*}

\begin{table*}
\caption{Fitting the kinematics of HD 62623.}
\label{tab:kin-hd62623}
\resizebox{\textwidth}{!}{ 
\begin{tabular}{l cccc cccc cccc}
\hline
Date (UT)    &  \multicolumn{4}{c}{[OI] \lam5577} & \multicolumn{4}{c}{[CaII] \lam7291} & \multicolumn{4}{c}{[OI] \lam6300} \\
             &   $v_{\rm rot}$  &  $v_g$  & ring & range & $v_{\rm rot}$  &  $v_g$  & ring & range & $v_{\rm rot}$  &  $v_g$  & ring & range \\
             & (\kms) & (\kms) & & (\kms) & (\kms) & (\kms) & & (\kms) & (\kms) & (\kms) & & (\kms)\\  
\hline
2008-12-21   & 57.5$\pm$1.5 & 13.0$\pm$2.0  & 0\degr-95\degr & [|57.5|,|-5.0|] & 41.0$\pm$2.0 & 9.5$\pm$0.5 & compl. & [-41.0,41.0] & 37.0$\pm$1.0 & 11.0$\pm$1.0 & compl. & [-37.0,37.0] \\
             &                 &                 & & &  60.5$\pm$0.5 & 11.0$\pm$1.0 & compl. & [-60.5,60.5] & 60.0$\pm$1.0 & 12.0$\pm$2.0 & compl. & [-60.0,60.0]\\

2010-05-03   & 61.0$\pm$2.0 & 10.0$\pm$1.0 & 0\degr-95\degr & [|61.0|,|-5.3|] &  40.5$\pm$0.5 & 8.5$\pm$0.5 & compl. & [-40.5,40.5] & 37.5$\pm$1.0 & 11.0$\pm$1.0 & compl. & [-37.5,37.5]\\
             &                 &                 & & &  60.0$\pm$1.0 & 11.0$\pm$1.0 & compl. & [-60.0,60.0] & 61.5$\pm$1.0 & 12.0$\pm$1.0 & compl. & [-61.5,61.5] \\

2013-05-09   & 57.0$\pm$3.0 & 8.5$\pm$1.5 & 0\degr-95\degr & [|57.0|,|-5.0|] &  41.0$\pm$0.5 & 7.5$\pm$0.5 & 20\degr--360\degr & [|38.5|,|41.0|] & 38.0$\pm$1.0 & 11.5$\pm$0.5 & compl. & [-38.0,38.0] \\
             &                 &          & & &  60.0$\pm$1.0 & 11.0$\pm$1.0 & compl. & [-60.0,60.0] & 62.0$\pm$1.0 & 12.0$\pm$1.0 & compl. & [-62.0,62.0] \\

2014-11-29   & 55.0$\pm$5.0 & 11.0$\pm$2.0 & 0\degr-95\degr & [|55.0|,|-4.8|]  &  42.0$\pm$1.0 & 8.5$\pm$0.5 & 20\degr--360\degr & [|39.5|,|42.0|]& 43.0$\pm$2.0 &     10.0$\pm$1.0 & 20\degr--360\degr & [|40.4|,|43.0|]\\
             &                 &                 & & &  61.0$\pm$1.0 & 10.0$\pm$0.5 & 20\degr--360\degr & [|57.3|,|61.0|] & 62.5$\pm$1.5 & 11.0$\pm$1.0 & 20\degr--360\degr & [|58.7|,|62.5|]\\

2015-05-10   & 55.0$\pm$3.0 & 11.0$\pm$1.0 & 0\degr-95\degr & [|55.0|,|-4.8|]  &  42.5$\pm$0.5 & 8.5$\pm$0.5 & 20\degr--360\degr & [|39.9|,|42.5|] & 41.5$\pm$1.5 & 11.5$\pm$0.5 & 20\degr--360\degr & [|39.0|,|41.5|] \\
             &                 &                 & & &  61.5$\pm$0.5 & 11.0$\pm$1.0 & 20\degr--360\degr  & [|57.8|,|61.5|] & 60.0$\pm$1.0 & 13.0$\pm$1.0 & 20\degr--360\degr  & [|56.4|,|60.0|]\\

2015-10-12   & \multicolumn{4}{c}{too noisy - not used} &  44.0$\pm$1.0 & 11.0$\pm$1.0 & 20\degr--360\degr & [|41.3|,|44.0|] & 40.0$\pm$3.0 & 11.5$\pm$1.5 & 20\degr--360\degr & [|37.6|,|40.0|] \\
             &                 &                 & & &  62.0$\pm$1.5 & 11.0$\pm$1.0 & 20\degr--360\degr & [|58.3|,|62.0|] & 64.0$\pm$2.0 & 13.0$\pm$1.0 & 20\degr--360\degr & [|60.1|,|64.0|]\\ 
\hline 
\end{tabular}
}
\end{table*}

\begin{table*}
\caption{Fitting the kinematics of GG Car.}
\label{tab:kin-ggcar}
\begin{tabular}{l cccc cccc}
\hline
Date (UT)    &  \multicolumn{4}{c}{[CaII] \lam7291} & \multicolumn{4}{c}{[OI] \lam6300} \\
             &  $v_{\rm rot}$  &  $v_g$  & ring & range &  $v_{\rm rot}$  &  $v_g$  & ring & range  \\
             & (\kms) & (\kms) &  & (\kms) & (\kms) & (\kms) & & (\kms)\\  
\hline
1999-04-18   & 96.0$\pm$2.0 & 6.5$\pm$0.5 & 0\degr--165\degr & [|96.0|,|-92.7|] & 35.5$\pm$1.5 & 12.0$\pm$2.0 & complete & [-35.5,35.5]\\
             &               &             & & & 91.5$\pm$3.5 & 9.0$\pm$2.0 & complete & [-91.5,91.5] \\
2000-02-23   & 96.0$\pm$2.0 & 6.5$\pm$0.5 & 0\degr--165\degr & [|96.0|,|-92.7|] & 30.0$\pm$1.5 & 13.0$\pm$1.0 & complete & [-30.0,30.0]\\
             &               &             & & & 90.5$\pm$2.5 & 9.0$\pm$2.0 & complete & [-90.5,90.5] \\
2008-12-22   & 94.0$\pm$2.0 & 8.5$\pm$1.0 & 5\degr--180\degr & [|93.6|,|-94|] & 29.0$\pm$1.0 & 13.0$\pm$1.0 & complete & [-29.0,29.0]\\
             &               &             & & & 51.0$\pm$3.0 & 14.0$\pm$2.0 & complete & [-51.0,51.0] \\
             &               &             & & & 91.5$\pm$8.5 & 12.0$\pm$2.0 & complete & [-91.5,91.5]\\
2009-06-09   & 97.0$\pm$1.0 & 8.5$\pm$1.5 & 15\degr--180\degr & [|93.7|,|-97|] & 28.5$\pm$1.5 & 13.0$\pm$1.0 & complete & [-28.5,28.5]\\
             &               &             & & & 79.0$\pm$5.0 & 12.0$\pm$2.0 & complete & [-79.0,79.0]\\
2011-03-23   & 97.0$\pm$2.0 & 8.0$\pm$1.0 & 15\degr--180\degr & [|93.7|,|-97|] & 30.0$\pm$1.0 & 13.0$\pm$1.0 & complete & [-30.0,30.0]\\
             &               &             & & & 82.0$\pm$2.0 & 12.0$\pm$2.0 & complete & [-82.0,82.0]\\
2015-05-13   & 92.5$\pm$2.5 & 8.5$\pm$1.5 & 10\degr--180\degr & [|91.1|,|-92.5|] & 29.0$\pm$1.0 & 13.0$\pm$1.0 & complete & [-29.0,29.0]\\
             &               &             & & & 82.5$\pm$7.5 & 14.0$\pm$2.0 & complete & [-82.5,82.5]\\
2015-11-23   & 97.0$\pm$2.0 & 8.0$\pm$1.0 & 20\degr--180\degr & [|91.2|,|-97|] & 28.0$\pm$1.0 & 13.0$\pm$1.0 & complete & [-28.0,28.0]\\
             &               &             & & & 77.0$\pm$8.0 & 14.0$\pm$2.0 & complete & [-77.0,77.0]\\
2015-11-26   & 96.5$\pm$1.5 & 9.5$\pm$1.5 & 20\degr--180\degr & [|90.7|,|-97|] & 29.0$\pm$1.0 & 13.0$\pm$1.0 & complete & [-29.0,29.0]\\
             &               &             & & & 80.0$\pm$5.0 & 14.0$\pm$2.0 & complete & [-80.0,80.0]\\

\hline& & & & 
\end{tabular}
\end{table*}

\begin{table*}
\caption{Fitting the kinematics of MWC 137.}
\label{tab:kin-mwc137}
\begin{tabular}{l cccc}
\hline
Date (UT)    & \multicolumn{3}{c}{[OI] \lam6300} & \\
             &  $v_{\rm rot}$  &  $v_g$  & ring & range \\
             & (\kms) & (\kms) & & (\kms) \\  
\hline
2015-12-05   & 20.0$\pm$0.5 & 7.5$\pm$0.5 & complete & [-20.0,20.0] \\
             & 31.0$\pm$1.0 & 10.0$\pm$1.0 & complete & [-31.0,31.0] \\
             & 47.0$\pm$1.0 & 7.0$\pm$1.0 & complete & [-47.0,47.0] \\
             & 67.0$\pm$2.0 & 15.0$\pm$2.0 & complete & [-67.0,67.0] \\
2016-02-28   & 20.5$\pm$0.5 & 7.5$\pm$0.5 & complete & [-20.5,20.5] \\
             & 31.0$\pm$1.0 & 10.0$\pm$1.0 & complete & [-31.0,31.0] \\
             & 46.5$\pm$1.0 & 8.0$\pm$1.0 & complete & [-46.5,46.5] \\
             & 69.0$\pm$2.0 & 9.0$\pm$2.0 & complete & [-69.0,69.0] \\
\hline
\end{tabular}
\end{table*}

\begin{table*}
\caption{Fitting the kinematics of HD 87643.}
\label{tab:kin-hd87643}
\begin{tabular}{ c cccc cccc}
\hline
Date (UT)    &  \multicolumn{4}{c}{[CaII] \lam7291} & \multicolumn{4}{c}{[OI] \lam6300} \\
             & $v_{\rm rot}$  &  $v_g$  & ring & range & $v_{\rm rot}$ &  $v_g$ & ring & range \\
             & (\kms) & (\kms) &  & (\kms) & (\kms) & (\kms) &  & (\kms)\\  
\hline
1999-04-18   & 12.0$\pm$0.5 & 11.0$\pm$1.0 & complete & [-12.0,12.0] & $<3$* & 12.0$\pm$1.0 & 0\degr--155\degr & [|3|,|-2.7|]\\
             &                 &             & & & 23.0$\pm$1.0 & 16.0$\pm$2.0 & 0\degr--155\degr & [|23.0|,|-20.8|]\\
             &                 &             & & & 62.0$\pm$4.0 & 16.0$\pm$2.0 & 0\degr--155\degr & [|62.0|,|-56.2|]\\
             &                 &             & & & 90.0$\pm$5.0 & 9.0$\pm$2.0 & 0\degr--155\degr & [|90.0|,|-81.6|]\\

2000-02-23   & 11.0$\pm$1.0 & 12.0$\pm$2.0 & complete & [-11.0,11.0] & 9.0$\pm$1.0 & 9.0$\pm$2.0 & 0\degr--150\degr & [|9|,|-7.8|]\\
             &                &             & & & 24.5$\pm$1.5 & 15.5$\pm$2.5 & 0\degr--150\degr & [|24.5|,|-21.2|]\\
             &                 &             & & & 54.0$\pm$2.0 & 16.0$\pm$2.0 & 0\degr--150\degr & [|54.0|,|-46.8|]\\
             &                 &             & & & 85.5$\pm$2.5 & 9.0$\pm$2.0 & 0\degr--150\degr & [|85.5|,|-74.0|]\\

2015-05-12   & \multicolumn{8}{c}{data of too low SNR or compromised - not used}  \\

2015-10-13   & 10.0$\pm$0.5 & 10.0$\pm$2.0 & complete & [-10.0,10.0] & 10.0$\pm$1.0 & 12.0$\pm$1.0 & 0\degr--140\degr & [|10.0|,|-7.7|]\\
             &                &             & & & 32.0$\pm$2.0 & 14.0$\pm$2.0 & 0\degr--140\degr & [|32.0|,|-24.5|]\\
             &                &             & & & 62.0$\pm$2.0 & 16.0$\pm$2.0 & 0\degr--140\degr & [|62.0|,|-47.5|]\\
             &                &             & & & 91.0$\pm$5.0 & 9.0$\pm$2.0 & 0\degr--140\degr & [|91.0|,|-69.7|]\\

2016-04-13   & 9.5$\pm$0.5 & 10.0$\pm$1.0 & complete & [-9.5,9.5] & 10.0$\pm$1.0 & 12.0$\pm$1.0 & 0\degr--140\degr & [|10.0|,|-7.7|]\\
             &                &             & & & 33.0$\pm$1.0 & 14.0$\pm$2.0 & 0\degr--140\degr & [|33.0|,|-25.3|]\\
             &                &             & & & 66.0$\pm$2.0 & 16.0$\pm$2.0 & 0\degr--140\degr & [|66.0|,|-50.6|]\\
             &                &             & & & 90.0$\pm$5.0 & 9.0$\pm$2.0 & 0\degr--140\degr & [|90.0|,|-68.9|]\\
\hline 
\end{tabular}

\flushleft{ 
Note: * We are not certain about the existence of this ring as it is at the limit of what we can fit. It could also be due to the presence of the sky emission line (at \lam6300.3). Unfortunately, there is no sky spectrum available to help us remove or identify the contribution of the sky line. 
}

\end{table*}

\begin{table*}
\caption{Fitting the kinematics of Hen 3-298.}
\label{tab:kin-hen3-298}
\begin{tabular}{l ccc ccc ccc}
\hline
Date (UT)    &  \multicolumn{3}{c}{\oi \lam5577} & \multicolumn{3}{c}{\caii \lam7291} & \multicolumn{3}{c}{\oi \lam6300} \\
             & $v_{\rm rot}$  &  $v_g$  & range & $v_{\rm rot}$ &  $v_g$ & range & $v_{\rm rot}$  &  $v_g$ & range\\
             & (\kms) & (\kms) & (\kms) & (\kms) & (\kms) & (\kms) & (\kms) & (\kms) & (\kms)\\  
\hline
2005-04-19   & 22.5$\pm$1.5 & 8$\pm$1.5  & [-22.5,22.5] &  22.0$\pm$0.5 & 8.0$\pm$0.5 & [-22.0,22.0] & 19.0$\pm$0.5 & 12.0$\pm$0.5 & [-19.0,19.0]\\
2015-05-11   & \multicolumn{3}{c}{too noisy - not used} & 21.5$\pm$0.5 & 7.75$\pm$0.5 & [-21.5,21.5] & 19.0$\pm$0.5 & 12.0$\pm$1.0 & [-19.0,19.0]\\
2015-11-26   & \multicolumn{3}{c}{too noisy - not used} & 21.0$\pm$0.5 & 7.0$\pm$0.5 & [-21.0,21.0] & 18.5$\pm$0.5 & 12.0$\pm$1.0 & [-18.5,18.5]\\
2015-12-06   & 22.5$\pm$1.0 & 10.5$\pm$1.5 & [-22.5,22.5] & 21.5$\pm$0.5 & 7.25$\pm$0.25 & [-21.5,21.5] & 18.5$\pm$0.5 & 12.5$\pm$0.5 & [-18.5,18.5]\\
2016-01-12   & 24.5$\pm$2.0 & 10.5$\pm$1.5 & [-24.5,24.5] & 21.5$\pm$0.5 & 7.75$\pm$0.25 & [-21.5,21.5] & 18.5$\pm$0.5 & 12.25$\pm$0.25 & [-18.5,18.5]\\
\hline
\end{tabular}

\flushleft{
Note: All rings are complete rings.
}

\end{table*}


\bsp	
\label{lastpage}
\end{document}